\documentclass[12pt]{article}
\pdfoutput=1
\usepackage{epsf,amsmath,amssymb,graphicx,dcolumn}
\usepackage{scalefnt,cite,hyperref}
\usepackage{cleveref,etoolbox,color,soul}
\usepackage{listings}
\usepackage[utf8]{inputenc}
\usepackage{booktabs, colortbl}
\usepackage{multirow,dsfont}
\usepackage{tensor}
\usepackage{units}

\usepackage{subcaption}

\Crefname{figure}{Fig.}{Figs.}
\usepackage{placeins}
\usepackage{soul}

\newcommand{\citere}[1]{Ref.\,\cite{#1}}
\newcommand{\citeres}[1]{Refs.\,\cite{#1}}
\newcommand{\Tr}{\text{Tr}}
\newcommand{\abbrev}{\scalefont{1}}
\newcommand{\eqn}[1]{Eq.\,(\ref{#1})}
\newcommand{\eqns}[1]{Eqs.\,(\ref{#1})}
\newcommand{\fig}[1]{Fig.\,\ref{#1}}

\newcommand{\sct}[1]{Section~\ref{#1}}
\newcommand{\eft}{{\abbrev EFT}}
\newcommand{\sm}{{\abbrev SM}}
\newcommand{\lep}{{\abbrev LEP}}

\newcommand{\MM}{\mathcal{M}}
\newcommand{\Wop}[1]{{\rm {#1}}}
\newcommand{\onehalf}{\nicefrac{1}{2}}

\newcounter{notecount}
\def\bra{\langle}
\def\ket{\rangle}

\begin{document}
\thispagestyle{empty}
\begin{flushleft}
\today
\end{flushleft}

\begin{flushright}
  {\tt KA-TP-06-2021}\\
  {\tt P3H-21-024}
\end{flushright}

\long\def\symbolfootnote[#1]#2{\begingroup%
\def\thefootnote{\fnsymbol{footnote}}\footnote[#1]{#2}\endgroup}

\vspace{0.1cm}

\begin{center}
\Large\bf\boldmath
Effective field theory versus UV-complete model: \\
vector boson scattering as a case study
\unboldmath
\end{center}
\vspace{0.05cm}
\begin{center}
Jannis Lang, %$^a$,
Stefan Liebler, %$^a$,
Heiko Sch\"afer-Siebert, %$^{a}$,
Dieter Zeppenfeld%$^a$
\\[0.4cm]
{\small
Institute for Theoretical Physics (ITP), Karlsruhe Institute of Technology,\\ D-76131 Karlsruhe, Germany\\[0.2em]
}
\end{center}
\vspace*{1mm}
\begin{abstract}
\noindent
Effective field theories (EFT) are commonly used to parameterize effects of
BSM physics in vector boson scattering (VBS).
For Wilson coefficients which are large enough to produce presently 
observable effects, the validity range of the EFT represents only a
fraction of the energy range covered by the LHC, however. In order to
shed light on possible extrapolations into the high energy region, a class of
UV-complete toy models, with extra SU(2) multiplets of
scalars or of fermions with vector-like weak couplings, is considered.
By calculating
the Wilson coefficients up to energy-dimension eight, and full
one-loop contributions to VBS due to the heavy multiplets,
the EFT approach, with and without unitarization at high energy, is compared
to the perturbative prediction. For high multiplicities, e.g. nonets of
fermions, the toy models predict sizable effects in
transversely polarized VBS, but only outside the validity range of the EFT.
At lower energies, dimension-eight operators are needed for an adequate
description of the models, providing another example that dimension-eight 
can be more important than dimension-six operators.
A simplified VBFNLO implementation is used to estimate sensitivity of VBS
to such BSM effects at the LHC. Unitarization captures qualitative features
of the toy models at high energy but significantly underestimates signal
cross sections in the threshold region of the new particles.
\end{abstract}

\setcounter{footnote}{0}

\newpage

\pagenumbering{arabic}

%%%%%%%%%%%%%%%%%%%%%%%%%%%%%%%%%%%%%%%%%%%%%%%%%%%%%%%%%%%%%%%%%%%%%%%%%%%%%%%
%%%%%%%%%%%%%%%%%%%%%%%%%%%%%%%%%%%%%%%%%%%%%%%%%%%%%%%%%%%%%%%%%%%%%%%%%%%%%%%

\section{Introduction}
\label{sec:introduction}
%\vspace*{-2.5mm}

With over 300~fb$^{-1}$ of data collected at the LHC, vector boson scattering
processes (VBS) have become important testing grounds for the
non-abelian structure of electroweak interactions, as described by the
Standard Model (\sm{}). While early measurements of the CMS and ATLAS
collaborations firmly established the presence of these electroweak
$qq\to qqVV$ signals~\cite{Khachatryan:2014sta,Aaboud:2016ffv},
at a level as predicted by the \sm{}, the focus is now shifting to precise
comparisons between data and theoretical predictions and the search for
possible signals beyond the \sm{}. Possible deviations in the three- and
four-vector-boson-couplings as compared to \sm{} predictions, so called
anomalous triple gauge couplings (aTGC) or anomalous quartic gauge couplings
(aQGC) have received particular attention~\cite{Aaboud:2016ffv,Sirunyan:2020gyx}.

Anomalous couplings are conveniently described within an effective field
theory (EFT)
framework~\cite{Buchmuller:1985jz,Hagiwara:1993ck,Grzadkowski:2010es,Eboli:2006wa}
with operators of energy dimension $d=6$ and 8 such as 
\begin{align}\label{eq:EFTintro}
\mathcal{L}_{EFT} &= \sum_{d=6}^\infty \sum_{i} \frac{{f_i}^{(d)}}{\Lambda^{d-4}}{O_i}^{(d)} = \sum_{i} \frac{{f_i}^{(6)}}{\Lambda^2}{O_i}^{(6)} +\sum_{i} \frac{{f_i}^{(8)}}{\Lambda^4}{O_i}^{(8)} + ... \nonumber\\
 &=\frac{f_{WWW}}{\Lambda^2}\Tr\left(\tensor{\hat{W}}{^\mu_\nu}\tensor{\hat{W}}{^\nu_\rho}\tensor{\hat{W}}{^\rho_\mu}\right) + \dots \nonumber\\
 &+\frac{f_{T_0}}{\Lambda^4}\Tr\left(\hat{W}^{\mu\nu}\hat{W}_{\mu\nu}\right) \Tr\left(\hat{W}^{\alpha\beta}\hat{W}_{\alpha\beta}\right)+ \dots \nonumber\\
&+\frac{f_{M_0}}{\Lambda^4} \hbox{Tr}\left [ {\hat{W}}_{\mu\nu} {\hat{W}}^{\mu\nu} \right ]
  \times  \left [ \left ( D_\beta \Phi \right)^\dagger
    D^\beta \Phi \right ]  
  + \dots \nonumber\\
&+\frac{f_{S_0}}{\Lambda^4}\left [ \left ( D_\mu \Phi \right)^\dagger
      D_\nu \Phi \right ] \times
    \left [ \left ( D^\mu \Phi \right)^\dagger
      D^\nu \Phi \right ] \, + \dots .
\end{align}
Here $\Phi$ represents the Higgs doublet field ($\Phi=(0,v+h)/\sqrt{2}$
in the unitary gauge), ${\hat{W}}_{\mu\nu}$ is
the SU(2) field strength tensor, and the scale of new physics is generically
called $\Lambda$. For a process with typical momentum transfer $Q^2$, deviations
from SM amplitudes will generally scale like $(Q/\Lambda)^{d-4}$
or  $(v/\Lambda)^{d-4}$ within the validity range of the EFT, which is set
by $Q^2<\Lambda^2$, and, thus, dimension-6 operators are expected to dominate
at the LHC. The dynamics of the new physics may, however, suppress
the Wilson coefficients $f_i/\Lambda^2$ of dimension-6 operators 
such that the leading effects arise from dimension-8 operators. A simple
example is the Euler-Heisenberg Lagrangian~\cite{Heisenberg:1935qt},
where a term analogous to $f_{T_0}$ in \eqn{eq:EFTintro} exists, due
to the electron loop producing a 4-photon
interaction, while a $f_{WWW}$-like term is forbidden by Furry's theorem, i.e.
the fact that the photon is C-odd. As we will see later, such a suppression
of aTGCs vs. aQGCs, due to Furry's theorem, does indeed occur for simple
extensions of the SM.

A second consideration is whether the operators in the EFT Lagrangian are loop
induced or whether they may occur due to tree level effects of the underlying
UV-complete model. While terms such as $f_{S_0}$ in \eqn{eq:EFTintro} can
arise from tree level scalar exchange in e.g. a two-Higgs-doublet model, the
appearance of field strength tensors in the effective Lagrangian generally
points to a loop origin of the effective interaction. The concomitant
suppression by loop factors $g^2/(16\pi^2)$ tends to substantially reduce
the phenomenological importance of such operators~\cite{Arzt:1994gp}, at least
when they are competing with SM tree level effects.\footnote{The impact of
  tree-level competition is
  exemplified by the $f_{WWW}$ aTGC term in  \eqn{eq:EFTintro}, which
  has not been observed to date, as compared to the Higgs coupling to gluons,
  which, in spite of being loop induced, is responsible for the bulk of Higgs
  production at the LHC.}
One might thus be tempted to ignore any loop induced operators in the study
of VBS, arguing that loop induced effects are too small to be observed
above sizable SM contributions to VBS. We will show that this assumption, while
warranted for a large class of models, can be false when high multiplicities
of the additional fields occur in the UV physics.

New charged particles are expected to first appear in direct pair production
at colliders, or, indirectly, in the simplest $n$-point functions which can
be probed. Already three decades ago, de Rujula and collaborators
asked ``Does LEP-1 obviate LEP-2?''~\cite{DeRujula:1991ufe}, i.e. does the
non-observation of new physics in 2-point functions, via the $S$, $T$, $U$
parameters of LEP-1, exclude the observation of anomalies in $W$-pair
production at LEP-2. Translated to the present, the question arises whether
the measurement of quartic couplings in VBS can lead to a first BSM signal,
even though the $\bar qq\to VV$ vector boson pair production processes are
observed at the LHC with much higher statistics and are also known better
theoretically, with the availability of e.g. NNLO QCD
corrections~\cite{Cascioli:2014yka,Gehrmann:2014fva,Grazzini:2016swo,Grazzini:2016ctr,Grazzini:2017ckn}.
An additional scalar resonance would be an obvious
example for the power of VBS. However, we will show that for large isospin
multiplets also purely transverse aQGCs, like the $f_T$ term
in  \eqn{eq:EFTintro}, may be more sensitive probes of new physics
than aTGCs, because aQGCs are enhanced by an additional factor of
isospin-squared. Of course, very large isospin representations will 
push models out of the perturbative domain, a tension which we will probe by
unitarity considerations.

Typically, Wilson coefficients of dimension-8 operators, which are large
enough to be observable in present data, will produce scattering amplitudes
which exceed unitarity bounds within the energy range of the LHC. Not taking
into account the resulting unitarity limits on cross sections results in
overly stringent constraints on Wilson coefficients. Accordingly, recent LHC
analyses take unitarity constraints into account in their extraction of limits
on aQGC (see e.g. Refs.~\cite{Aaboud:2016ffv,Sirunyan:2020gyx}). Unfortunately,
unitarization procedures for VBS amplitudes such as the T-matrix approach of
Refs.~\cite{Kilian:2014zja,Kilian:2015opv,Brass:2018hfw} or the $T_u$-model
of Ref.~\cite{Perez:2018kav} are arbitrary to a significant degree, introducing
model dependence in the measurement of the Wilson coefficients of an EFT.
A comparison of unitarized cross sections with predictions of UV-complete
toy models can help to make the unitarization procedures more realistic. 

In this paper we address the above questions by confronting the EFT
description of the transverse, dimension-6 $f_{WWW}$- and dimension-8 $f_T$-type
operators
in \eqn{eq:EFTintro} with a fairly generic class of UV-complete models
with additional matter fields in arbitrary weak isospin representations.
The model, given by the addition to the SM of $(2J_R+1)$-dimensional SU(2)
multiplets of heavy scalars and/or heavy non-chiral Dirac fermions, is further
specified in Section~\ref{sec:model}. At the one-loop level, which is
summarized in Section~\ref{sec:oneloop}, its low energy manifestation is
described by an EFT Lagrangian involving (derivatives of) SU(2) field
strengths, i.e. only transverse operators contribute.
The relevant operators are introduced in Section~\ref{sec:Wilson68}.
In Section~\ref{sec:coefficients} we present the
Wilson coefficients of all EFT terms up to energy dimension eight. Many
of them have been measured in various experiments and the existing bounds
are translated into exclusions of $(J_R,M_R)$ combinations
in \sct{sec:WCbounds}, with $M_R$ denoting the mass of the heavy
scalars or fermions.
The resulting amplitudes for VBS processes diverge at high energy,
proportional up to $s^2/M_R^4$ where $\sqrt{s}$ is the c.m. energy of
the $VV\to VV$ VBS process.
In order to assess the validity range of the EFT description,
we also calculate, in Section~\ref{sec:fulloneloop}, the full
one-loop corrections
to the $VV\to VV$ scattering amplitudes which arise from the heavy new
matter fields. Performing a partial wave analysis, their trajectory
around the Argand circle is analyzed in \sct{sec:hel_proj}
and we are thus able to specify for
which isospin $J_R$ a perturbative description is still justified.
At the limit of this perturbative range, we compare the full
one-loop corrected cross sections for $VV\to VV$ scattering with the
EFT approximation in \sct{sec:VBS-XS}. We find
good agreement of the two approaches only for $\sqrt{s}\lesssim 1.3M_R$,
as is to be expected.
  
While the above results are obtained in a simplified setting, neglecting
the hypercharge coupling $g'$ and considering on-shell scattering only,
an approximate implementation for the full $qq\to qqVV$ processes within
the {\tt VBFNLO} Monte Carlo program~\cite{Arnold:2008rz,Baglio:2014uba}
is presented in Section~\ref{sec:VBFNLO}. On the theoretical side, this
allows for a more realistic comparison of the full model with its EFT
approximation, at dimension-8 level, and with the unitarized replacement
of the EFT in the $T_u$ model of Ref.~\cite{Perez:2018kav}. Also here,
good agreement between the three
approaches is found for $\sqrt{s}<1.3M_R$, very large deviations
are seen in the threshold region for pair production of the heavy
particles ($M_R<\sqrt{s}<2M_R$), and qualitatively acceptable agreement
of the $T_u$-model with the full one-loop corrected results occurs
well above pair production threshold. The {\tt VBFNLO} implementation
also allows to compare the expected VBS signals of high isospin
fermion or scalar multiplets with LHC data. Finally, conclusions are
presented in Section~\ref{sec:conclusions}.

\section{The toy model and its EFT}
\label{sec:EFT}

As a set of minimal, ultraviolet-complete models, we consider the addition 
of heavy fermionic or scalar SU$(2)_L$ multiplets to the \sm{}. These new
degrees of freedom are assumed to couple through the SU$(2)_L$ gauge
interaction only, i.e. they are color singlets and have vanishing hypercharge.
In addition, they are assumed to have negligible Yukawa or $\Phi^4$ couplings
to the \sm{} Higgs doublet field and, thus, they do not mix with the other
particles of the \sm{}. 

\subsection{Definition and description of the toy model}
\label{sec:model}

To be specific, and in order to fix the notation, we consider Dirac
fermions~$\Psi$ with mass $M_F$ or complex scalars~$\Phi$ with mass $M_S$,
which transform under a generic SU$(2)_L$ representation~$R$ described by
the generators $t_R^a$. %, which obey $[t_R^a,t_R^b]=i\epsilon^{abc}t_R^c$. 
The representation $R$ is specified by its isospin, $J_R$, i.e. each
multiplet has $(2J_R+1)$ components and $t_R^a t_R^a=J_R(J_R+1)$. 
Since the covariant derivatives of the new, hypercharge-zero fields contain
SU$(2)_L$ gauge fields only, 
\begin{align}
 D_\mu= \partial_\mu + i g t_R^a W^a_\mu\,,
\end{align}
no relevant information will be lost in our loop calculations and EFT
considerations by simplifying the model to a pure SU$(2)_L$ gauge theory,
i.e. we restrict most of the discussion to the SU$(2)_L$ limit of the
electroweak sector of the \sm{}, where the hypercharge coupling $g'$
of U$(1)_Y$ is set to zero.

Within this approximation, the covariant derivative of the Higgs doublet
field is
$ \hat{D}_\mu = \partial_\mu +ig\frac{\tau^a}{2}W_\mu^a$, 
with the Pauli matrices $\tau^a$, and the SU$(2)_L$ field-strength tensor
is defined through
\begin{align}
  \hat{W}^{\mu\nu}=\left[\hat{D}^\mu,\hat{D}^\nu\right]=
  ig\frac{\tau^a}{2}\left(\partial^\mu W^{a\nu}-\partial^\nu W^{a\mu}-
  g\epsilon^{abc}W^{b\mu}W^{c\nu}\right) =
  ig\frac{\tau^a}{2}  W^{a\mu\nu}\, .
 \label{eq:fieldstrengthtensor}
\end{align}
Suppressing the \sm{} fermions and the Higgs potential, the relevant toy-model
Lagrangian can be written as
\begin{align}
\mathcal{L}=&\frac{1}{2}\left(\partial_\mu H\right)^2-\frac{m_H^2}{2}H^2
-\frac{1}{2} \Tr\left(\hat{W}^{\mu\nu}\hat{W}_{\mu\nu}\right) + \frac{m_W ^2}{2}
 \left(\sum_{a=1}^3 W^a_\mu W^{a\mu}\right)\left(1+\frac{H}{v}\right)^2\nonumber\\ 
 &+\bar{\Psi}\left(i \gamma_\mu D^\mu -M_F\right) \Psi +
 (D^\mu \Phi)^\dagger (D_\mu \Phi) - M_S{}^2 \Phi^\dagger \Phi \,.
 \label{eq:lag_model}
\end{align}
The first line corresponds to the well-known \sm{} Lagrangian of the gauge and
Higgs sector in the unitary gauge, albeit with equal $W$ and $Z$
masses,\footnote{The absence of a massless photon in our toy model
  and the common $W$ and $Z$ masses significantly simplify the analytical
  results for VBS amplitudes as well as the unitarity considerations in
  Section~\ref{sec:hel_proj}.}
$m_W=m_Z=gv/2$.
The second line of \eqn{eq:lag_model} comprises the new particles and
their interactions with the gauge fields, which give rise to familiar
expressions for the Feynman rules. 
Our loop calculations will involve internal lines of only
the new fermions or scalars. Thus, results in the unitary gauge, outlined above,
are identical to those of a more general $R_\xi$ gauge.
The new fermions' gauge couplings are assumed to be non-chiral and therefore their
mass, $M_F$, can be chosen arbitrarily (and large) prior to spontaneous symmetry
breaking of the SU$(2)_L$. In principle our toy model can host various fermionic
and scalar multiplets without mass mixing. However, this will be irrelevant for our
phenomenological discussion.

Our toy model is closely related to new-physics solutions of problems that remain
open in the \sm{}, like an explanation of neutrino masses or dark matter:
Large fermionic and scalar SU$(2)_L$ multiplets up to quintets that couple among each other and
to the lepton sector of the \sm{} are known to provide an explanation of neutrino
masses~\cite{Kumericki:2012bh,Law:2013gma,Yu:2015pwa,Nomura:2016jnl,Wang:2016lve,Nomura:2017abu,Nomura:2018cle}.
However, we are interested in the impact of such SU$(2)_L$ multiplets on vector boson scattering only and thus
allow the new particles to couple to the \sm{} only through the gauge coupling of SU$(2)_L$.
Such large multiplets can also contain a dark matter candidate, as the lightest
particle in the multiplet spectrum can be stable. 
Our model coincides with a class of minimal dark matter
models \cite{Cirelli:2005uq,AbdusSalam:2013eya,Hambye:2009pw}, which range from
SU$(2)_L$ triplets up to septets (see \citere{Chao:2018xwz} for a review).

Embedding the heavy matter fields into the full \sm{}, i.e. including the U$(1)_Y$ gauge
field, the mass degeneracy of the additional SU$(2)_L$ multiplets at lowest-order in perturbation
theory is lifted at the one-loop level, due to the mass splitting between photon, $W$ and $Z$.
As an example, and in agreement with the literature~\cite{Cirelli:2005uq}, for a $J_F=4$ fermion
nonet of zero hypercharge, the differences in mass of the charged states with respect to the neutral
state are given by
$\Delta m^{4\pm} = m_{\Psi^{4\pm}}-m_{\Psi^0}=2.7$\,GeV, $\Delta m^{3\pm}=1.5$\,GeV,
$\Delta m^{2\pm} = 0.66$\,GeV and $\Delta m^{\pm}=0.166$\,GeV.
Thus, cascade decays from the heavier, charged states to the neutral, stable
state proceed through far off-shell gauge bosons that result in very soft leptons and pions.
If on the one hand the decay length of the charged states is short enough that they decay within
the interaction region and do not result in a charged track in the detector, but on the other hand
the final state leptons and pions are still soft enough, then such dark matter models are very hard to 
be probed at the LHC.

A thorough discussion of the current exclusions of a minimal dark matter
quintet, based on LHC searches, can be found in
\citere{Ostdiek:2015aga}, including projections for the high-luminosity runs.
Expected exclusion bounds range to quintet masses up to $750$\,GeV for
integrated luminosities of $3$\,ab$^{-1}$. Going beyond quintets, i.e. beyond
$J_R=2$, results in the existence of additional charged states, that are
pair-produced with large cross sections that rise proportional to
$J_R(J_R+1)$~\cite{Cirelli:2005uq}.
It is also known that the minimal dark matter model needs masses of the dark
matter candidate in the range of $10$\,TeV~\cite{Cirelli:2009uv} in order to
achieve the correct relic abundance, such that for lighter
masses, as discussed in this paper, only a fraction of the observed
abundance can be explained by the neutral, stable particle
of the minimal dark matter model.
Alternatively, extensions with additional singlet
states~\cite{Bharucha:2018pfu} are discussed in the literature,
which, however, change the collider phenomenology substantially without
impacting too much the discussion carried out in this paper.
Because the collider constraints can always be avoided by somewhat increasing
the mass parameters $M_F$ or $M_S$, we omit a more detailed discussion of
direct exclusion limits from existing collider searches, as this
is not the focus of the present paper.

\subsection{One-loop corrections to gauge-boson vertices, isospin factors
  and renormalization}
\label{sec:oneloop}

The additional SU$(2)_L$ multiplets contribute to various observables at
the one-loop level. Here we focus on contributions to VBS, which enter
through propagator, three- and four-boson vertex corrections. In practice
we use {\tt FeynCalc}~\cite{Mertig:1990an,Shtabovenko:2016sxi} for the handling
of Dirac structures, we regularize ultraviolet divergences through dimensional
regularization and express our findings in terms of Passarino-Veltman
integrals~\cite{tHooft:1978jhc,Passarino:1978jh} to obtain numerical results.
We refrain from showing the explicit, rather lengthy expressions for the
corrections to the three- and four-gauge-boson vertices.\footnote{The full
  amplitudes are implemented in a code named {\tt VeBoS} that is described
  in \citere{Master-Thesis} and can be obtained upon request from JL.}
Instead, we here concentrate on the general structure of the one-loop
contributions, including their representation dependence.

We start with the vacuum polarization. For the fermion case only a single
diagram
contributes. For the scalar case a second diagram involving the four-particle
vertex needs to be added. As the propagator corrections are purely transverse,
they generically take the form
\begin{align}
  \vcenter{\hbox{\includegraphics[page=2,scale=0.8]{figures/Renormalization2W}}} + \vcenter{\hbox{\includegraphics[page=3,scale=0.8]{figures/Renormalization2W}}}=
  i \delta^{ab}\left(p^2 g^{\mu\nu}-p^\mu p^\nu\right)T_S \Pi_{S}(p^2,M_S^2)\,
  \label{eq:propagator}
\end{align}
for the scalar case and similarly, replacing $T_S \Pi_{S}(p^2,M_S^2)$ by
$T_F \Pi_{F}(p^2,M_F^2)$ for the fermion loop.
For the diagrams involving two three-particle vertices the isospin factor
is $\Tr\left(t_R ^a t_R ^b\right)=T_R \delta^{ab}$ 
while the scalar diagram with the four-particle vertex has
$\Tr\left(\{t_R ^a ,t_R ^b\}\right)=2 T_R \delta^{ab}$.
Therein $T_R=T_F$ or $T_S$ is the index of the representation, given by
$T_R=\frac{1}{3}\left[J_R(J_R+1)(2J_R+1)\right]$ for an
isospin $J_R$ representation.
The explicit form of the propagator corrections is given by
\begin{align}\nonumber
 \Pi_{F}(p^2,M_F^2)=-\frac{g^2}{16\pi^2}\frac{4}{9p^2}\left[6M_F^2-p^2-6A_0(M_F^2)+3(2M_F^2+p^2)B_0(p^2,M_F^2,M_F^2)\right]\,,\\
  \Pi_{S}(p^2,M_S^2)=\frac{g^2}{16\pi^2}\frac{1}{9p^2}\left[12M_S^2-2p^2-12A_0(M_S^2)+3(4M_S^2-p^2)B_0(p^2,M_S^2,M_S^2)\right]\,.
  \label{eq:VacPol}
\end{align}
These expressions can be expanded in small momenta $p^2$ motivated
by $p^2\sim m_W^2\ll M_F^2,M_S^2$. As we employ dimensional regularization,
working in $d=4-2\epsilon$ dimensions, we parameterize the divergences
in terms of $\Delta_\epsilon=\frac{1}{\epsilon}-\gamma_E+\log(4\pi)$.
The propagator corrections then turn into
\begin{align}\nonumber
 \Pi_{F}(0,M_F^2)=-\frac{g^2}{16\pi^2}\frac{4}{3}\left[\Delta+\log\left(\frac{\mu^2}{M_F^2}\right)\right]\,,\\
 \Pi_{S}(0,M_S^2)=-\frac{g^2}{16\pi^2}\frac{1}{3}\left[\Delta+\log\left(\frac{\mu^2}{M_S^2}\right)\right]\,,
 \label{eq:vacuumpol}
\end{align}
where $\mu$ denotes the renormalization scale. It is therefore clear that
choosing an $\overline{\text{MS}}$ scheme with $\mu=M_F,M_S$ leads
practically to an on-shell scheme with almost vanishing propagator corrections,
such that electroweak precision data are not significantly impacted
by the extra degrees of freedom.

We now turn our attention to the three-boson vertex.
The result is generically of the form
\begin{align}
  \left(\vcenter{\hbox{\includegraphics[page=1,scale=0.7]{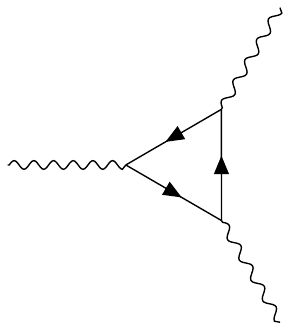}}}+\text{permutations}\right)=
  i \epsilon^{abc}T_F\Gamma_{3,F} ^{\mu\nu\rho}(p_1,p_2,p_3,M_F ^2)\, 
\end{align}
for heavy fermions, and analogously for scalars. Here ``permutations''
of the external gauge bosons refers to the Feynman graph with
the opposite direction of the fermion arrow.
For the three-boson vertex correction in the scalar case, all diagrams
involving the $WW\Phi\Phi$ seagull vertex vanish individually because,
among other reasons, their isospin factor
$\Tr\left(t_R ^a \{t_R ^b,t_R ^c\}\right)$ is zero. 
The sum of the depicted diagrams, on the other hand, carries an isospin
factor $\Tr\left(t_R ^a [t_R ^b, t_R ^c]\right)=iT_R \epsilon^{abc}$. 

For the scalar case, the one-loop corrections to the four-gauge boson vertex
take the form
\begin{align}
&\left(\vcenter{\hbox{\includegraphics[page=3,scale=0.8]{figures/Renormalization4W}}}
 +\vcenter{\hbox{\includegraphics[page=4,scale=0.8]{figures/Renormalization4W}}}
 +\vcenter{\hbox{\includegraphics[page=2,scale=0.8]{figures/Renormalization4W}}}
 +\text{permutations}\right)\nonumber\\
 &+\left(\vcenter{\hbox{\includegraphics[page=5,scale=0.8]{figures/Renormalization4W}}}+\text{$t$- and $u$-type}\right)=
 \tilde{\Gamma}_{4,S} ^{\mu\nu\alpha\beta,abcd}(p_1,p_2,p_3,p_4,M_S ^2,J_S)\,,\label{eq:boxSc}
\end{align}
while the corresponding heavy fermion contribution,
$\tilde{\Gamma}_{4,F} ^{\mu\nu\alpha\beta,abcd}(p_1,p_2,p_3,p_4,M_F ^2,J_F)$
is given by a box-diagram analogous to the one shown, and its permutations.
Above, the three permutations for the scalar one-loop graphs involving two
four-particle vertices are labeled $s$-, $t$- and $u$-type diagram.
For all diagrams we get traces over four generators, which generically result in
\begin{align}
  \Tr\left(t_R ^a t_R ^b t_R ^c t_R ^d\right) = &\frac{1}{15}T_R(3 C_{2,R} -1)
  \left(\delta^{a b}\delta^{c d}+\delta^{a c}\delta^{b d}+\delta^{a d}\delta^{b c}\right)\nonumber\\
  &+\frac{1}{6} T_R \left(\delta^{a b}\delta^{c d}-2\delta^{a c}\delta^{b d}+\delta^{a d}\delta^{b c}\right)\,.
 \label{eq:box_RepFactor}
\end{align}
Here, $T_R=\frac{1}{3}\left[J_R(J_R+1)(2J_R+1)\right]$ is the index, and
$C_{2,R}=J_R(J_R+1)$ is the quadratic casimir of a representation.
\eqn{eq:box_RepFactor} separates the isospin symmetric part proportional to
$T_R(3 C_{2,R} -1)$, which is the only contribution to
pure $ZZ\rightarrow ZZ$ scattering, from terms coming from products of $\epsilon$ tensors.
Alternatively, all one-loop corrections can be sorted into terms proportional to $T_R$ only
and terms proportional to $C_{2,R}T_R$, which provides the full isospin dependence.
As a consequence, numerical evaluations can be performed for a fixed mass choice and can then easily be
reweighted to any representation~$R$. We use this procedure in
{\tt VeBoS} and our implementation in {\tt VBFNLO}.

We now turn our attention to the renormalization of ultraviolet divergences.
Since we are only considering $n$-point functions of the SU$(2)_L$ gauge fields
which are transverse, as in \eqn{eq:propagator}, only the
$\Tr\left(\hat{W}^{\mu\nu}\hat{W}_{\mu\nu}\right)$ term in \eqn{eq:lag_model}
can produce the necessary counter terms, via the renormalization of
the gauge fields and the coupling constant. Denoting the renormalization
constants between the bare Lagrangian parameters
(with subscript $0$) and the physical parameters as 
\begin{align}\nonumber
 W_0^{a\mu}=\sqrt{Z_3}W^{a\mu}, \qquad g_0=Z_gg\,,
\end{align}
the renormalization constants of the three and four gauge-boson vertices are
given by $Z_{3W}=Z_gZ_3^{3/2}$ and $Z_{4W}=Z_g^2Z_3^2$, respectively.
Writing $Z_i=1+\delta_i$, we obtain $\delta_{3W}=\delta_g+\frac{3}{2}\delta_3$
and $\delta_{4W}=2\delta_g+2\delta_3$, which defines the counter-terms for the three
and four gauge-boson vertices, respectively. The counter-term of the propagator
reads $\delta_{2W}=\delta_3$. It directly follows from the vacuum polarization
in \eqn{eq:vacuumpol} that in the $\overline{\text{MS}}$ scheme the field
renormalization constant $\delta_3$ is given by
\begin{align}
  \delta_3 = -g^2 \Delta_\epsilon \left(\sum_F n_F \frac{T_F}{12\pi^2} +
  \sum_S n_S \frac{T_S}{48\pi^2}\right)\,,
\end{align}
assuming $n_F$ and $n_S$ copies of a fermionic and scalar multiplet with isospin $J_F$
and $J_S$, respectively.
From the ultraviolet divergence of the three gauge-boson vertex we obtain
\begin{align}
  \left(\delta_g+\frac{3}{2}\delta_3\right)
  &= -g^2 \Delta_\epsilon \left(\sum_F n_F \frac{T_F}{12\pi^2}  +
                               \sum_S n_S \frac{T_S}{48\pi^2}\right) = \delta_3
\end{align}
and, therefore, the renormalization constant of the SU$(2)_L$ coupling is given
by $\delta_g=-\frac{1}{2}\delta_3$. It is easy to check that indeed the now fully
determined counter term to the four gauge-boson vertex, namely
$\delta_{4W}=2(\delta_g+\delta_3)=\delta_3$, also cancels the divergence of the
corresponding one-loop corrections to gauge-boson four-point functions.

\subsection{Effective-field theory description}
\label{sec:Wilson68}

Considering the low-energy expansion of the now finite two-, three- and
four-point functions of the gauge bosons, we can derive the
contributions of additional heavy SU$(2)_L$ multiplets to the EFT
Lagrangian 
\begin{align}
  \mathcal{L}_{EFT}=\sum_{d=6}^\infty \sum_{i} \frac{{f_i}^{(d)}}{\Lambda^{d-4}}{O_i}^{(d)}
    \label{eq:EFT_series}\,,
\end{align}
i.e. we can now determine the Wilson coefficients $f_i/\Lambda^2$ and $f_i/\Lambda^4$
of the dimension-6 and dimension-8 operators which are relevant for VBS
in our concrete model.

By construction, the SU$(2)_{L}$ gauge bosons are the only \sm{} particles which couple via 
the heavy fermions or scalars and which can, at the one-loop level, appear
as fields in the effective Lagrangian. Gauge invariance then implies that we
can restrict the EFT Lagrangian to operators which contain the SU$(2)_{L}$ field strength
tensor $\hat{W}^{\mu\nu}$ of \eqn{eq:fieldstrengthtensor} or its commutator with covariant
derivatives. For a description of vector boson scattering we need vacuum polarization
effects (dimension-4 operators and higher), anomalous triple-gauge
couplings (dimension-6 and -8 operators) and anomalous quartic-gauge couplings
(dimension-8 operators).

The dimension-4 operator is unique and given by 
\begin{align}
O_{WW}&=\Tr\left(\hat{W}^{\mu\nu}\hat{W}_{\mu\nu}\right)\,.
\end{align}
It is already part of the \sm{} kinetic term. Since
it receives contributions from the additional SU$(2)_L$ multiplets
it is explicitly listed here. However, as discussed in the last section,
it disappears as a result of renormalization.
As a basis for the dimension-6 operators, following~\citere{Hagiwara:1993ck}, we choose 
\begin{align}
  O_{WWW}&=\Tr\left(\tensor{\hat{W}}{^\mu_\nu}\tensor{\hat{W}}{^\nu_\rho}\tensor{\hat{W}}{^\rho_\mu}\right)\,,  \nonumber
\\
  O_{DW}&=\Tr\left([\hat{D}_\alpha,\hat{W}^{\mu\nu}][\hat{D}^\alpha,\hat{W}_{\mu\nu}]\right)\,.
\label{eq:dim6operators}
\end{align}

Our focus is on the dimension-8 operators which first appear as aQGC in
vector boson scattering, and which involve four field-strength tensors, namely 
\begin{align}\nonumber
O_{T_0}&=\Tr\left(\hat{W}^{\mu \nu}\hat{W}_{\mu\nu}\right)\Tr\left(\hat{W}^{\alpha \beta}\hat{W}_{\alpha\beta}\right)\,,\\\nonumber
O_{T_1}&=\Tr\left(\hat{W}^{\mu \nu}\hat{W}_{\alpha\beta}\right)\Tr\left(\hat{W}^{\alpha \beta}\hat{W}_{\mu\nu}\right)\,,\\\nonumber
O_{T_2}&=\Tr\left(\hat{W}^{\mu\nu}\hat{W}_{\nu\alpha}\right)\Tr\left(\hat{W}^{\alpha\beta}\hat{W}_{\beta\mu}\right)\,,\\
O_{T_3}&=\Tr\left(\hat{W}^{\mu \nu}\hat{W}^{\alpha\beta}\right)\Tr\left(\hat{W}_{\nu \alpha}\hat{W}_{\beta\mu}\right)\,.
\label{eq:toperators}
\end{align}
Here we follow the notation of \citeres{Eboli:2006wa,Eboli:2016kko,Perez:2018kav,Rauch:2016pai}.
However, since our definition of $\hat{W}^{\mu \nu}$ includes the coupling factor $ig$,
the operators in \eqn{eq:toperators} contain an overall factor $g^4$ as compared
to the {\' E}boli convention of \citeres{Eboli:2006wa,Eboli:2016kko}.
Please note that $O_{T_3}$ is a non-vanishing, linearly independent operator which was
missing in previous discussions and, therefore, is also not experimentally
constrained explicitly. We refer to the four operators in \eqn{eq:toperators}
as $T$-operators.\footnote{Similarly, the operators named
  $O_{T_5}$, $O_{T_6}$ and $O_{T_7}$ in \citere{Eboli:2006wa} should be supplemented by
  $O_{T_4}=\Tr\left(\hat{W}^{\mu \nu}\hat{W}^{\alpha\beta}\right)\hat{B}_{\nu \alpha}\hat{B}_{\beta\mu}$
  when discussing mixed SU$(2)_{L}$ and U$(1)_Y$ gauge boson scattering.}

For three field-strength tensors in combination with covariant derivatives,
seven operators can be written down which are, however, related through 
integration-by-parts for the covariant derivative, the Jacobi identity for
covariant
derivatives, and $\hat{W}^{\mu\nu}=\left[\hat{D}^\mu,\hat{D}^\nu\right]$.
Our set of independent operators is then given by
\begin{align}\nonumber
O_{DWWW_0}&=\Tr\left([\hat{D}_\alpha,\tensor{\hat{W}}{^\mu _\nu}][\hat{D}^\alpha,\tensor{\hat{W}}{^\nu_\rho}]\tensor{\hat{W}}{^\rho_\mu}\right)\,,\\
O_{DWWW_1}&=\Tr\left([\hat{D}_\alpha,\tensor{\hat{W}}{^\mu^\nu}][\hat{D}_\beta,\tensor{\hat{W}}{_\mu_\nu}]\tensor{\hat{W}}{^\alpha^\beta}\right)\,.
\label{eq:DWWWoperators}
\end{align}
For the case of two field-strength tensors in combination with covariant derivatives we pick
\begin{align}
O_{D2W}=\Tr\left([\hat{D}_\alpha,[\hat{D}^\alpha,\hat{W}^{\mu\nu}]][\hat{D}_\beta,[\hat{D}^\beta,\hat{W}_{\mu\nu}]]\right)\,,
\label{eq:D2Woperator}
\end{align}
as, again, other combinations are not linearly independent.

The above choice of basis is by no means unique but simply driven by the
physics of the underlying theory. Looking at the dimension-6 operators
one could make an argument for using the Warsaw Basis~\cite{Grzadkowski:2010es}
as this could make comparisons easier and gives more direct access to
exclusion limits for the Wilson coefficients. Rewriting the $O_{DW}$ operator
using integration-by-parts, the commutator for the covariant derivatives
as well as  the Jacobi identity for covariant derivatives, one arrives
at~\footnote{The operator $O_{WWW}$ is equivalent to the corresponding
  operator in the Warsaw Basis up to a factor, namely,
  $\Wop{O}_W=\epsilon^{I,J,K}W_{\mu}^{I,\nu}W_{\nu}^{J,\rho}W_{\rho}^{K,\mu}=
  \frac{4}{g^3}O_{WWW}$.}
\begin{equation}
  O_{DW}=2 \Tr\left([\hat{D}_\alpha,\hat{W}^{\alpha\mu}][\hat{D}^\nu,\hat{W}_{\nu\mu}]\right)-4\, O_{WWW} \, .
  \label{eq:DWtoWWW}
\end{equation}
At this point we can use the equations of motion for the $W$-field twice:
\begin{equation}
[\hat{D}^\alpha,\hat{W}_{\alpha\mu}]=\frac{i g^2\,\tau^{I}}{4}\left(i \phi^{\dagger} \tau^I D_{\mu} \phi - i \left(D_{\mu}\phi \right)^{\dagger}\tau^I \phi+\bar{l}\gamma_\mu \tau^I l + \bar{q}\gamma_\mu \tau^I q\right) + \mathcal{O}(\frac{1}{\Lambda^2}).
\end{equation} 
Here the last term accounts for corrections to the equations of motion due
to the presence of $O_{DW}$ and $O_{WWW}$ which are of order $\frac{1}{\Lambda^2}$.
This leaves us with four types of terms: operators containing four Higgs fields,
the ones containing fermionic currents, and also terms of order $\frac{1}{\Lambda^2}$
and $\frac{1}{\Lambda^4}$ of which only the latter may be neglected as they are of
order $\frac{1}{\Lambda^6}$ when including the Wilson coefficient of $O_{DW}$.
The terms with four Higgs fields can again be rewritten by using relations
for the SU$(2)_L$ generators as well as the equations of motion for the Higgs
field and one finds
\begin{equation}
  O_{DW}=-g^3\, \Wop{O}_{W}-\frac{3g^4}{4}\Wop{O}_{\phi\square}+
  \frac{g^4 m_H^2}{2}\left(\phi^\dagger \phi\right)^2-
  2g^4\lambda \Wop{O}_\phi+\Wop{O}_{\psi^2\phi^2D}+\Wop{O}_{\psi^4}+
  \mathcal{O}(\frac{1}{\Lambda^2})\, ,
\label{eq:DWtoWS}
\end{equation}
where the operators on the r.h.s., with $\phi$ representing the SM Higgs doublet
field, are written in the Warsaw basis. Here
$\psi$ stands for either leptons or quarks. The last two operators are
just symbolizing the class of additional operators that can appear following
the conventions in Ref.~\cite{Grzadkowski:2010es}. Individual terms in this
new set now give local corrections
to Higgs and fermionic observables that, at the one-loop level, are not present
for the BSM heavy matter fields which we consider. Therefore, these corrections
have to cancel each other when calculating amplitudes for physical processes
in our new physics model. This also implies that any exclusion limits for 
Wilson coefficients are only useful when provided with full error correlations.
With this in mind it becomes apparent that there is no advantage in using the
Warsaw basis over the basis defined
by Eqs.~(\ref{eq:dim6operators})-(\ref{eq:D2Woperator}).

\subsection{Wilson coefficients due to additional SU$(2)_L$ multiplets}
\label{sec:coefficients}
After presenting the full EFT Lagrangian we can map our concrete model
onto the Wilson coefficients at low energies by integrating out the heavy
new degrees of freedom. Again we assume to have $n_F$ and $n_S$ copies of
fermionic and scalar multiplets with isospin $J_F$ and $J_S$, respectively.
We expand the one-loop corrections in small momenta with respect to the
mass scale of the new heavy degrees of freedom, i.e.  $\frac{p_i\cdot p_j}{M ^2}\ll 1$, which 
for the Wilson coefficients originating from propagator corrections results in
\begin{align}
f_{WW}&=%\left[-g^2\right] 
\sum_F n_F \frac{T_F}{24\pi^2}\left(\Delta_\epsilon+\log\left(\frac{\mu^2}{M_F^2}\right)\right) 
+ \sum_S n_S   \frac{T_S}{96\pi^2}\left(\Delta_\epsilon+\log\left(\frac{\mu^2}{M_S^2}\right)  \right)\,,\\
\frac{f_{DW}}{\Lambda^2}&=%\left[-g^2\right]
 \sum_F n_F \frac{T_F}{120\pi^2M_F^2} +\sum_S n_S \frac{T_S}{960\pi^2M_S^2} \,,\\
\frac{f_{D2W}}{\Lambda^4}&=%\left[-g^2\right] 
\sum_F n_F \frac{T_F}{1120\pi^2M_F^4} + \sum_S n_S \frac{T_S}{13440\pi^2M_S^4}  \,.
\end{align}
%The square brackets with powers of $[i g]$ are written such that 
%it is easier to change to a different sign convention for the gauge coupling and the
%field-strength tensor convention $\hat{W}^{\mu\nu}=[\hat{D}^\mu,\hat{D}^\nu]$,
%which in contrast to our definition \eqn{eq:fieldstrengthtensor} comes without $\frac{1}{-ig}$ normalization.
We continue to expand the three gauge boson-vertex corrections in a similar manner in small momenta, which leads to
\begin{align}
\frac{f_{WWW}}{\Lambda^2}&=%\left[i\,g^3\right] 
\sum_F n_F \frac{13 T_F}{360\pi^2 M_F^2} +\sum_S n_S \frac{T_S}{360\pi^2 M_S^2} \,,\\
\frac{f_{DWWW_0}}{\Lambda^4}&=%\left[i\,g^3\right] 
\sum_F n_F \frac{2 T_F}{105\pi^2 M_F^4} + \sum_S n_S \frac{T_S}{1120\pi^2 M_S^4} \,,\\
\frac{f_{DWWW_1}}{\Lambda^4}&=%\left[i\,g^3\right] 
\sum_F n_F \frac{T_F}{630\pi^2 M_F^4} + \sum_S n_S \frac{T_S}{4032\pi^2 M_S^4} \,.
\end{align}
Finally, the Wilson coefficients of the $T$-operators are given by
\begin{align}
\frac{f_{T_0}}{\Lambda^4}&=%\left[g^4\right]
 \sum_F n_F \frac{\left(-14C_{2,F}+1\right) T_F}{10080\pi^2M_F^4} + \sum_S n_S \frac{\left(7C_{2,S}-2\right) T_S}{40320\pi^2M_S^4}  \,,\\
\frac{f_{T_1}}{\Lambda^4}&=%\left[g^4\right]
 \sum_F n_F \frac{\left(-28C_{2,F}+13\right) T_F}{10080\pi^2M_F^4} + \sum_S n_S \frac{\left(14C_{2,S}-5\right) T_S}{40320\pi^2M_S^4}    \,,\\
\frac{f_{T_2}}{\Lambda^4}&=%\left[g^4\right]
 \sum_F n_F \frac{\left(196C_{2,F}-397\right) T_F}{25200\pi^2M_F^4} + \sum_S n_S \frac{\left(14C_{2,S}-23\right) T_S}{50400\pi^2M_S^4}    \,,\\
\frac{f_{T_3}}{\Lambda^4}&=%\left[g^4\right]
 \sum_F n_F \frac{\left(98C_{2,F}+299\right) T_F}{25200\pi^2M_F^4} + \sum_S n_S \frac{\left(7C_{2,S}+16\right) T_S}{50400\pi^2M_S^4}   \,.
\label{eq:fT_coefficients}
\end{align}
As discussed previously, all UV divergences of propagator-, triangle-, and
box-diagrams are contained in the coefficient of the field-strength-square
operator $O_{WW}$ and, hence, disappear upon renormalization of the SU$(2)_{L}$
gauge coupling $g$ and the gauge boson fields.

Some additional remarks are in order:
\begin{itemize}
\item The isospin factors of propagator and three-boson-vertex
  corrections are given by $T_R$ and thus rise with $J_R^3$.
  The Wilson coefficients of the $T$-operators, on the other hand, are
  enhanced by an additional casimir factor, i.e. they rise as $J_R^5$,
  enhancing their importance over the dimension-6 operators for
  high isospin.
\item Individual operators in the scalar case are more suppressed than
  for heavy fermions of the same mass and isospin. 
\item The Wilson coefficients of the $T$-operators $f_{T_0}$ and
  $f_{T_1}$ receive opposite sign contributions from heavy scalars vs.
  heavy fermions, which can induce cancellations between $T$-operator
  contributions to e.g. VBS cross sections. $f_{T_2}$ changes its sign
  at $J_R\sim 1$ for both the fermionic and scalar multiplets.
\end{itemize}

\subsection{Confronting Wilson coefficients with experiment}
\label{sec:WCbounds}

Having obtained the low-energy EFT of the general model, we next investigate
whether specific model realizations are compatible with experimental limits
on four-fermion contact terms, anomalous triple- and anomalous
quartic-gauge couplings. 
Since we are only interested in rough estimates for our toy model, we
restrict the discussion to the single multiplet realization and determine
bounds in the $(J_R,M_R)$ parameter space.

\begin{figure}[tb!]
\begin{center}
\includegraphics[page=2,width=0.50\textwidth]{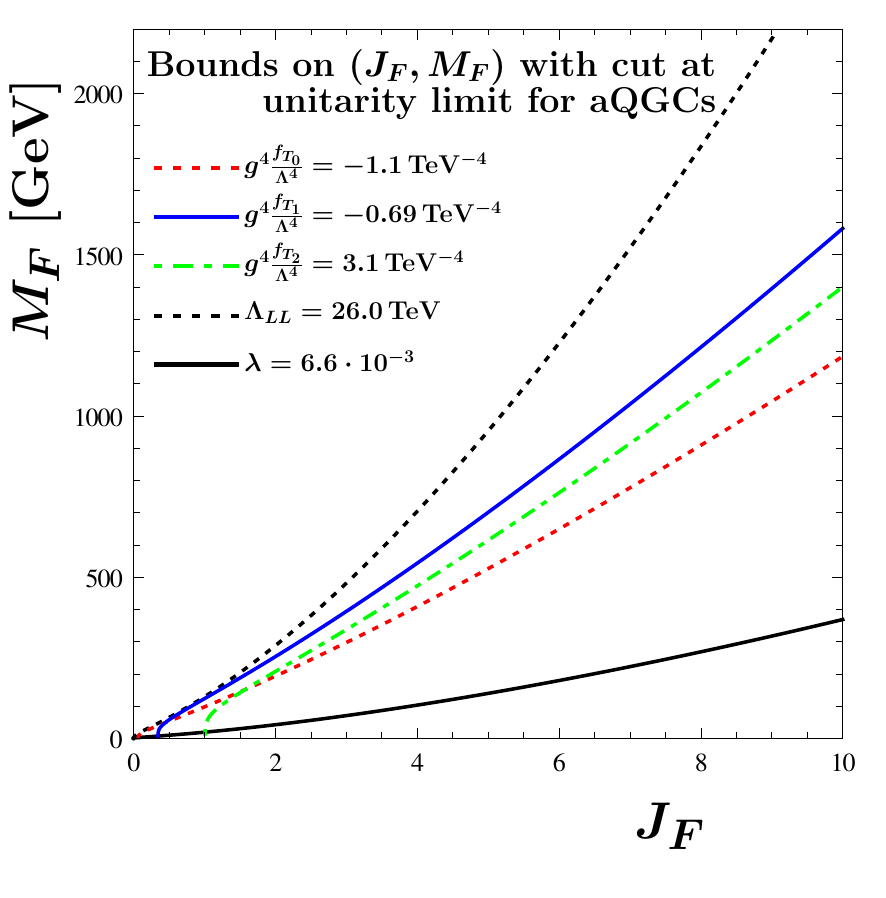}\hfill
\includegraphics[page=4,width=0.50\textwidth]{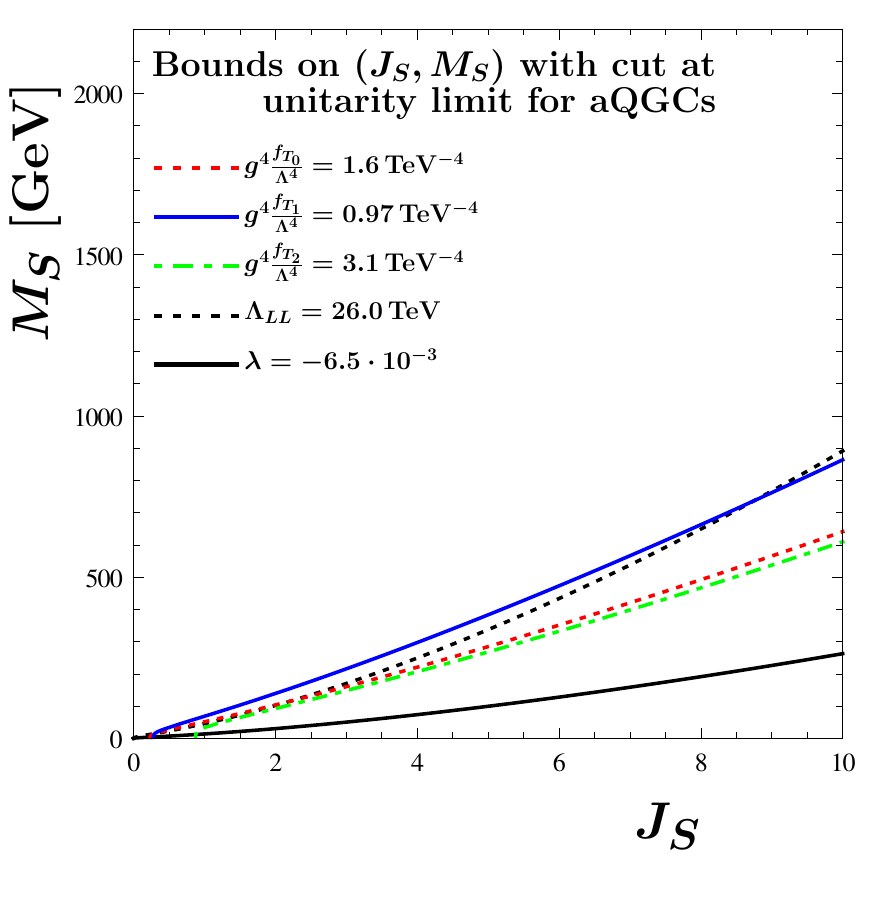}
\end{center}
\caption[Bounds in $(J_R,M)$ parameter space]{Translation of the experimental
  bounds on Wilson coefficients as in Table~\ref{tab:Wilson_bounds} to the
  $(J_R,M_R)$ parameter space: masses above the curves are allowed by the
  individual measurements. Shown are the cases of a fermion multiplet (left
  panel) and a scalar multiplet (right panel). For aQGCs, experimental results
  are derived with a step-function form-factor, cutting off the EFT prediction
  at the unitarity limit (see \citere{Sirunyan:2020gyx} for details). }
        \label{fig:Wilson_bounds}
\end{figure}

The insertion of $O_{DW}$ into the gauge boson propagator of a
$q\bar q\to l^+l^-$ amplitude scales like
\begin{equation}
  \frac{1}{p^2-m_W^2}\times p^4\frac{f_{DW}}{\Lambda^2}\times \frac{1}{p^2-m_W^2}
  \quad\sim\quad \frac{f_{DW}}{\Lambda^2}\nonumber
\end{equation}
at high energy and, thus, reduces to a flavor universal, dimension-6
four-fermion interaction which affects left-chiral amplitudes only. Writing 
the coefficient of the $\bar Q_L\gamma^\mu Q_L\, \bar L\gamma_\mu L$ EFT
operator as $2\pi\,\eta_{LL}/\Lambda_{LL}^2$ one finds $\eta_{LL}=-1$ and
\begin{equation}
  \Lambda^2_{LL} =2\pi \, \frac{240\pi^2 M_F^2}{g^4 T_F} \qquad {\rm or}
  \qquad \Lambda^2_{LL} = 2\pi \, \frac{1920\pi^2 M_S^2}{g^4 T_S}\, ,
\end{equation}
for a single fermion or scalar multiplet in the loop, respectively.
In the EFT approximation, experimental bounds on four-fermion contact terms,
such as recent ATLAS measurements~\cite{Aad:2020otl}, can thus be turned
into the lower limits on $M_R^2/T_R$ which are shown in \fig{fig:Wilson_bounds}.

aTGCs are typically measured in the \lep{} parameterization, where the effective
$WWV$-coupling parameters $\Delta g_1$, $\Delta\kappa$ and $\lambda$ are
used~\cite{Hagiwara:1986vm}. $\lambda$ parameterizes
transverse weak boson interactions and thus is the one of primary interest
here. The loop-corrections of our model contribute both to propagators, via the
dimension-6 operator $O_{DW}$, and to triple gauge-boson vertices, via both
$O_{DW}$ and $O_{WWW}$. In the $\bar qq\to WW$ amplitude, in the SU$(2)_L$
limit, the former is equivalent to an energy dependent
$\Delta g_1=\Delta\kappa$ contribution, while the latter are best
captured after the operator redefinition in \eqn{eq:DWtoWWW}. One finds 
\begin{align}
  \lambda =\frac{3  m_W^2 g^2}{2}\left(\frac{f_{WWW}}{\Lambda^2}-
  4 \frac{f_{DW}}{\Lambda^2}\right)\, &= g^2\,m_W^2 \, 
  \left(\sum_F n_F \frac{T_F}{240\pi^2 M_F^2} -
  \sum_S n_S \frac{T_S}{480\pi^2 M_S^2} \right) \, ,\\
  \Delta\kappa =\Delta g_1= 2 g^2 m_W^2 \frac{f_{DW}}{\Lambda^2}
  \,  \frac{s-2 m_W^2}{s-m_W^2} &\approx g^2\,m_W^2 \,
   \left(\sum_F n_F \frac{T_F}{60\pi^2M_F^2} +
   \sum_S n_S \frac{T_S}{480\pi^2M_S^2}\right)
   \, .
  \label{eq:lambda_par}
\end{align}
Going beyond the SU$(2)_L$ limit, the $\lambda$-aTGC can easily be compared
to existing data, via $\lambda_\gamma=\lambda_Z=\lambda$. The $\Delta g_1$ and
$\Delta\kappa$ contributions would appear as process dependent form-factors,
however, which are constrained at a similar level as the induced
$\lambda$-aTGC.

\renewcommand{\arraystretch}{1.3}
\begin{table}[tb]
\centering
\begin{tabular}{c c c c c}
Coefficient & limit without unitar. & channel & limit with unitarity cut & channel
\\\hline
$\eta_{LL}=-1$: $\Lambda_{LL}$ & $26.0$ TeV & $pp\to l^+l^-X$ & -- & --
\\
$\lambda$ & $[-6.5,6.6]\cdot 10^{-3}$ & $WV\to l\nu J$ & -- & --
\\
$g^4\frac{f_{T_0}}{\Lambda^4}$ & $[-0.24,0.22]\,$TeV$^{-4}$ & $ZZ$ & $[-1.1,1.6]\,$TeV$^{-4}$ & $ssWW,WZ$
\\
$g^4\frac{f_{T_1}}{\Lambda^4}$ & $[-0.12,0.14]\,$TeV$^{-4}$ & $ssWW,WZ$ & $[-0.69,0.97]\,$TeV$^{-4}$ & $ssWW,WZ$
\\
$g^4\frac{f_{T_2}}{\Lambda^4}$ & $[-0.35,0.48]\,$TeV$^{-4}$ & $ssWW,WZ$ & $[-1.6,3.1]\,$TeV$^{-4}$ & $ssWW,WZ$
\end{tabular}
\caption[Wilson Bounds]{Experimental 95\% CL bounds on four-fermion
  couplings~\cite{Aad:2020otl}, aTGCs~\cite{CMS-SMP-18-008} and
  aQGCs~\cite{Sirunyan:2017fvv,Sirunyan:2020alo}. The values for aQGCs from
  the CMS analyses are derived from leptonic decay channels in VBS processes
  only. aQGC bounds are presented with and without a unitarization in form of
  a cut at the perturbative unitarity limit of the EFT calculation.
  In addition, we name the channels from which the presented bounds
  are derived.
}
\label{tab:Wilson_bounds}
\end{table}
\renewcommand{\arraystretch}{1.}

Table \ref{tab:Wilson_bounds} summarizes relevant bounds
on four-fermion couplings, aTGCs, and aQGCs. 
For the bounds on $T$-operators, we use the recent CMS results 
of~\citeres{Sirunyan:2017fvv,Sirunyan:2020alo} on leptonic decay
modes in VBS processes, which also take into account limitations from
unitarity.
Since these aQGC bounds are only specified for insertions of
single EFT operators, \fig{fig:Wilson_bounds} depicts
the limits on the $(J_R,M_R)$ parameter
space for each Wilson coefficient separately. Although the strong correlations
between the Wilson coefficients in our model are not taken into account
by this procedure, it suffices here, since we are only aiming at a rough
estimate of allowed regions in the $(J_R,M_R)$ plane.\footnote{
  The problem is best exemplified by the deviation of the $T_1$ operator
  prediction from the full EFT result in the case of same-sign $WW$
  scattering and a heavy fermion multiplet, depicted
  in \fig{fig:XSonshellFe} (lower left). }
The figures indicate that a measurement of aQGCs has higher potential to find
effects of the BSM physics considered here than aTGC measurements.
To be more concrete, we consider the cases
of a fermion with $J_F=4$ and of a scalar with $J_S=6$ in the following,
values which represent maximal multiplet sizes compatible with 
unitarity and perturbativity limits to be discussed in \sct{sec:hel_proj}.

For the fermion multiplet, the strongest bound is associated with 
searches for $qqll$ contact terms, disfavoring masses below $700\,$GeV.
Somewhat weaker, at $M_F>540\,$GeV, are bounds derived from limits on the
dimension-8 $T_1$-operator, which do take into account unitarity constraints.
Restrictions via the other $T$-operators are qualitatively similar.
The aQGC bounds in turn are much more stringent than
the bound from the $\lambda$-aTGC, which only excludes
masses below $100\,$GeV in the \eft{} dimension-6 approximation.

The discussion is comparable in the scalar case. The bounds in
the $(J_S,M_S)$ parameter plane are less stringent, however, as
anticipated from the higher numerical suppression in the Wilson coefficients. 
Again all $T$-operator bounds appear to be stronger than the aTGC
bounds, even when including unitarity considerations only for the former
(see \fig{fig:Wilson_bounds}, right), making VBS more promising
for the investigation of this model than vector boson pair production. For
$J_S=6$, the bound on the coefficient of the $T_1$ operator disfavors masses
%below $760\,$GeV without considering the unitarity limit and
below $470\,$GeV, while searches for $qqll$ contact terms appear
to imply a slightly weaker bound of $M_S>430$~GeV.

Guided by these constraints, we will use $J_F=4,\,M_F=600$~GeV and
$J_S=6,\,M_S=600$~GeV as benchmark points in the remainder of the paper
to illustrate consequences for VBS at the LHC. For fermion multiplets,
$M_F=600$~GeV
appears to be in slight tension with $qqll$ contact term limits.
However, additional BSM effects which we do not consider, like an extra
$Z'$, might ameliorate the tension. Since we are not explicitly proposing
a viable BSM model, but are rather interested in generic features for VBS,
we prefer a common mass for fermions and scalars in our subsequent discussions.

Another reason for taking the $(J_R,M_R)$ bounds derived from \eft{}
considerations only as indicative is the concern that the experimental
\eft{} analyses, especially for the $qqll$ contact terms and the aQGC,
are largely based on data with dilepton or $VV$ invariant masses
  above 1~TeV. For our large multiplet model the \eft{} approximation cannot
be expected to hold for invariant masses of the order of the threshold energy
of $2\,M_R$ or larger.

\begin{figure}[tb!]
\begin{center}
\includegraphics[page=1,width=0.48\textwidth]{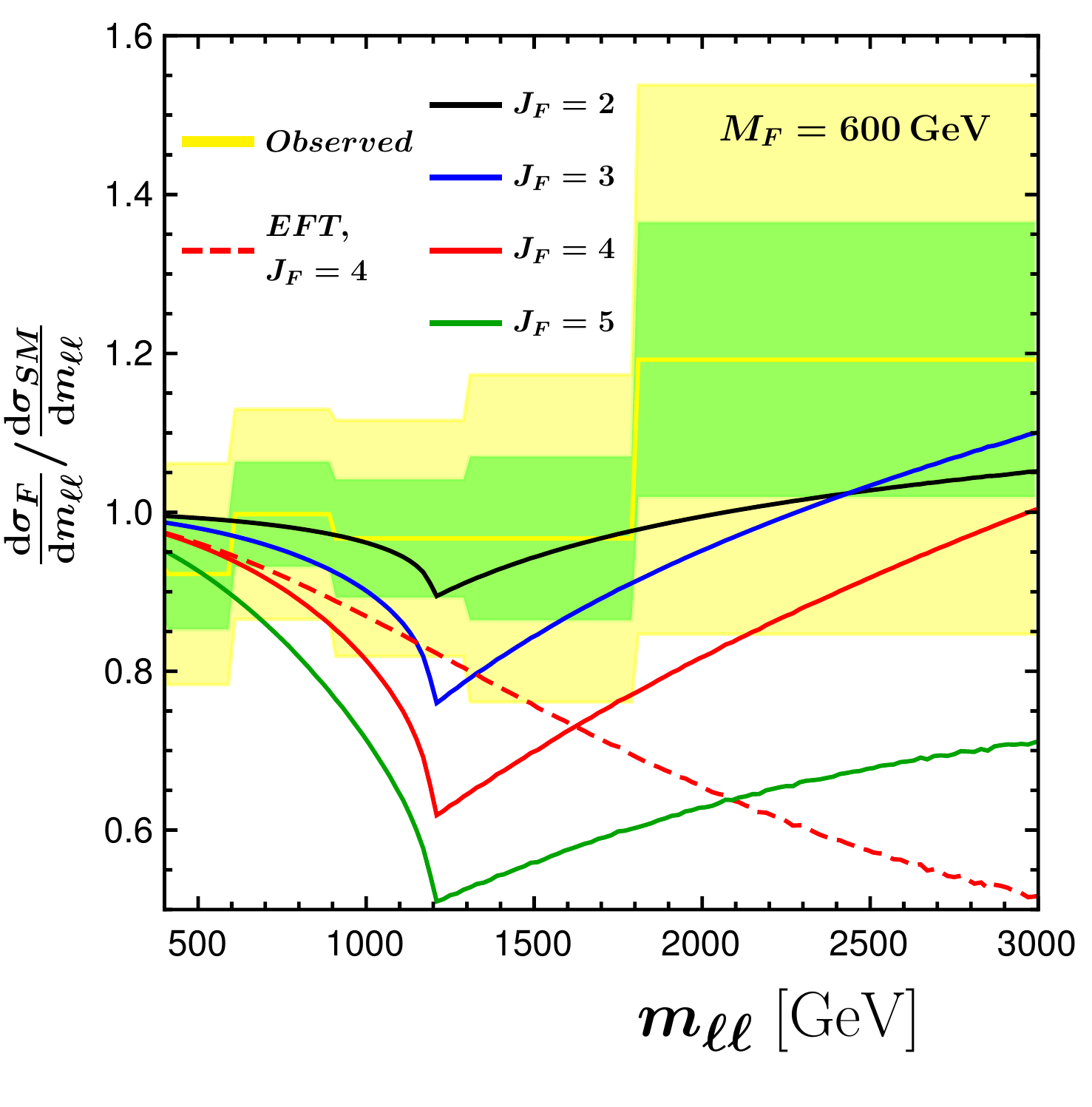}
\includegraphics[page=2,width=0.48\textwidth]{figures/WilsonCoefficients/PropagatorCorr.pdf}
\end{center}
\caption{Ratio of the modified Drell-Yan invariant mass distribution to the
  SM expectation at the LHC. For a multiplet mass of $M_R=600$~GeV, the
  left (right) panel considers the effect of four different isospin choices
  of a fermion (scalar) multiplet. The colored bands, based on the results of
  \citere{Sirunyan:2021khd},  provide an estimate of present experimental sensitivity
  at the $1\sigma$ and $2\sigma$ level. }        \label{fig:Drell-Yan}
\end{figure}

As a case in point, let us have a brief look at $q\bar q\to l^+l^-$
production at the LHC, from which $qqll$ contact term bounds are derived.
In this process and at 1-loop level, our model only influences the vector
boson propagator. The Dyson-resummation of the on-shell renormalized propagator
corrections, which follow from Eqs.~(\ref{eq:propagator}) and (\ref{eq:VacPol}),
leads to
\begin{align}
  D(p^2)
=\frac{-i}{p^2-m_{W,pole}^2-p^2 \Pi_{R,pole}}\,  \label{eq:prop-resum}
\end{align}
for $W^\pm$ exchange between purely left-chiral fermions. For comparison
with $l^+l^-$ data, $\gamma$-$Z$ propagator mixing needs to be
considered and the BSM contributions are somewhat diluted by U$(1)_Y$
hypercharge contributions which, in our toy model, are left unchanged with
respect to the SM.
The results are shown in \fig{fig:Drell-Yan} where the ratio of the dilepton
invariant mass distribution in our toy model as compared to the SM is
displayed for $M_R=600$~GeV and a selection of isospins. Also shown, for the
$J_F=4$ and $J_S=6$ benchmark points, is the
BSM to SM ratio in the \eft{} approximation, considering the
equivalent dimension-6 $qqll$ contact term only.

While BSM effects are negligible in the $m_{ll}<m_R/2$ region (below 1.5\% for
$J_F=4$ and well below 1\% for $J_S=6$), sizable deviations from the SM are
expected at higher energies, for large isospin representations. The \eft{}
dimension-6 approximation reproduces the full one-loop result considerably below
production threshold, but it fails spectacularly above threshold. Since the
bounds on $qqll$ contact terms in Table~\ref{tab:Wilson_bounds} heavily depend
on data at $m_{ll}>2.5$~TeV, one concludes that the contact term bounds
on the $(J_R,M_R)$ parameter plane cannot be taken at face value. However,
around the production threshold, the deviations from the SM are even bigger
than suggested by the \eft{} approximation, and a full analysis of the model
should be performed in this region to extract bounds on $M_R$.

As a first rough approximation, we have translated the results of a
recent CMS analysis of dilepton production~\cite{Sirunyan:2021khd}
to 1- and 2-$\sigma$ error
bands in \fig{fig:Drell-Yan}, assuming fully correlated systematic errors
between electron and muon channels and adding systematic and statistical
errors in quadrature, otherwise. Inspection of \fig{fig:Drell-Yan}
then suggests that for $M_R=600$~GeV and at 95\% CL the scalar benchmark
point with $J_S=6$ is still allowed, while strong evidence for the fermionic
case with $J_F=4$ should have been seen already. Since we consider our setup
as a toy model, and since we are primarily interested in its qualitative
predictions for VBS, we postpone a more detailed analysis 
to future work and use the benchmark point merely for illustration in the
following. 

\section{On-shell vector boson scattering at one-loop}
\label{sec:fulloneloop}

The large deviations between \eft{} results and the full
model prediction for lepton pair-production provide a strong motivation
to also consider the full model for VBS predictions.
For a study of its main phenomenological features, we start
with a discussion of on-shell VBS. In the following we neglect the normal
\sm{} electroweak corrections, which are parametrically small compared to the
contributions from extra SU$(2)_L$ multiplets, due to the $J_R^3$ to $J_R^5$
enhancement of the latter. Corrections from the new heavy degrees of freedom
originate from the previously discussed one-loop contributions
to the gauge-boson propagators and the three- and four-gauge-boson vertex
functions. In terms of Feynman diagrams, for external gauge bosons in the
adjoint basis ($W^\mu_{a_i}$ with $a_i=1,2,3$) the scattering amplitudes can
be depicted as
\begin{align}
&\vcenter{\hbox{\includegraphics[page=8,scale=0.75]{figures/VBS}}}
= \vcenter{\hbox{\includegraphics[page=1,scale=0.75]{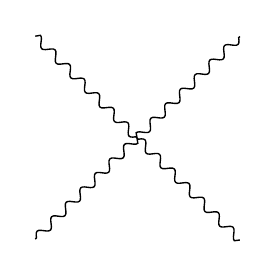}}}
+ \left(\vcenter{\hbox{\includegraphics[page=2,scale=0.75]{figures/VBS}}}+\vcenter{\hbox{\includegraphics[page=3,scale=0.75]{figures/VBS}}}+\text{$t$- and $u$-type}\right)\nonumber\\
&\qquad+\left(\vcenter{\hbox{\includegraphics[page=4,scale=0.75]{figures/VBS}}}
+\vcenter{\hbox{\includegraphics[page=5,scale=0.75]{figures/VBS}}}
+\vcenter{\hbox{\includegraphics[page=6,scale=0.75]{figures/VBS}}}+\text{$t$- and $u$-type}\right)+\vcenter{\hbox{\includegraphics[page=7,scale=0.75]{figures/VBS}}}\nonumber\\
&\qquad=i \left(\mathcal{M}_1 \delta^{ab}\delta^{cd}+\mathcal{M}_2 \delta^{ac}\delta^{bd}+\mathcal{M}_3 \delta^{ad}\delta^{bc}\right)\label{eq:VBF_fullAmp}\,.
\end{align}
The last equality follows from isospin symmetry, because all isospin
combinations for the four indices $a_i$ associated with the external gauge
bosons $W^\mu_{a_i}$ can be expressed by Kronecker deltas in the adjoint space.
Obviously, the reduced amplitudes $\mathcal{M}_1$, $\mathcal{M}_2$, and
$\mathcal{M}_3$ can be obtained from each other by crossing.

In this section we consider on-shell vector boson scattering in the SU$(2)_L$
limit, i.e. we set $g'=0$, which eliminates any photon
contributions\footnote{As a result one does not need to contend with any
  infrared or collinear singularities due to the massless photon. Also
  the unphysical Rutherford singularity (due to $t$-channel photon exchange)
  in charged $W$-scattering is eliminated, which in real life is regularized
  by the space-like virtuality of the incoming weak bosons.}
and leads to $Z=W^3$ and $m_Z=m_W=\frac{gv}{2}$, thus simplifying the scattering
kinematics. Projecting the amplitude of \eqn{eq:VBF_fullAmp} onto
$Z$ bosons (with a factor $\delta^{3 a_i}$) or $W$ bosons (with a factor
$\frac{1}{\sqrt{2}}\left(\delta^{1 a_i}\pm i \delta^{2 a_i}\right)$ for 
each incoming $W^\pm$ or outgoing $W^\mp$), one finds
\begin{subequations}
\begin{align}
\mathcal{M}\left(W^\pm W^\pm \rightarrow W^\pm W^\pm \right)&=\mathcal{M}_2+\mathcal{M}_3\,,\\
\mathcal{M}\left(W^\pm Z \rightarrow W^\pm Z\right)&=\mathcal{M}_2\,,\\
\mathcal{M}\left(W^\pm W^\mp \rightarrow W^\pm W^\mp\right)&=\mathcal{M}_1+\mathcal{M}_2\,,\\
\mathcal{M}\left(W^\pm W^\mp \rightarrow ZZ\right)&=\mathcal{M}_1\,,\\
\mathcal{M}\left(ZZ \rightarrow ZZ\right)&=\mathcal{M}_1+\mathcal{M}_2+\mathcal{M}_3\,.
\end{align}
\label{eq:process-amp}
\end{subequations}
As indicated by the absence of double vertex insertions in \eqn{eq:VBF_fullAmp},
our full amplitudes correspond to a coupling expansion to order $g^4$,
with $\overline{\text{MS}}$ renormalization. Before proceeding, we need to address 
whether such a perturbative expansion is justified for large multiplicities,
$2J_R+1$, which we do by a unitarity analysis of partial wave amplitudes.

\subsection{Partial wave analysis of VBS amplitudes}
\label{sec:hel_proj}

The notation for VBS partial wave amplitudes follows
\citere{Perez:2018kav}, the results of which are implemented in {\tt VBFNLO}.
However, the present
case is simplified by taking the SU$(2)_L$ limits and by considering on-shell
scattering only. Exploiting angular momentum conservation, the helicity
amplitudes $\mathcal{M}_{\lambda_1\lambda_2\lambda_3\lambda_4}$ of the various
$VV\to VV$ processes are decomposed into coefficients of Wigner d-functions
\begin{align}
\mathcal{M}_{\lambda_1\lambda_2\lambda_3\lambda_4}\left(s,\theta\right)=&8 \pi \mathcal{N}_{fi} \sum_{j=max\left(|\lambda_{12}|,|\lambda_{34}|\right)}^{\infty} (2 j+1)\mathcal{A}^j_{\lambda_1\lambda_2\lambda_3\lambda_4}\left(s\right) d^j_{\lambda_{12}\lambda_{34}}\left(\theta\right)\,,\label{eq:pw_exp}
\end{align}
where $\lambda_{ij}:=\lambda_i - \lambda_j$, and the normalization factor is
given by $\mathcal{N}_{fi}=\frac{1}{\beta\sqrt{S_f S_i}}$, with
$\beta=\sqrt{1-4m_W^2/s}$ and statistics factors $S_f$ or $S_i=1/2$ in case of
two identical weak bosons. $\mathcal{N}_{fi}$ is chosen such as to simplify the
constraints on the partial wave amplitudes
$\mathcal{A}^j_{\lambda_1\lambda_2\lambda_3\lambda_4}\left(s\right)$ which follow
from $S$-matrix unitarity, namely
\begin{align}
2 \text{Im}\left(\mathcal{A}_{\lambda_1\lambda_2\lambda_3\lambda_4} ^j\right)&\geq \sum_n  \sum_{\lambda_1 ',\lambda_2 '}\mathcal{A}_{\lambda_3\lambda_4\lambda_1 '\lambda_2 '} ^{j\ast}\mathcal{A}_{\lambda_1\lambda_2\lambda_1 '\lambda_2 '}^j\,.\label{eq:pw_unitarity}
\end{align}
Here $\text{Im}\left(\mathcal{A}_{\lambda_1\lambda_2\lambda_3\lambda_4} ^j\right)$
corresponds to the anti-hermitian part of the VBS scattering amplitude,
understood as a matrix in isospin and helicity space. The sum over $n$
on the right hand side includes all relevant di-boson intermediate states.
Equality in \eqn{eq:pw_unitarity} would only be reached when summing over a
complete set of intermediate states, including multi-particle states.
For any normalized linear combination of diboson states, $|n\ket$, the
expectation value of $\mathcal{A}^j$ will lie within the Argand circle.
In particular, this will be true for the eigenvector $|n_{\rm max}\ket$ of the
largest eigenvalue of $\text{Re}\left(\mathcal{A}^j\right)$ or
$\text{Im}\left(\mathcal{A}^j\right)$.  Setting
\begin{align}
a^j = a^j(s) = \bra n_{\rm max}|\mathcal{A}^j(s)|n_{\rm max}\ket \,,
\end{align}
which corresponds to the largest eigenvalue for a normal scattering matrix,
we thus get the Argand circle condition
\begin{align}
\text{Re}\left(a^j\right)^2+\left(\text{Im}\left(a^j\right)-1\right)^2\leq 1\,.
\end{align}
Since a finite perturbative expansion is only expected to fulfill this
requirement approximately, e.g. an on-shell tree-level calculation remains
always on the real axis and is thus never compatible with the strict
requirement, the perturbative unitarity bounds are given by
\begin{align}
  |a^j|^2\lesssim 2\,,\quad |\text{Re}\left(a^j\right)|\lesssim 1\,,\quad
  0\le \text{Im}\left(a^j\right)\lesssim 2\,, 
\label{eq:pert_unitarity}
\end{align}
which should be understood as demanding
that the eigenvalues of $\mathcal{A}^j$ remain close to the Argand circle
in a reasonable perturbative calculation.

\begin{figure}[tb!]
\begin{center}
\includegraphics[width=0.5\textwidth]{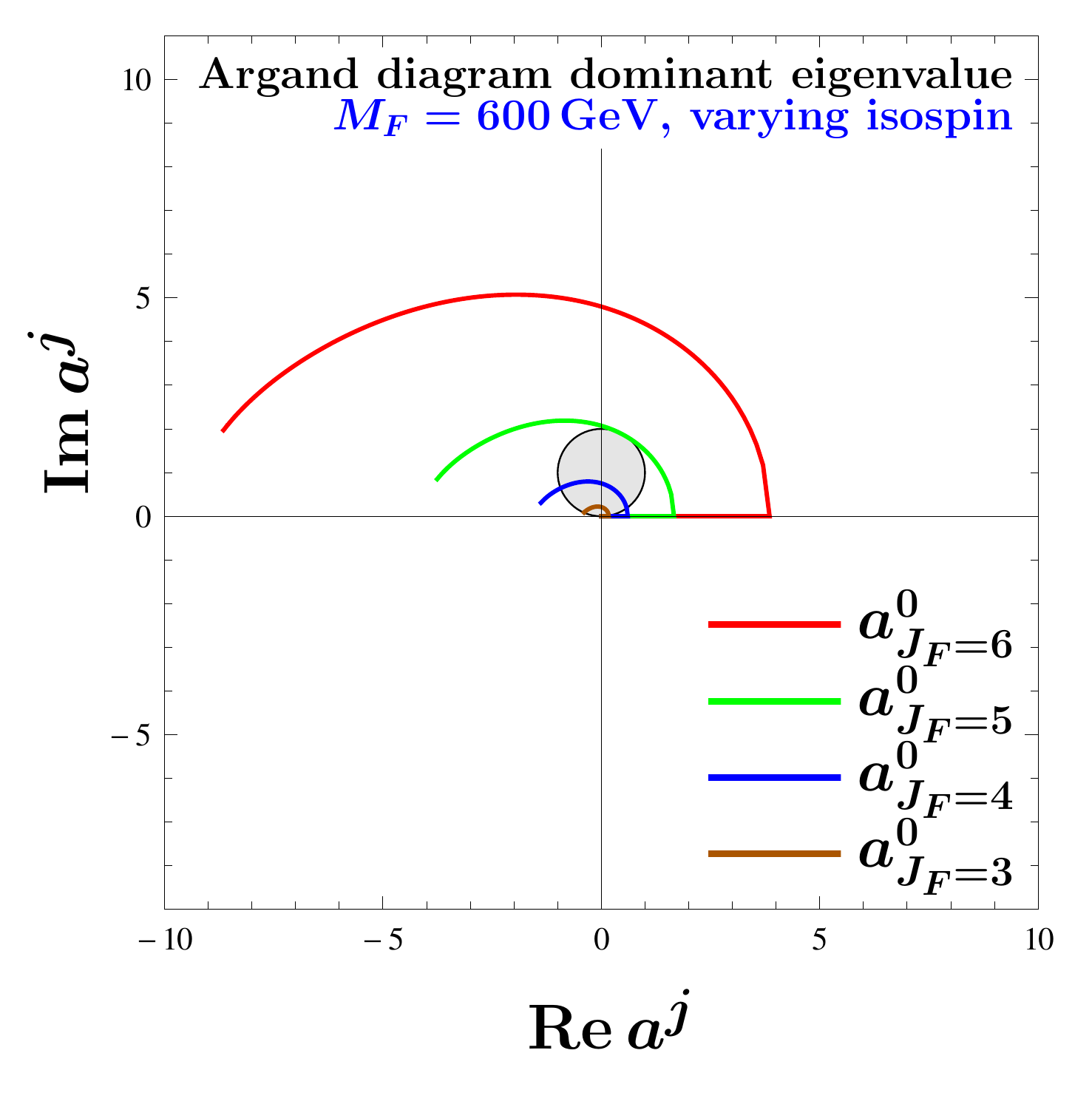}\hfill
\includegraphics[width=0.5\textwidth]{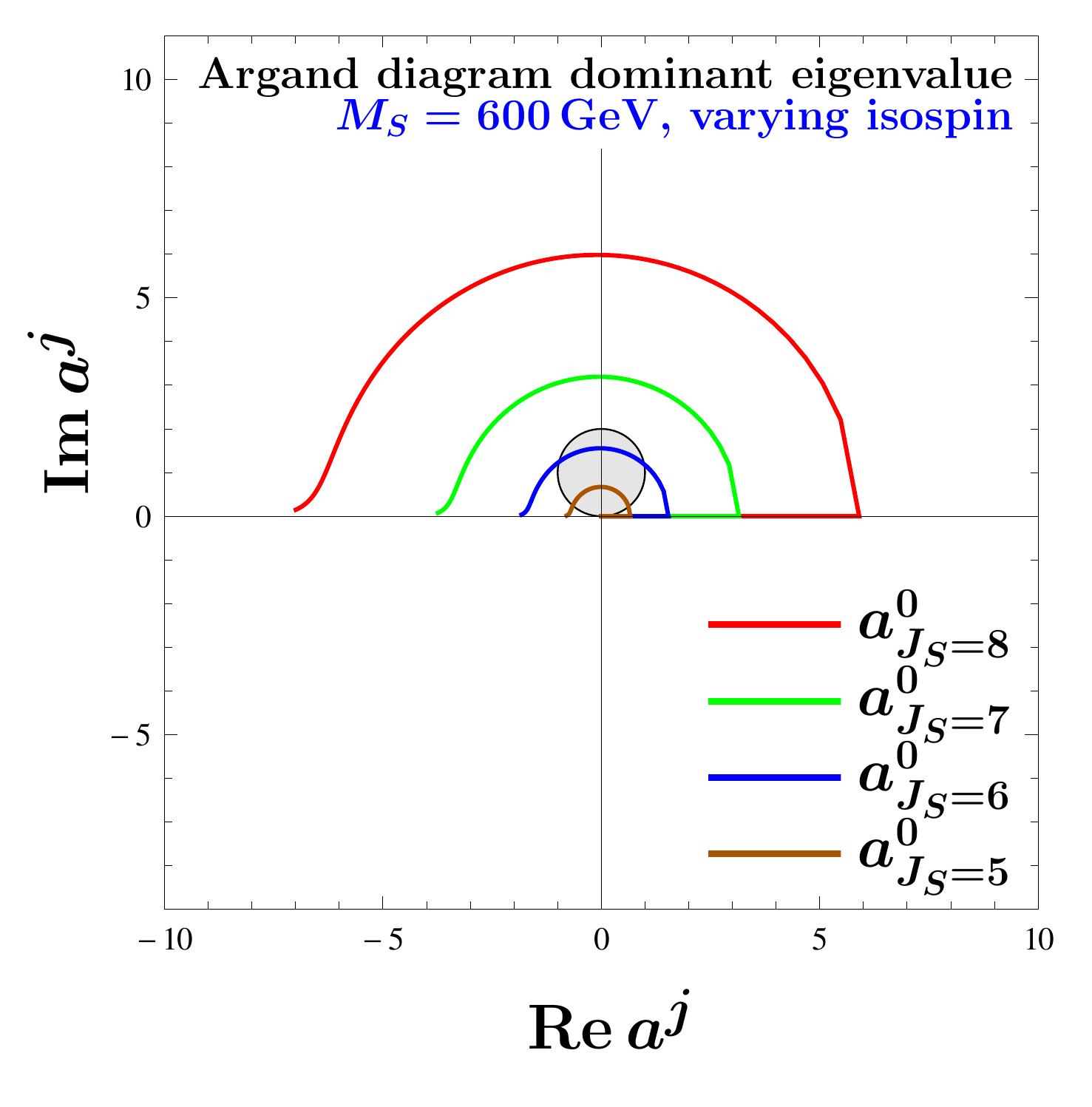}\hfill\\
\end{center}
\caption{Argand diagrams for the dominating iso-singlet, $j=0$ VBS partial
  wave amplitudes, $a^0_{J_R}$. The left (right) panel is for a single fermion
  (scalar) multiplet of isospin $J_F$ ($J_S$) of mass $M_R=600$~GeV. Only the
  BSM contribution to the amplitude is shown.}
        \label{fig:argand-circle}
\end{figure}
When investigating the behavior of the partial wave amplitude contributions
from single $(J_R,M_R)$ multiplets, one finds sizable effects only for the
transverse helicity amplitudes 
$\mathcal{M}_\text{1111}$,
$\mathcal{M}_\text{11-1-1}$, $\mathcal{M}_\text{1-11-1}$,
$\mathcal{M}_\text{1-1-11}$ (and with corresponding parity 
flipped helicities). Thus it is sufficient to diagonalize the partial wave
amplitudes in this restricted helicity space\footnote{A detailed
  construction of the eigenvalues may be found in \citere{Master-Thesis}.}.
This is made easier by the
fact that the VBS amplitudes are also block-diagonal in the basis of total
isospin of the two particle states, $|J_2,J_{2,z}\ket$, as we work in the
$g'=0$ limit of the \sm{} and our BSM contributions do not break the SU$(2)_L$
symmetry. The dominant eigenvalue is found in the $j=0$ partial wave and
corresponds to vanishing total isospin of the weak boson pair, $J_2=0$.
Its BSM contribution is called $a^0$ in the following.

\begin{figure}[tb!]
\begin{center}
\includegraphics[width=0.49\textwidth]{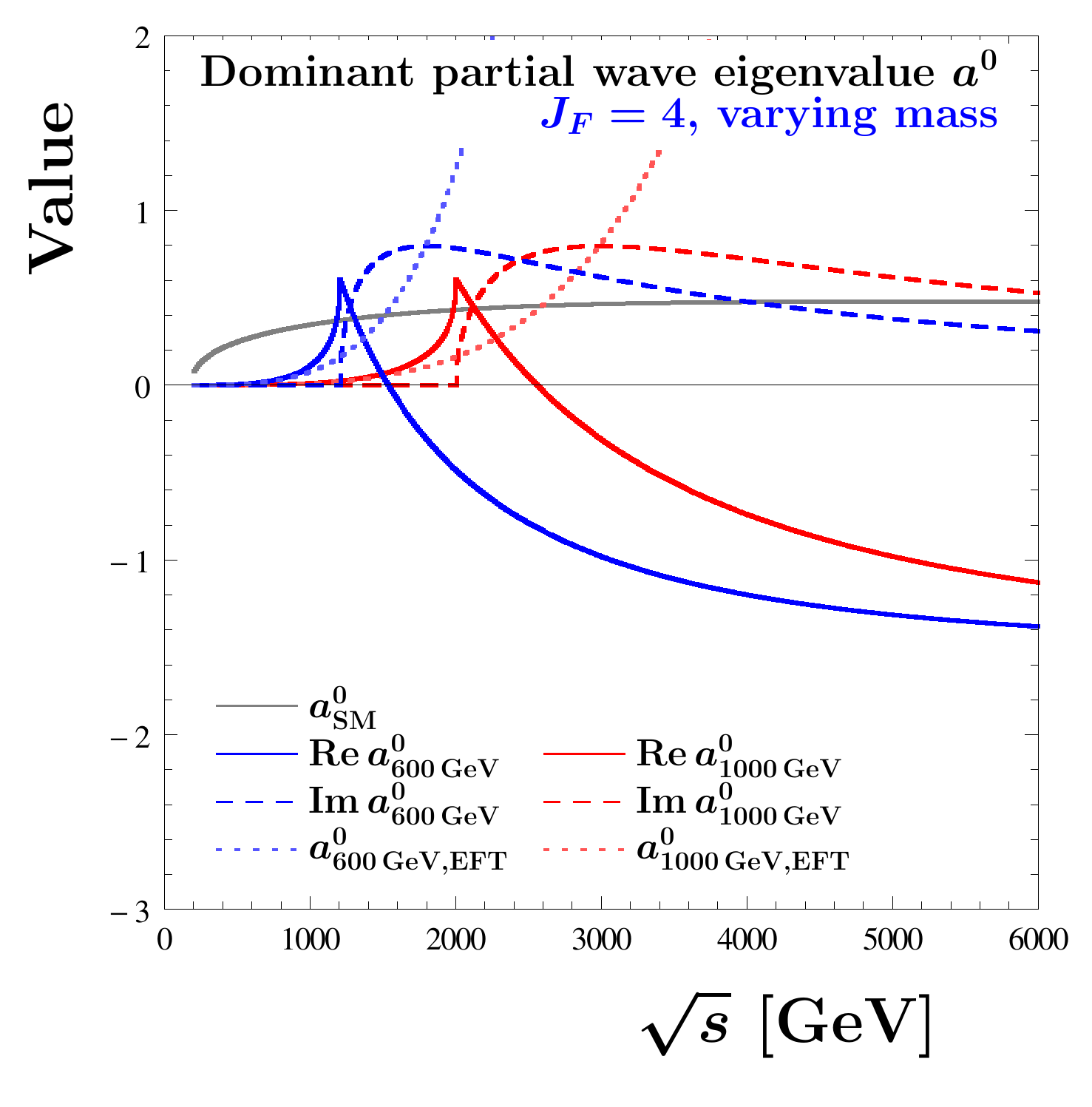}
\includegraphics[width=0.49\textwidth]{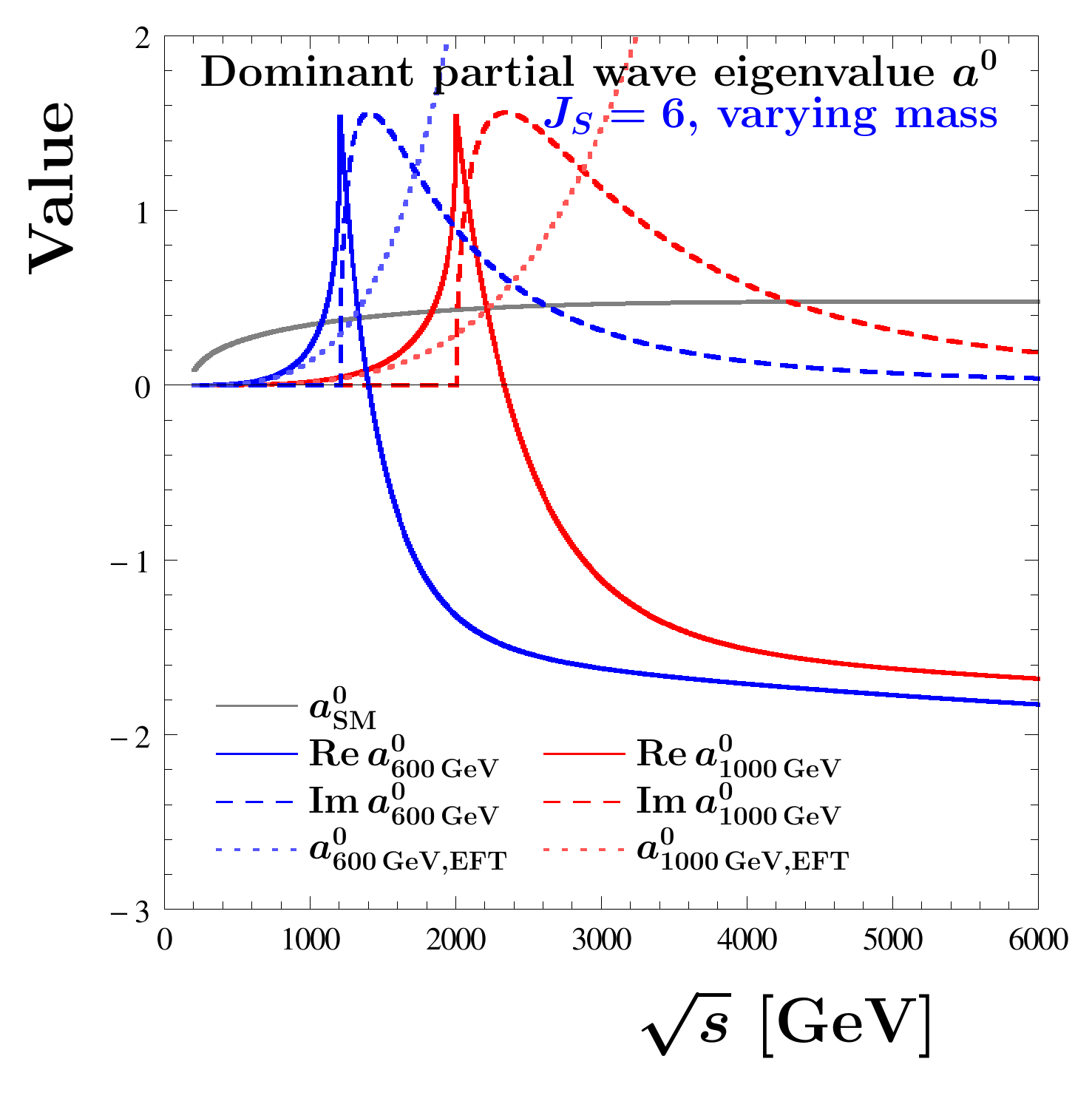}\\
\end{center}
\caption{Dominating iso-singlet, $j=0$ partial wave amplitudes, $a^0_{J_R}$,
  as a function of the diboson center of mass
  energy, $\sqrt{s}$. Results are shown for the pure contribution of a
  single $J_F=4$ fermion multiplet (left panel) and of a single $J_S=6$
  scalar multiplet (right panel), for two different masses of the extra
  particles, 600~GeV (blue curves) and 1000~GeV (red curves). Solid lines
  represent the real part of $a^0$, long dashed lines the imaginary part,
  and the short dashed lines give the purely real partial wave amplitudes
  in the EFT approximation. The \sm{} contribution, at tree level, is shown
  for comparison.}             \label{fig:argand-energy}
\end{figure}
Results are presented in \fig{fig:argand-circle} for both the fermionic
multiplet (left panel) and the scalar case (right panel), for a range of
isospin choices, $J_R$, for a multiplet mass of $M_F=600\,{\rm GeV}=M_S$,
and for $200\,{\rm GeV}\,<\sqrt{s}< 6000\,{\rm GeV}$.
For $J_F\leq 4$ and $J_S\leq 5$, the path of $a^0_{J_R}(s)$ in the complex plane
is compatible with the unitarity bounds of \eqn{eq:pert_unitarity}, while
the scalar case with $J_S=6$ can be considered marginally perturbative.
The strong dependence on multiplet size was to be expected because the
four-boson vertex function at 1-loop order grows like $J_R^5$. 
In the following we would like to estimate maximal deviations from the SM
which can be expected in VBS. For this purpose, the multiplet
realizations of $(J_F=4,M_F=600\,$GeV$)$ and $(J_S=6,M_S=600\,$GeV$)$, chosen
earlier, appear well suited for a qualitative discussion.

\fig{fig:argand-energy} shows the dependence of both the real and imaginary
parts of the dominant partial wave eigenvalue, $a^0_{J_R}(s)$  on the
center of mass energy of the $VV\to VV$ process. Here we fix $J_F=4$ for the
fermion model (left panel) and $J_S=6$ for the extra scalar multiplet (right
panel). Also the EFT prediction
for $a^0_{J_R}(s)$ is shown, which is purely real and which, sufficiently
below production threshold at $\sqrt{s}=2\,M_R$, agrees well with
the full calculation. The LO SM contribution to the dominant partial wave
amplitude is also depicted (gray solid line), for comparison, and because it
should be added to the anomalous contribution, $a^0_{J_R}(s)$.
Adding these two contributions, the resulting full partial wave amplitude
stays fully within the perturbative unitarity bounds of
\eqn{eq:pert_unitarity} for the $J_F=4$ fermion multiplet and agreement
for the $J_S=6$ scalar case is also improved at high energy. Violation of the
$|\text{Re}\left(a^j\right)|\lesssim 1$ bound for the scalar case is
actually limited to a small region around threshold, while the imaginary part
of the amplitude, which dominates somewhat above threshold, is consistent
with the unitarity limit. The problem with the real part of the amplitude
around $\sqrt{s}=2\,M_S$ could be ameliorated by SU$(2)_L$-breaking effects,
i.e. by mass splitting the multiplet and thus distributing its threshold
effects over a larger energy range, by binding effects due to additional
(strong) interactions of the scalar multiplet(s) or by other modifications of
our toy model. This provides another argument to not discard the $J_S=6$ case.

In order to demonstrate the effect of different mass values,
\fig{fig:argand-energy} shows results for $M_R=600$~GeV and $M_R=1000$~GeV.
One finds excellent scaling: the amplitude effectively only depends on
$s/M_R^2$ because the electroweak
scale merely enters via the mass, $m_W$, of the external gauge bosons, and
this gives tiny $(m_W/M_R)^2$ corrections. This scaling behavior implies that
the effects which we will demonstrate with $M_R=600$~GeV in the following
can simply be shifted to higher energy for larger multiplet masses, with an
approximately invariant ratio of BSM to SM VBS cross sections.

\subsection{Cross sections for on-shell vector boson scattering}
\label{sec:VBS-XS}

With limited statistics and sizable backgrounds for VBS events at the LHC,
integrated, unpolarized VBS cross sections are sufficient for a first
survey of LHC capabilities. According to \eqn{eq:process-amp}, the information
is contained in three crossing related helicity amplitudes which can be
disentangled by measuring cross sections for different combinations of weak
bosons: same-sign $WW$ scattering, i.e. $W^\pm W^\pm\to W^\pm W^\pm$ gives access
to $\mathcal{M}_2+\mathcal{M}_3$,  $WZ\to WZ$ depends on $\mathcal{M}_2$ only,
and the $ZZ$ production processes  $WW\to ZZ$ and  $ZZ\to ZZ$ provide separate
information on $\mathcal{M}_1$. 

Separating SM and new physics (NP) contributions, the squared amplitude,
which determines the cross section, is given by 
\begin{align}
\scriptstyle|\mathcal{M}_{tot}|^2=\underbrace{\scriptstyle|\mathcal{M}_{\text{SM}}|^2+2 \text{Re}\left(\mathcal{M}_{\text{SM}}{\mathcal{M}_{\text{NP}}}^\ast\right)}_{\text{(int)}}+|\mathcal{M}_{\text{NP}}|^2\,. \label{eq:def-interference}
\end{align}
Our new physics amplitude is the 1-loop contribution due to the additional
heavy fermion or scalar multiplet. In a typical NLO calculation, only
the interference term of the NP would be considered, together with the \sm{}
Born contribution. On the other hand, when considering anomalous couplings,
frequently also the $|\mathcal{M}_{\text{NP}}|^2$ term is included in the cross section
prediction, and this is indeed the default {\tt VBFNLO} implementation of VBS.
Since there are no infrared divergences in the present case, and because the
purely anomalous part is necessary for a prediction of the $ZZ\to ZZ$
cross section, which due to the small Higgs boson mass has a tiny \sm{}
amplitude, we do include the $|\mathcal{M}_{\text{NP}}|^2$ term in our
calculation, but we also show the NLO-type result in the following,
dubbed ``full model (int)'' in the figures. An analysis of the parametrically
leading terms in a perturbative expansion support the inclusion of the
quadratic NP term. At large $J_R$, the dominant two-loop contribution is
given by diagrams with an additional virtual vector boson propagator within
the the multiplet loop of the four-particle vertex functions, e.g.
$\vcenter{\hbox{\includegraphics[page=1,scale=0.6]{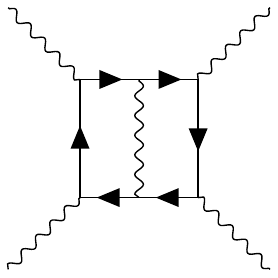}}}$ and
$\vcenter{\hbox{\includegraphics[page=2,scale=0.6]{figures/2loop}}}$ in the
fermion case, which have representation factors of $T_R C_{2,R}^2\sim J_R ^{7}$. 
On the other hand, the square of the one-loop four-particle vertex function,
i.e. the $|\mathcal{M}_{\text{NP}}|^2$ term above, is enhanced by a factor 
$T_R^2 C_{2,R}^2 \sim J_R ^{10}$ (see \eqn{eq:box_RepFactor}), which would not
arise until the three-loop level in the interference term.

Subsequently we present the integrated, unpolarized cross sections for
scattering angles $5^\circ<\theta<175^\circ$ in the center of mass frame
for the still academic $W^\pm W^\pm\to W^\pm W^\pm$, $WZ\to WZ$, $WW\to ZZ$, 
and $ZZ\to ZZ$ scattering processes. A more complete LHC simulation of
the full $qq\to qqVV$ processes will be postponed until \sct{sec:VBFNLO}.
The limited angular range reduces the forward and backward enhancement in
the \sm{} contribution, which is due to $t$- or $u$-channel Higgs or $W,Z$
exchange. For the new-physics contribution, we use the single multiplet
realization with parameter choices discussed in \sct{sec:coefficients}. 

\begin{figure}[tb!]
\begin{center}
\includegraphics[page=3,width=0.47\textwidth]{figures/XSplots/XS_Fe.pdf}\hfill
\includegraphics[page=2,width=0.47\textwidth]{figures/XSplots/XS_Fe.pdf}\\
\includegraphics[page=1,width=0.47\textwidth]{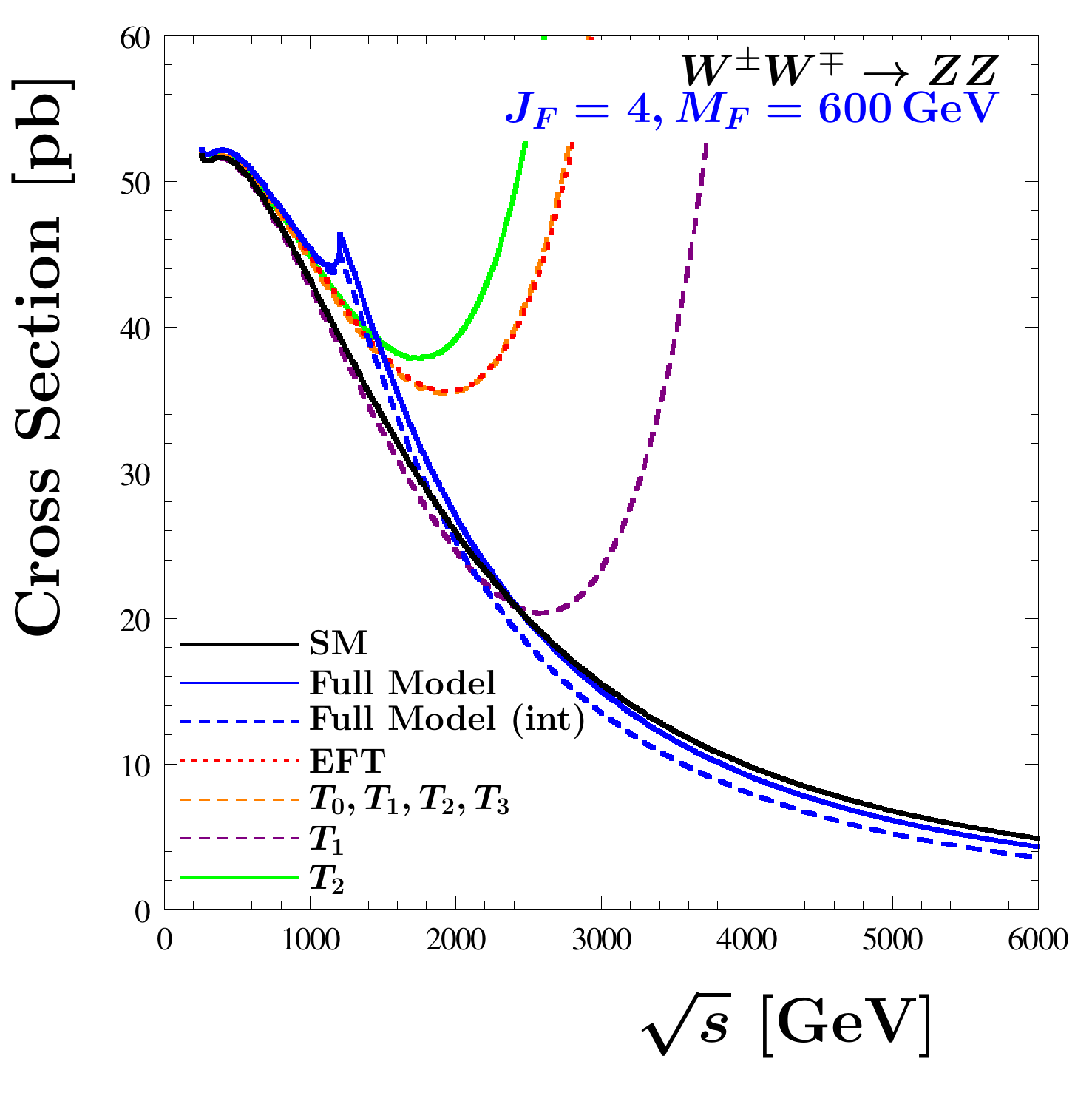}\hfill
\includegraphics[page=4,width=0.47\textwidth]{figures/XSplots/XS_Fe.pdf}
\end{center}
\caption{Unpolarized $VV\to VV$ cross sections as a function of the COM
  energy $\sqrt{s}=m_{VV}$ for same-sign $WW$ scattering (upper left),
  $WZ$-scattering (upper right), the $WW\rightarrow ZZ$ process (lower left)
  and $ZZ$-scattering (lower right). BSM curves are for the case of a single
  fermion multiplet with $J_F=4,M_F=600\,$GeV. The diverging curves
  represent several EFT approximations, up to dimension-8 operators (see
  text for details). The ``full model'' with full fermion 1-loop contributions
  is represented by the solid blue line while ``full model (int)'' (dashed blue)
  includes the interference term in \eqn{eq:def-interference} only.
  The solid black line represents the SM expectation, in the $g'=0$ limit.
  }
        \label{fig:XSonshellFe}
\end{figure}

\begin{figure}[tb!]
\begin{center}
\includegraphics[page=3,width=0.47\textwidth]{figures/XSplots/XS_Sc.pdf}\hfill
\includegraphics[page=2,width=0.47\textwidth]{figures/XSplots/XS_Sc.pdf}\\
\includegraphics[page=1,width=0.47\textwidth]{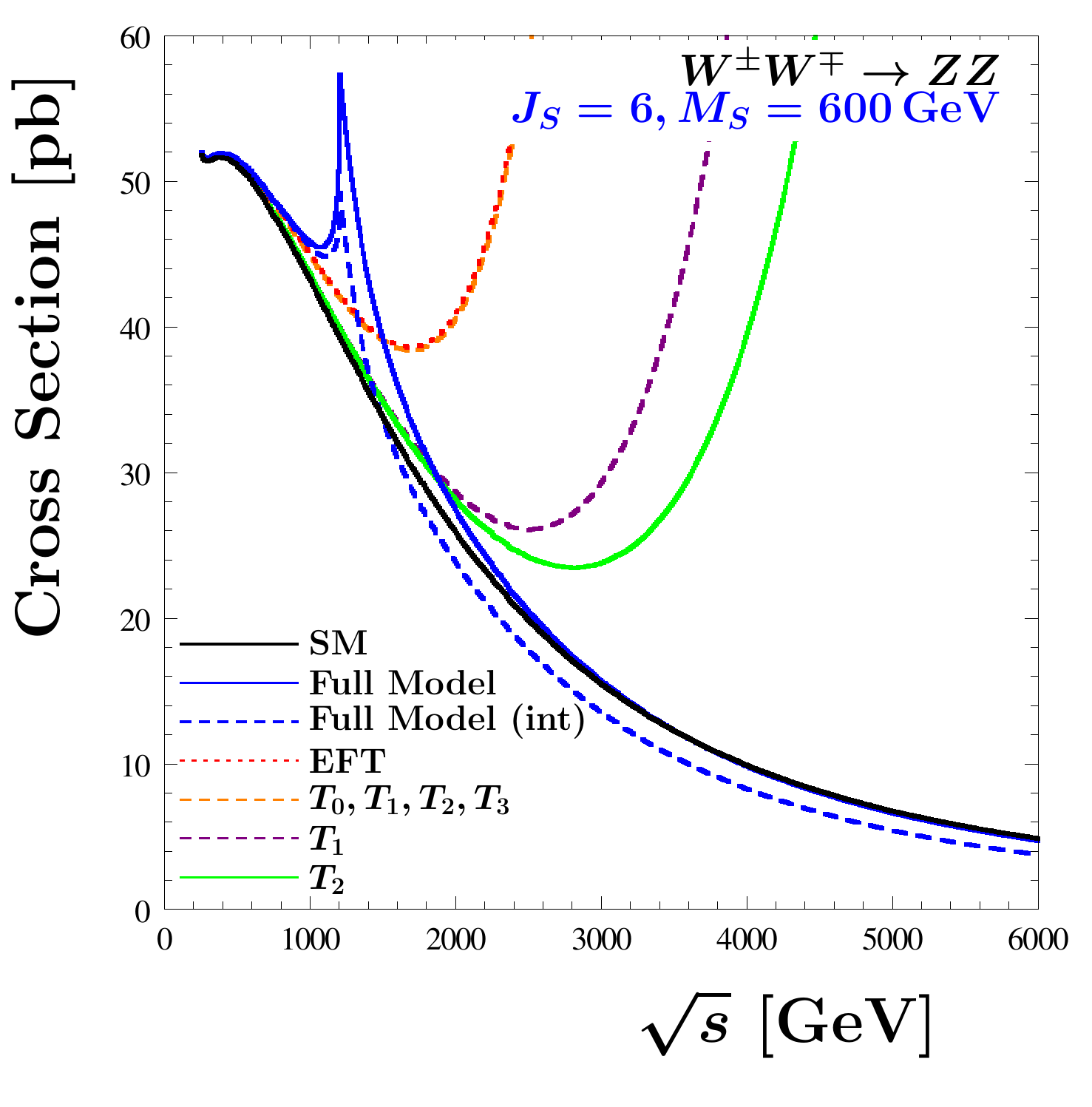}\hfill
\includegraphics[page=4,width=0.47\textwidth]{figures/XSplots/XS_Sc.pdf}
\end{center}
\caption{Same as \fig{fig:XSonshellFe}, but for a single scalar multiplet with
  $J_S=6,M_S=600\,$GeV. }
        \label{fig:XSonshellSc}
\end{figure}

Cross section results, as a function of the center-of-mass energy
$\sqrt{s}=m_{VV}$, are shown in \fig{fig:XSonshellFe} for the case of a
fermion nonet and in \fig{fig:XSonshellSc} for a single scalar multiplet
with $J_S=6$. 
A striking and well-known feature of the \sm{} prediction is the strong
suppression of the $ZZ\to ZZ$ cross section, due to the small value of
$m_H=125$~GeV. According to \eqn{eq:process-amp} this implies
\begin{align}
  \mathcal{M}_{2;\text{SM}}+\mathcal{M}_{3;\text{SM}}
  \approx -\mathcal{M}_{1;\text{SM}} \, ,
\end{align}
a relation which does not carry over to the NP contributions. Since
$\mathcal{M}_1$ describes $W^+W^-\to ZZ$ while $\mathcal{M}_2+\mathcal{M}_3$
is the amplitude for same-sign $WW$ scattering, one finds opposite signs
for the interference terms, $\mathcal{M}_{\text{SM}}{\mathcal{M}_{\text{NP}}}^\ast$, when
comparing these two processes. $\mathcal{M}_2$ is the amplitude for
$WZ$ scattering, which is related to $\mathcal{M}_3$ by $\theta\to \pi-\theta$
crossing. As a result one finds the same interference characteristics for $WZ$
scattering as for same-sign $WW$ scattering. Effects are larger for the latter,
however, by roughly a factor of two, because the important NP amplitudes are
fairly independent of scattering angle and thus add up, while the \sm{}
$t$- and $u$-channel amplitudes $\mathcal{M}_{2;\text{SM}}$ and $\mathcal{M}_{3;\text{SM}}$
peak in opposite hemispheres.

Comparing the $J_F=4$ fermion case in \fig{fig:XSonshellFe} with the
$J_S=6$ scalar multiplet in \fig{fig:XSonshellSc}, the effect of the scalar
multiplet on cross-sections is more pronounced due to the higher representation.
The constructive interference between the \sm{} and the anomalous 
contribution in the $WW\to ZZ$ process (lower left) is particularly striking
around the threshold for fermion or scalar pair production, which results
in a peak structure centered around $\sqrt{s}=2M_R$. 
For same-sign $WW$ and for $WZ$ scattering  (upper row) we see destructive
interference between the \sm{} and the anomalous contribution in the
low-energy regime.
This leads to a dip in the cross section at pair production threshold which,
however, is soon eclipsed by the strong cross section enhancement above
threshold, in particular for the scalar case. This peak above threshold
is driven to a substantial degree by a large contribution from the
imaginary part of the scattering amplitude, which was already evident in
\fig{fig:argand-energy}.

\begin{figure}[tb!]
\begin{center}
\includegraphics[page=3,width=0.45\textwidth]{figures/XSplots/XS_Fe_ratioSM.pdf}\hfill
\includegraphics[page=1,width=0.45\textwidth]{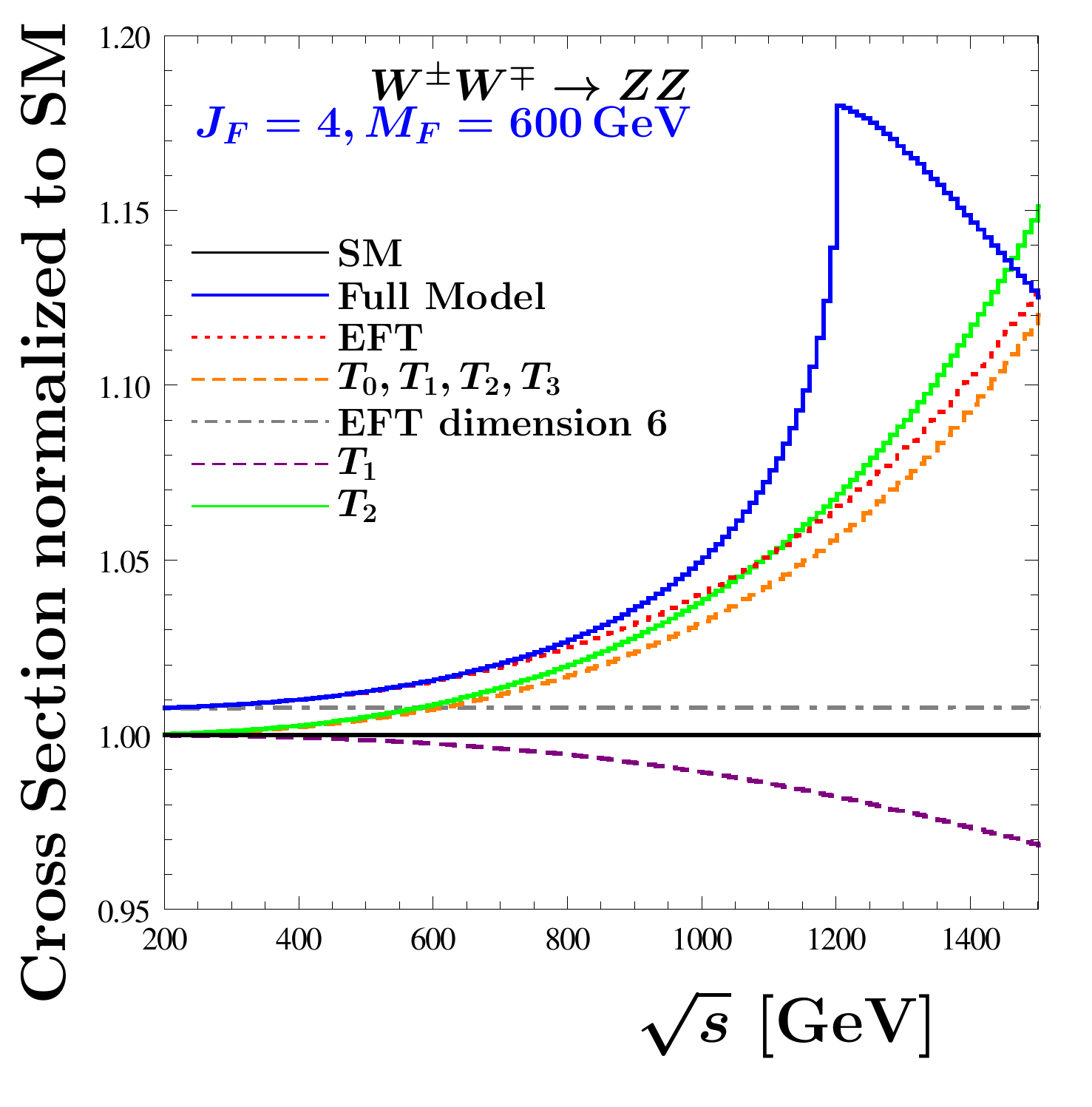}
\end{center}
\caption{Ratio of the unpolarized cross sections for same-sign $WW$ scattering
  (left panel) and the $WW\rightarrow ZZ$ process (right panel) in the single
  fermion multiplet model with $(J_F=4,M_F=600\,$GeV$)$ to the corresponding
  \sm{} cross section in the low energy region. In addition to the curves as
  in \fig{fig:XSonshellFe}, the almost horizontal line at $\approx 1.008$
  represents the result for the EFT at dimension-6 level. See text for details.}
        \label{fig:XSonshellFe_ratioSM}
\end{figure}
Comparing the behavior of the total amplitude squared with the NLO
approximation, we observe discrepancies in the high energy regime, starting
at threshold, especially for the cases of $WZ$ and same-sign $WW$ scattering.
This demonstrates that our model realizations, with large isospin
representations, are only marginally perturbative and might exhibit 
sizable higher order corrections. This is also evident for $ZZ$ scattering
(lower right panels) which is clearly dominated by the anomalous contribution,
i.e. the one-loop amplitudes effectively provide LO estimates, similar to
$gg\to ZZ$ in the \sm{}.

In all cases one finds good accordance between the full model and its
complete \eft{} realization in the low-energy regime.
For the EFT, four different results are shown. The complete ``EFT'' curves
(orange dotted) contain the contributions from all dimension-6 and dimension-8
operators in \sct{sec:coefficients}. They are virtually indistinguishable
from the orange dashed lines, dubbed $T_0,\,,T_1,\,,T_2,\,,T_3$, which only 
incorporate the contributions from the dimension-8 $T$-operators.
This agreement implies that the contributions from the dimension-6 operators
are tiny. In addition, the violet dashed and green solid lines show predictions
for an EFT, where only the $T_1$ or only the $T_2$ operator, with Wilson
coefficients as given by \eqns{eq:fT_coefficients} are included.
The single $T_1$ contribution in the fermion case exhibits the opposite
interference behavior with the \sm{} as compared to the full model and one
observes strong destructive interference between the individual $T$-operators.
For the scalar model, all $T$-operators interfere constructively, which
can be understood by the signs of individual Wilson coefficients 
in \eqn{eq:fT_coefficients}.

\begin{figure}[tb!]
\begin{center}
\includegraphics[page=3,width=0.45\textwidth]{figures/XSplots/XS_Sc_ratioSM.pdf}\hfill
\includegraphics[page=1,width=0.45\textwidth]{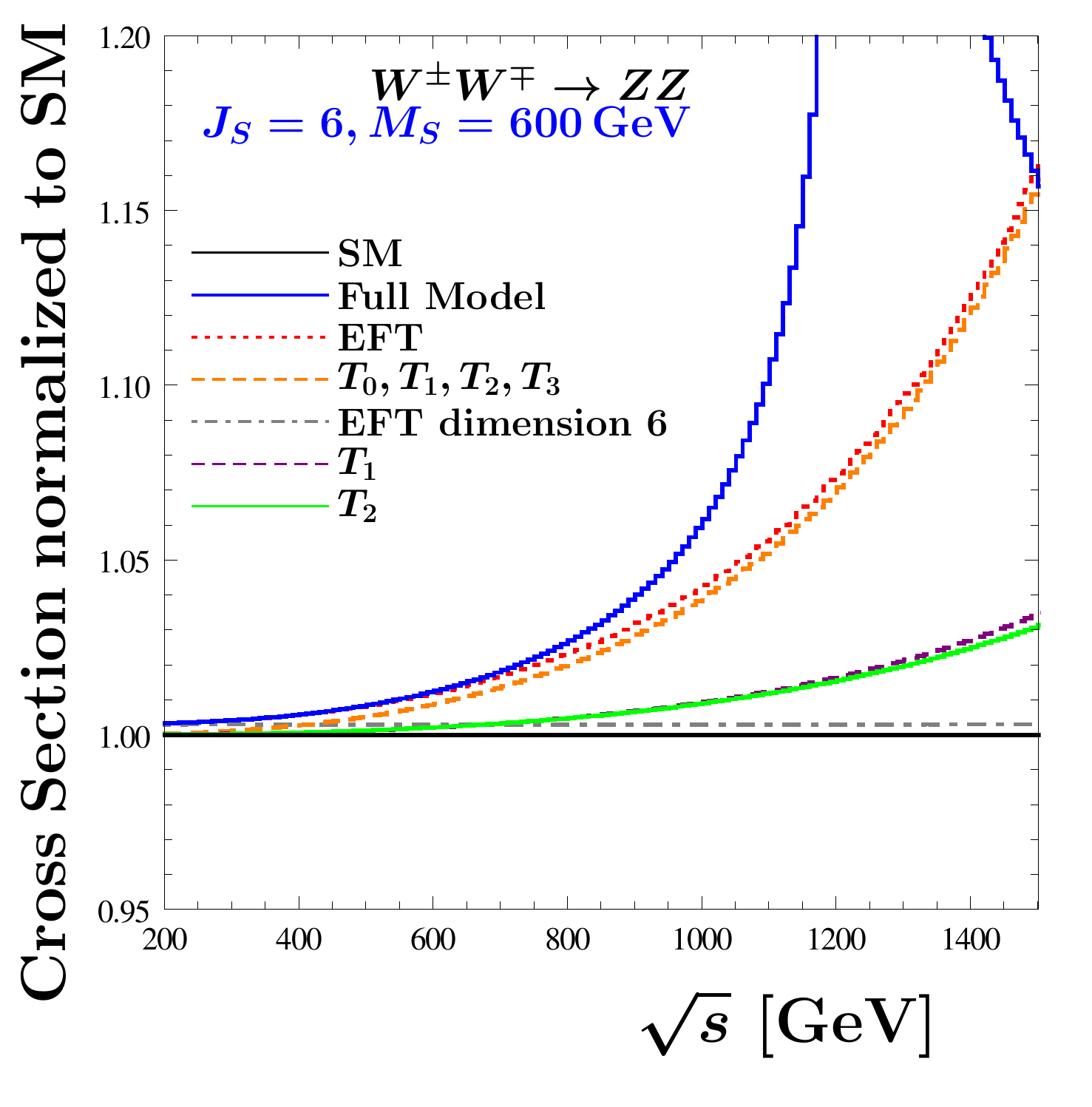}
\end{center}
\caption{Same as \fig{fig:XSonshellFe_ratioSM}, but for a single scalar
  multiplet with $(J_S=6,M_S=600\,$GeV$)$. }
        \label{fig:XSonshellSc_ratioSM}
\end{figure}
For a better evaluation of the validity of the \eft{} and its subsets
in the low-energy region, we present in \fig{fig:XSonshellFe_ratioSM}
and \fig{fig:XSonshellSc_ratioSM}
the cross sections, normalized to the \sm{} prediction, for same-sign $WW$
scattering and the $WW\rightarrow ZZ$ process. As discussed above,
results for $WZ$ scattering are qualitatively the same as for same-sign
$WW$ scattering.

The \eft{} agrees well with the full model for energies below
$\sim 800\,$GeV, which is consistent with its naive validity range, given by
the mass scale of the loop particle. In the fermion case, the discrepancy
stays below $\sim 1\,\%$ up to 1000~GeV, but would grow with $J_F^5$.
We also note that the \eft{} dimension-6 contribution only accounts for a
slowly varying offset of less than $1\,\%$ to the pure \sm{} cross section
(gray dash-dotted line).

\section{Implication for $VVjj$ events at the LHC}
\label{sec:VBFNLO}

The analysis of the last section provided cross sections for on-shell
$2\to 2$ VBS processes, in the $g'=0$ limit, and thus was largely academic.
A more realistic simulation for the
LHC needs to consider the full $qQ \to qQVV$ processes, including decay of
the produced electroweak gauge bosons, it needs to consider the off-shell
nature of both the space-like initial and the time-like final gauge bosons,
and it needs to go beyond the SU$(2)_L$ limit and incorporate that the $W^3$
is actually a linear combination of photon and $Z$ as mass eigenstates. 
While a full implementation of the model into {\tt VBFNLO} is foreseen for the
future, here we only make a first, approximate assessment of
consequences for the LHC.

A prominent feature of typical VBS kinematics consists of small virtualities
for all four electroweak bosons in the $V_1V_2\to V_3V_4$ subgraphs,
$|p^2_i|\ll s=m_{VV}^2$. At such small $p_i^2$, modifications of the weak boson
propagators are tiny, as evidenced by \fig{fig:Drell-Yan}. Also corrections to
$WWZ$ or $WW\gamma$ vertices, which appear for a high virtuality gauge boson
which is radiated off a quark line in $WZjj$ and $W^+W^-jj$ production,
are sub-dominant as was seen in the discussion of aTGCs in \sct{sec:WCbounds}.
The remaining BSM effects in our model then contain $V_1V_2\to V_3V_4$
subgraphs with off-shell helicity amplitudes
$\mathcal{M}^{VBS}_{\lambda_1\lambda_2\lambda_3\lambda_4}(\{p_i^2\};s,t,u)$.
Following \citere{Perez:2018kav}, the BSM part, which needs to be added to the
SM $qq\to qq\bar f_1 f_2 \bar f_3 f_4$ amplitude is then given by
\begin{align}
\label{eq:amplitude_decomposed}
\MM_{qq\rightarrow 4f qq}^{\rm BSM} = &
\prod_{i=1}^4 \frac{1}{p_i^2-m_{V_i}^2+i\,m_{V_i}\,\Gamma_{V_i}} 
\sum_{\{\lambda_i\}}
\epsilon^*_{J}(p_1,\lambda_1)\cdot J_{q \rightarrow qV_1} \,\,
\epsilon^*_{J}(p_2,\lambda_2)\cdot J_{q \rightarrow qV_2} \, \\ \nonumber
  & \MM^{VBS}_{\lambda_1\lambda_2\lambda_3\lambda_4}(\{p_i^2\};s,t,u) \,\,
  \epsilon_{J}(p_3,\lambda_3)\cdot J_{V_3 \rightarrow \bar{f_1}f_2} \,\,
  \epsilon_{J}(p_4,\lambda_4)\cdot J_{V_4 \rightarrow \bar{f_3}f_4} \, .
\end{align}
Here the $\epsilon_{J}(p_i,\lambda_i)$ are off-shell polarization vectors
(as defined in \citere{Perez:2018kav}) and the $J_{q \rightarrow qV_i}$ and 
$J_{V_i \rightarrow \bar{f_k}f_l}$ are quark- and lepton currents. A final
approximation now consists in replacing the off-shell helicity amplitudes
$\mathcal{M}^{VBS}_{\lambda_1\lambda_2\lambda_3\lambda_4}(\{p_i^2\};s,t,u)=
\mathcal{M}^{VBS}_{\lambda_1\lambda_2\lambda_3\lambda_4}(\{p_i^2\};s,\theta)$,
determined in the VBS center-of-mass frame, by their on-shell values 
$\mathcal{M}^{VBS}_{\lambda_1\lambda_2\lambda_3\lambda_4}(\{m_W^2\};s,\theta)=
\mathcal{M}_{\lambda_1\lambda_2\lambda_3\lambda_4}(s,\theta)$
of \eqn{eq:pw_exp}. We have verified that the above approximations are good
at the 10\% level by comparing on- and off-shell {\tt VBFNLO} results for the
dimension-8 $T$-operators~\cite{Master-Thesis}. 

In the next step we use our new {\tt VBFNLO} implementation of full on-shell
model amplitudes to study fiducial cross sections for the various VBS
processes. We want to find out to which degree ATLAS and CMS strategies for
measuring anomalous quartic gauge couplings\footnote{A nice summary on
  bounds on anomalous triple and quartic gauge couplings, from the analysis
  of diboson and triboson final states and from different experiments is
  provided in \citere{SummaryWebpage}.}
would also reveal high multiplicity extra matter multiplets, via their impact
on weak boson 4-point functions.
All results are produced for a proton-proton collider with a center-of-mass
energy of $13$\,TeV using the {\tt MMHT2014}~\cite{Harland-Lang:2014zoa}
parton distribution functions at leading order, which are linked through
{\tt LHAPDF}~\cite{Buckley:2014ana}. The Higgs-boson mass is set to
$125.09$\,GeV. For the $W$ and $Z$ boson mass {\tt VBFNLO} uses
$M_W=80.398$\,GeV and $M_Z=91.1876$\,GeV, respectively.
The weak mixing angle is set to $\sin^2\theta_W=0.222646$. However we remind
the reader that the inserted on-shell amplitudes are derived in the
SU$(2)_L$ limit, in which $M_W=M_Z=81.18$\,GeV. When we depict the number of
events we assume an integrated luminosity of $137$\,fb$^{-1}$.

\subsection{Destructive interference in same-sign $WW$ and $WZ$ scattering}

\begin{figure}[tb!]
\begin{center}
\includegraphics[width=0.5\textwidth]{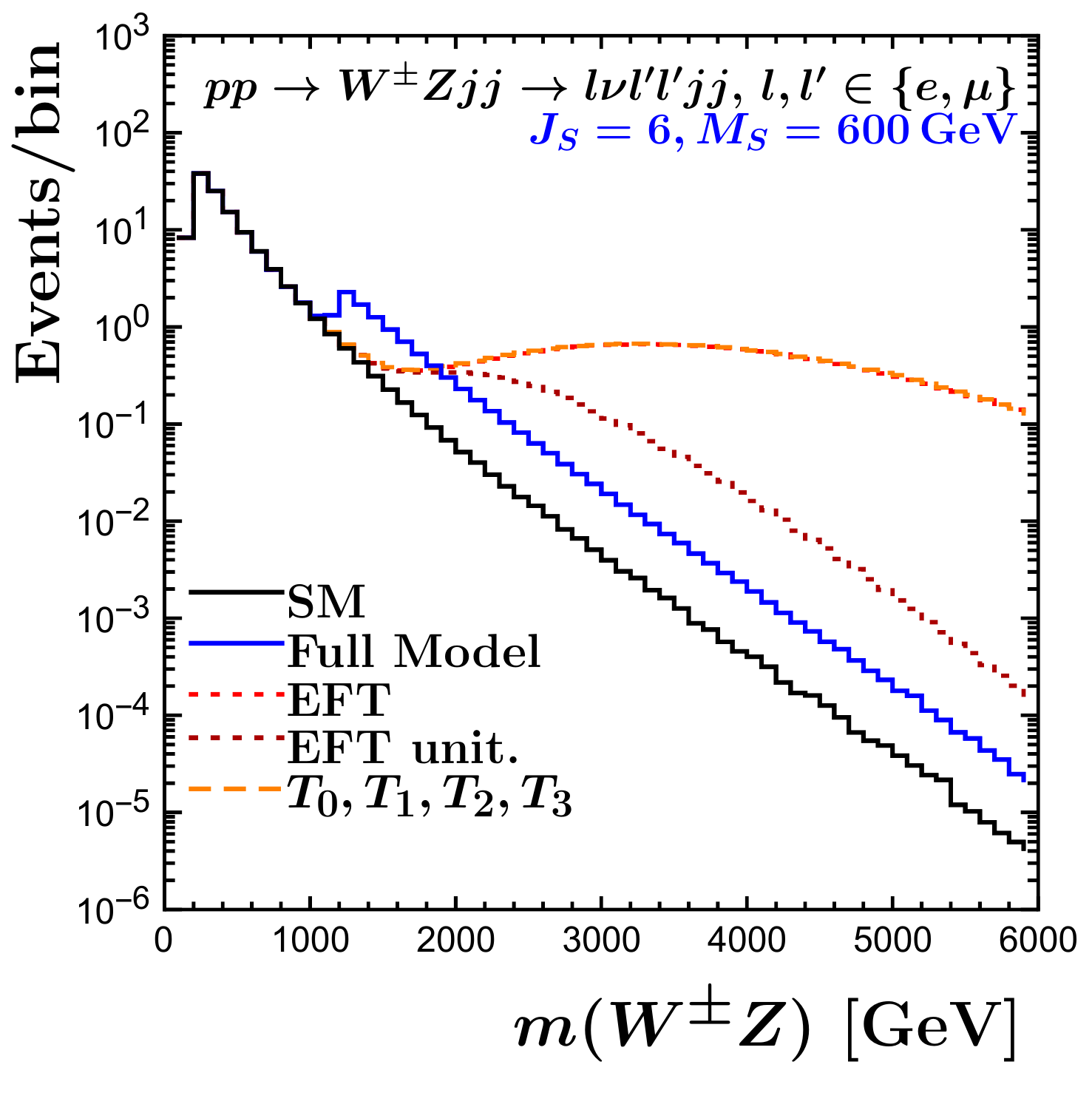}\hfill
\includegraphics[width=0.5\textwidth]{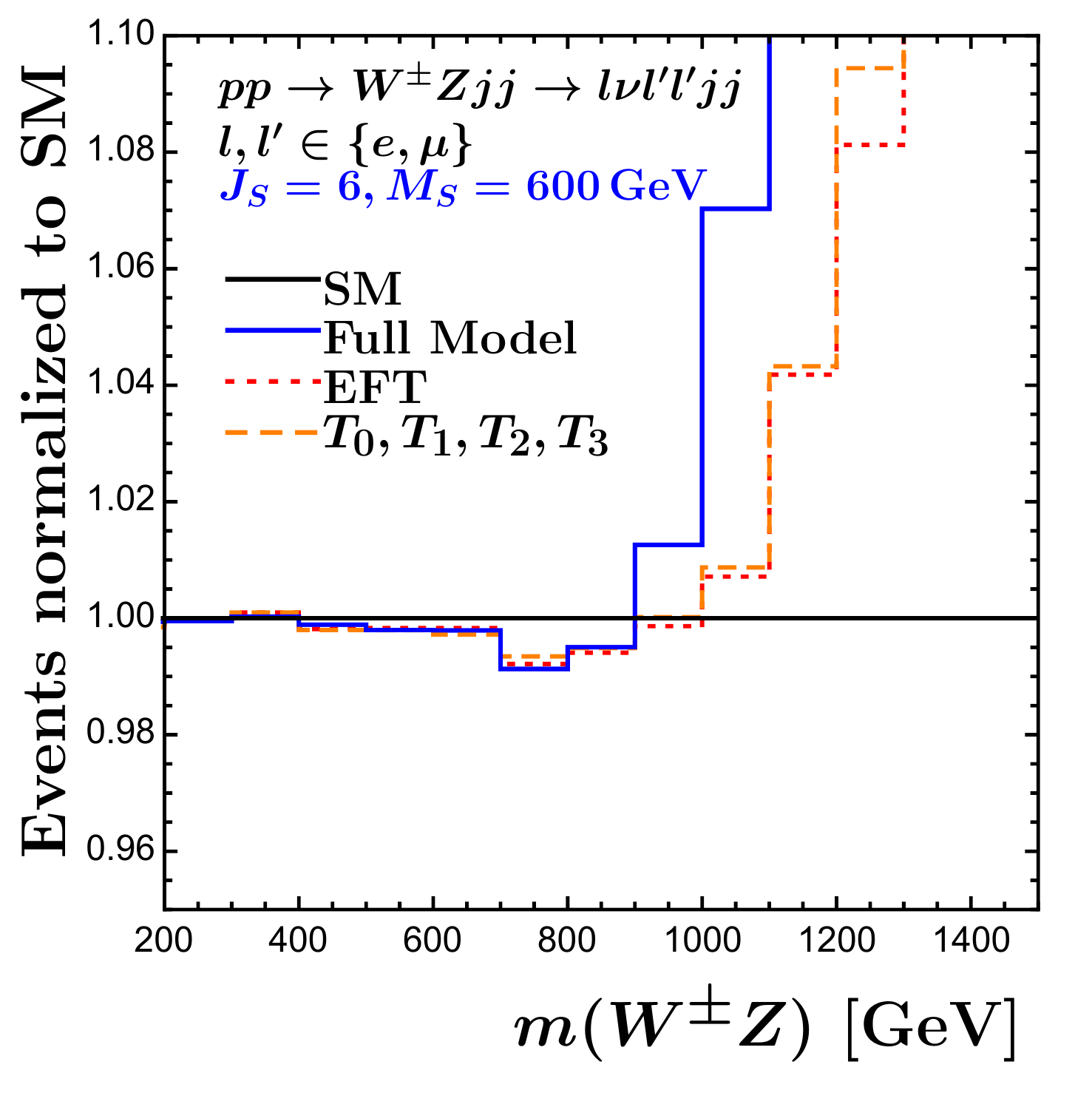}\\
\includegraphics[width=0.5\textwidth]{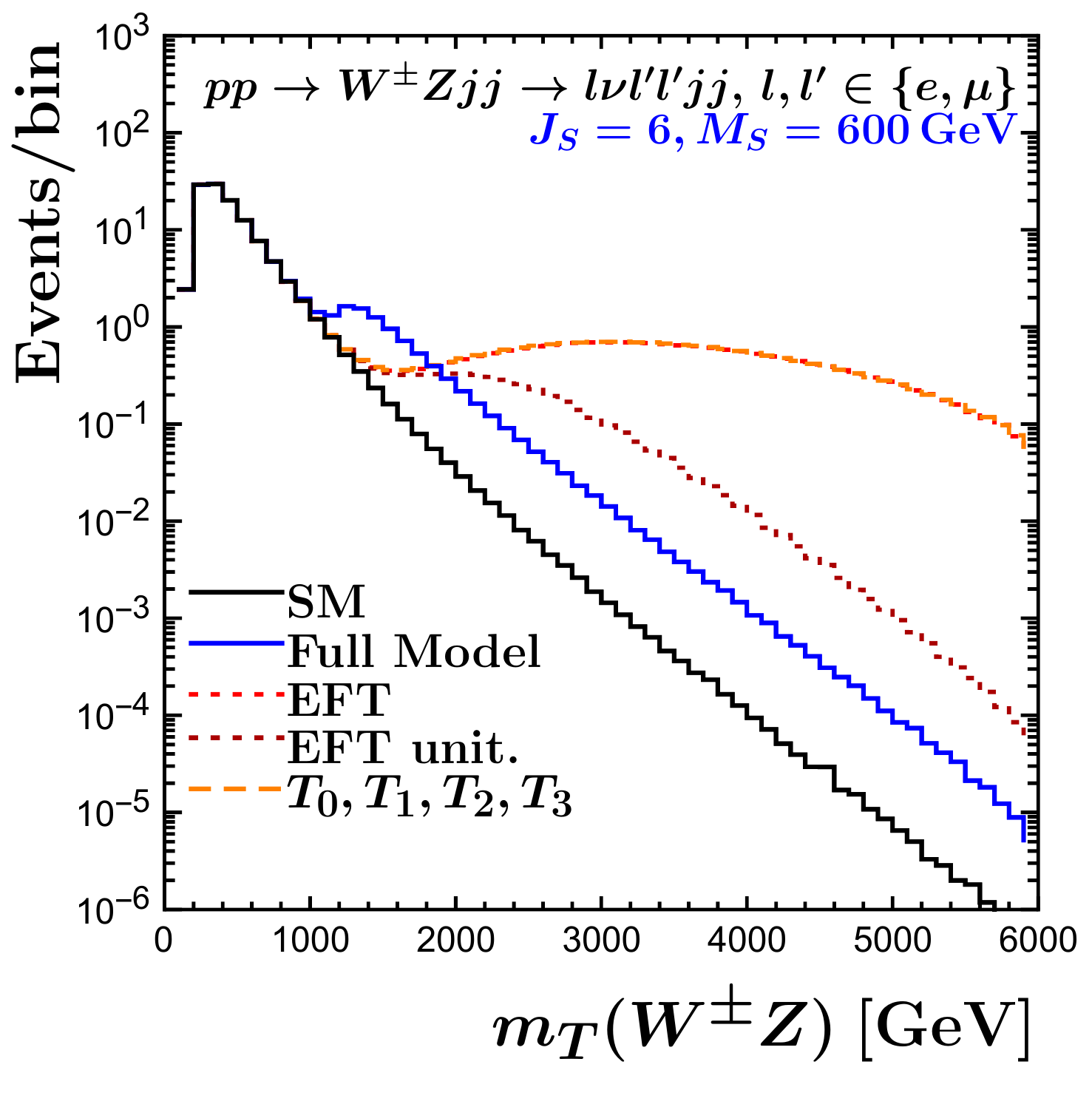}\hfill
\includegraphics[width=0.5\textwidth]{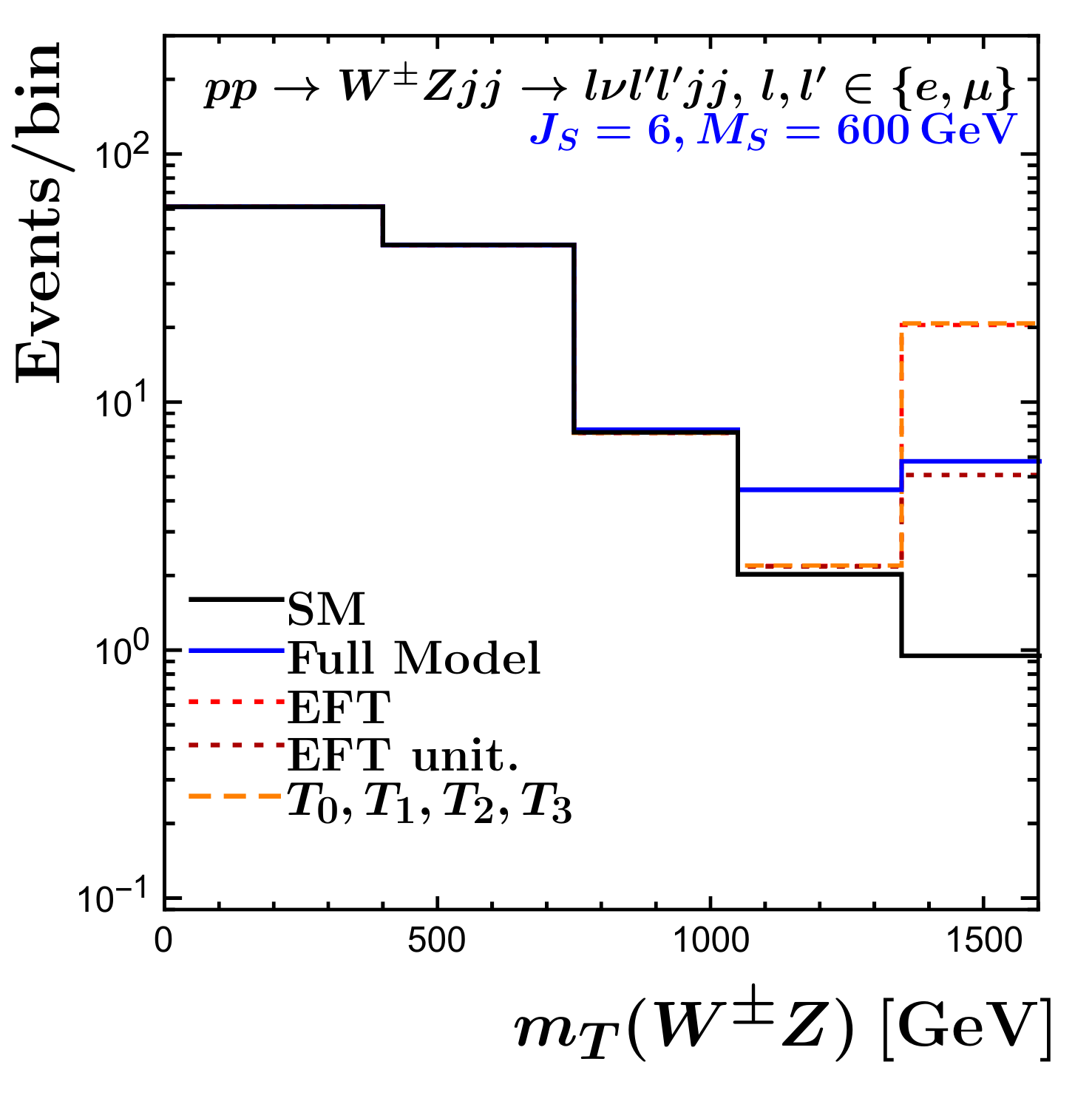}
\end{center}
\caption{$WZ$ invariant mass (upper row) and transverse mass (lower row)
  distributions for VBS production of $WZjj$ events at the LHC. Three panels
  show expected event numbers per bin for a run 2 integrated luminosity of
  $137$\,fb$^{-1}$ while the upper right panel is normalized to the SM. Bin
  width in the two left panels is $100$\,GeV, while on the lower right bin
  size is chosen as in \citere{Sirunyan:2020gyx}. The considered scenario is
  a scalar multiplet with $J_S=6$ and $M_S=600$\,GeV.}
\label{fig:VBFNLOWZs}
\end{figure}

\begin{figure}[tb!]
\begin{center}
\includegraphics[width=0.5\textwidth]{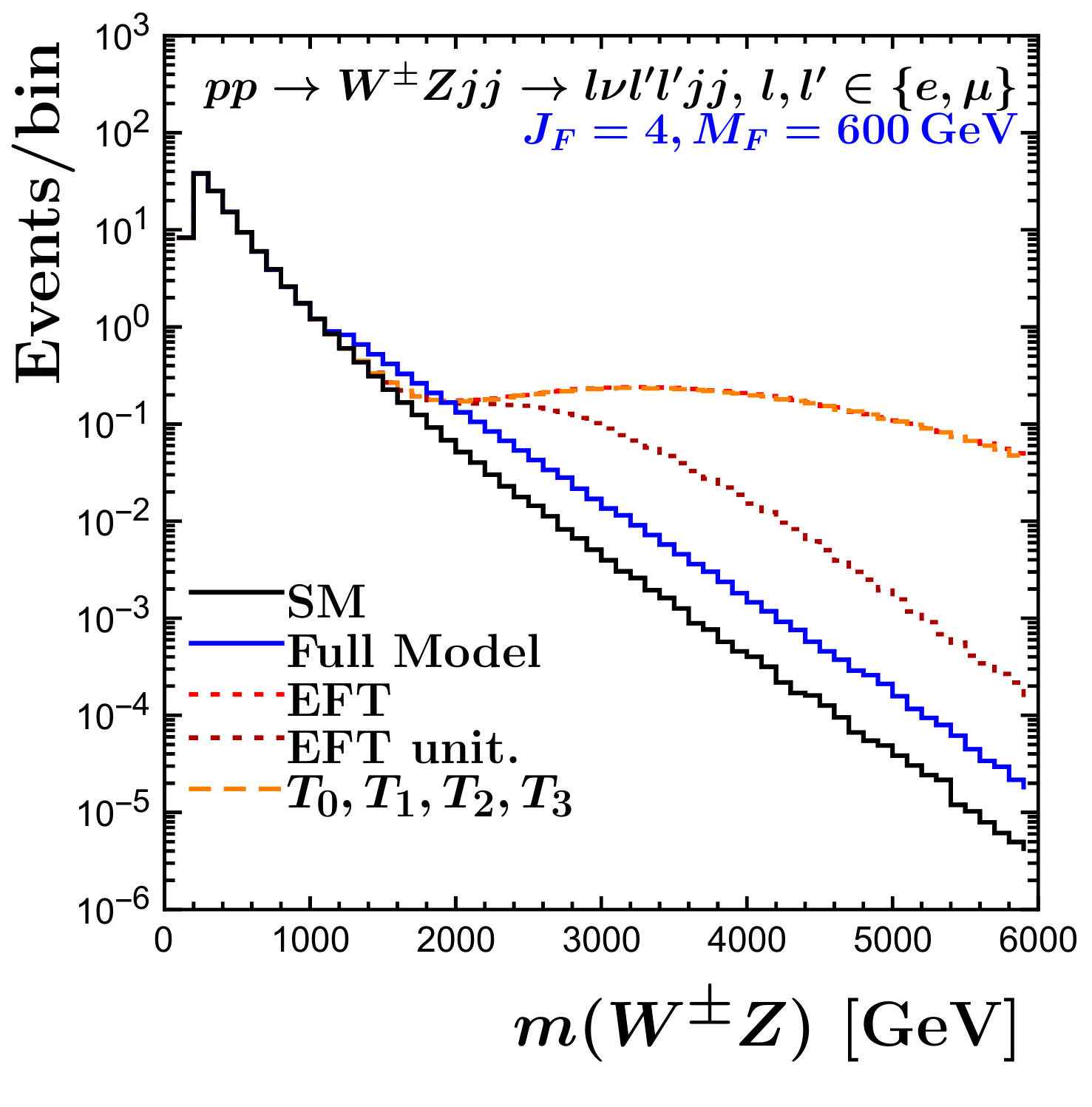}\hfill
\includegraphics[width=0.5\textwidth]{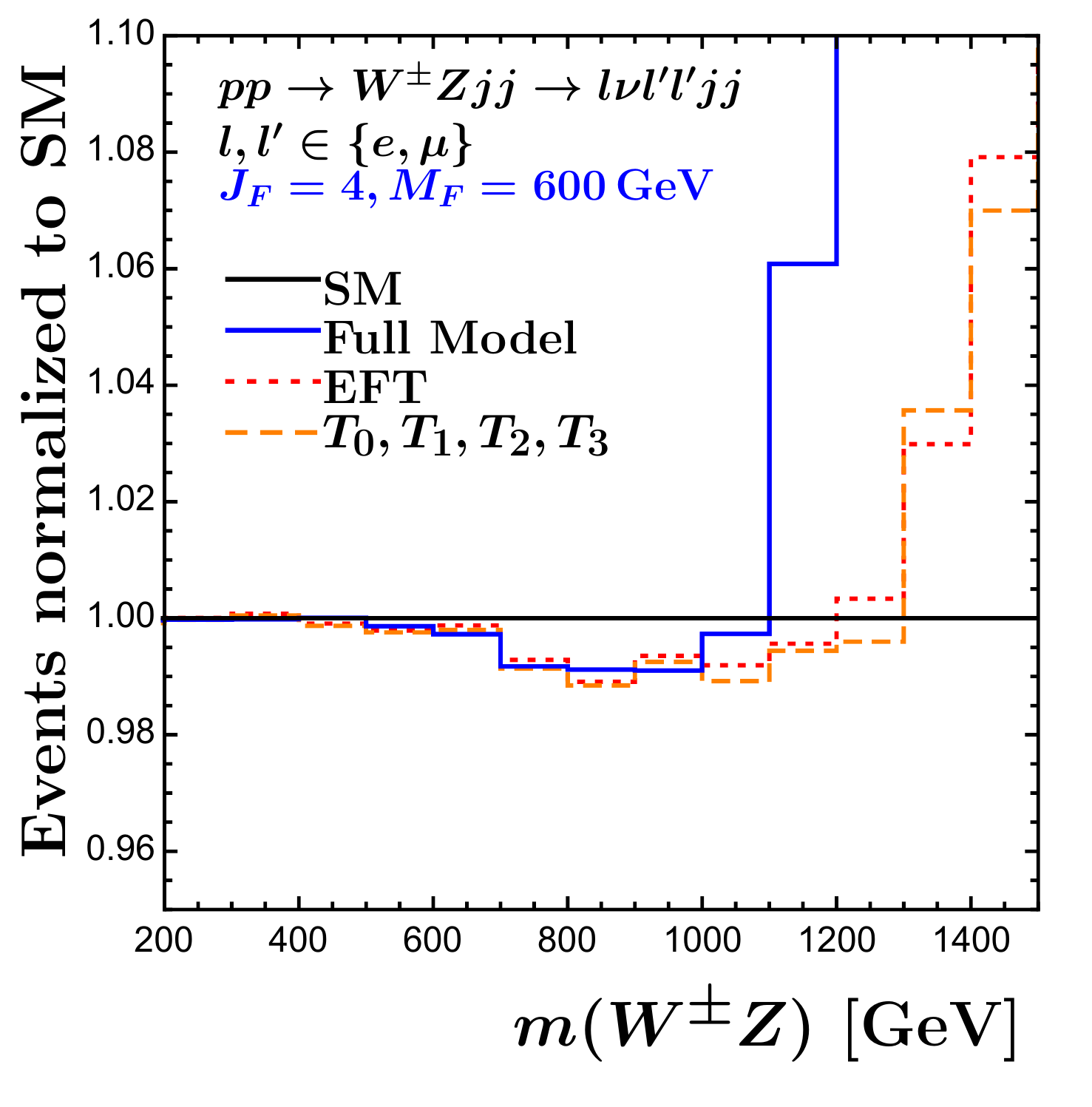}\\
\includegraphics[width=0.5\textwidth]{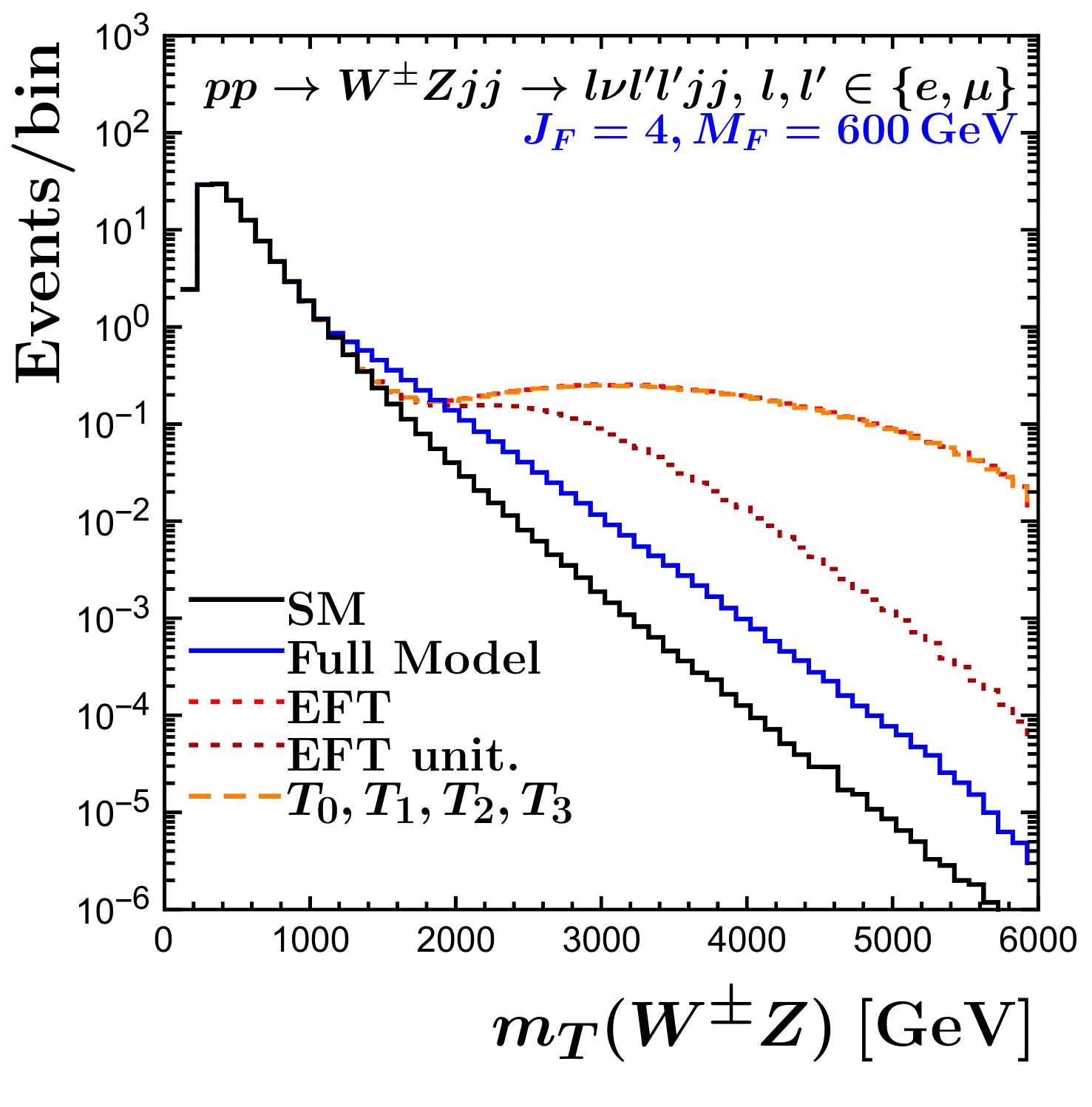}\hfill
\includegraphics[width=0.5\textwidth]{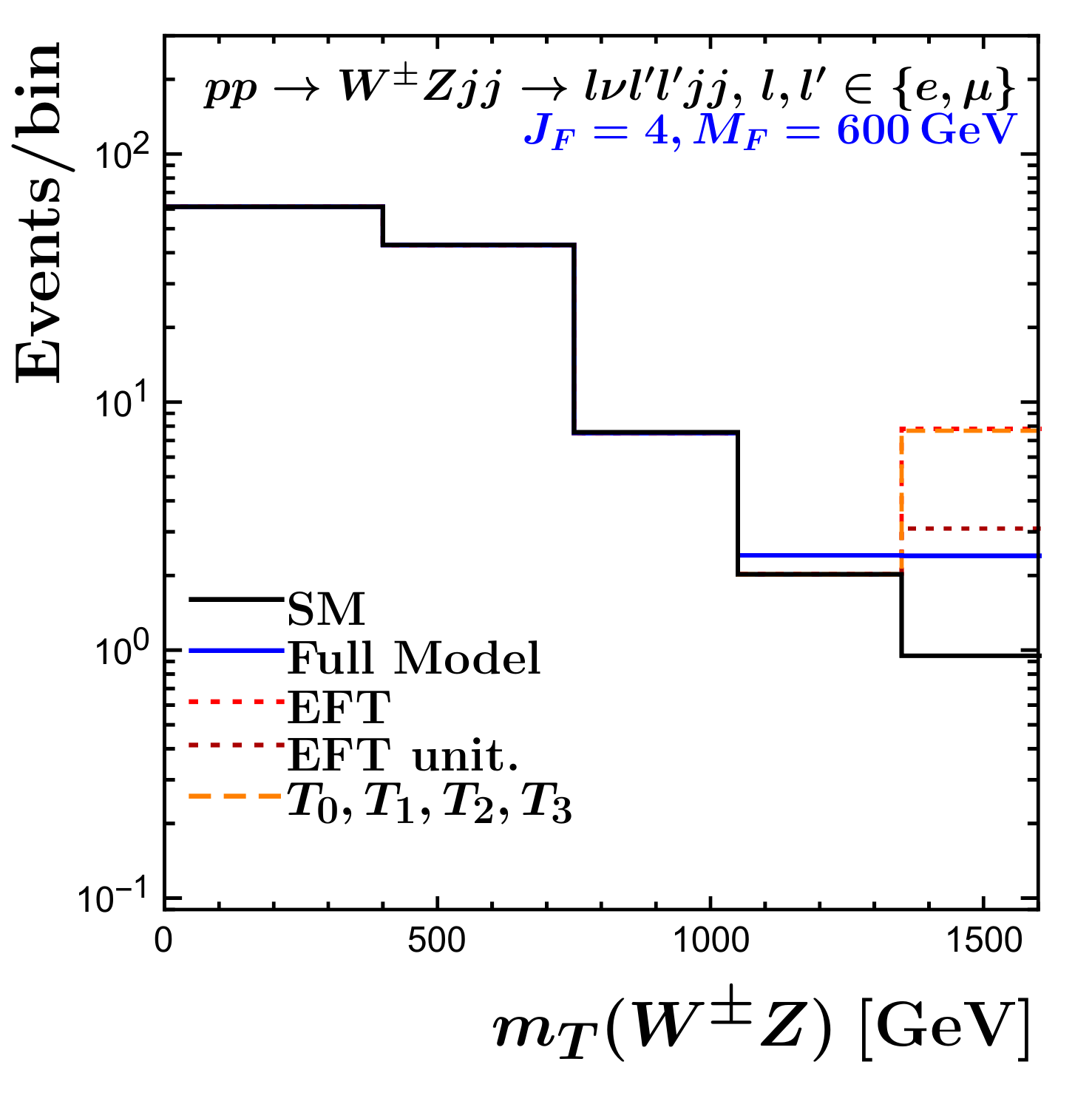}
\end{center}
\caption{Same as \fig{fig:VBFNLOWZs}, but for a fermionic multiplet with
  $J_F=4$ and $M_F=600$\,GeV.}
\label{fig:VBFNLOWZf}
\end{figure}

We use {\tt VBFNLO} to calculate the cross sections for the electroweak
processes
\begin{align}\nonumber
 &pp\to W^\pm Zjj\to l'^\pm \nu_{l'} l^\pm l^\mp jj\\
 &pp\to W^\pm W^\pm jj\to l^\pm \nu_{l} l'^\pm \nu_{l'}
\end{align}
with $l,l'\in \{e,\mu\}$, which were exploited in
\citeres{Sirunyan:2019ksz,Sirunyan:2020gyx} to set bounds on anomalous quartic gauge
couplings. In order to make a qualitative comparison with
the results of \citere{Sirunyan:2020gyx}, in particular their Fig.~6,
we have implemented a very similar cut-flow.
For the $WZ$ final state this cut-flow includes:
\begin{align}\nonumber
 p_T^l&>20\,\text{GeV}\,, &\qquad |\eta^e|&<2.5\,, &\qquad |\eta^\mu|&<2.4\\\nonumber
 |m_{ll}-m_Z|&<15\,\text{GeV}\,, & \qquad m_{3l}&>100\,\text{GeV}\,,&\qquad p_T^{\text{miss}}&>30\,\text{GeV}\\\nonumber
 |\eta^j|&<4.7\,, &\qquad p_T^j&>50\,\text{GeV}\,,&\qquad |\Delta R(j,l)|&>0.4\\
 m_{jj}&>500\,\text{GeV}\,,&\qquad |\Delta\eta_{jj}|&>2.5\,,&\qquad \text{max}(z_l^*)&<1.0\,.
 \label{eq:cutflowwz}
\end{align}
Therein the abbreviation ``$3l$'' denotes the three-lepton system, i.e. the
sum of the three lepton 4-momenta, and its invariant mass $m_{3l}$.
In contrast to the CMS analysis we only simulate the flavor combination
$l'=e,l=\mu$ and as an estimate for all flavor combinations multiply our
results by a factor of $4$. As a consequence we have just one cut on the
transverse momentum $p_T^l$, which is chosen equal for all leptons.
The cuts depicted in the last two lines of \eqn{eq:cutflowwz} are typical
vector boson scattering cuts, which enhance
the contribution of electroweak VBS over QCD-induced $VVjj$ events and
other SM backgrounds.
\begin{align}
  z_l^*=|\eta^l-\tfrac{1}{2}(\eta^{j_1}+\eta^{j_2})|/|\Delta \eta_{jj}|
  \label{eq:Zeppenfeld}
\end{align}
denotes the variable introduced in~\citere{Rainwater:1996ud}.

The cut-flow for the same-sign $WW$ final state is very similar: for charged
leptons we merely replace the $m_{3l}$ cut by $m_{ll}>20$\,GeV. Furthermore
we set $\text{max}(z_l^*)<0.75$. In contrast to \citere{Sirunyan:2020gyx} we
do not have a cut on $|m_{ee}-m_Z|$, as we generate our result
for the flavor combination $l'=e,l=\mu$ and multiply by $2$.

We present our results primarily for the scalar case with the combination
$J_S=6$ and $M_S=600$\,GeV which, according to the discussion in the previous
sections, may be considered a maximal plausible signal within our framework.
The results for the fermionic case with $J_F=4$ and $M_F=600$\,GeV, which
is disfavored by Drell-Yan data, do not provide much additional insight and
differ visibly only in the vicinity of the threshold peak, which is more
pronounced in the scalar case. We present the number of events for an
integrated luminosity of $137$\,fb$^{-1}$ as a function of the invariant
mass of the final state $WW$ and $WZ$ system which, however, is not
accessible experimentally. Therefore, also for a direct comparison with
Fig. 6 of \citere{Sirunyan:2020gyx}, we add figures showing the transverse
mass as defined there. 

For the $W^\pm Z$ final state the results for the scalar case are depicted
in \fig{fig:VBFNLOWZs} and for the fermionic case in \fig{fig:VBFNLOWZf}.
The $m(W^\pm Z)$ invariant mass distribution in \fig{fig:VBFNLOWZs} (upper left)
and its ratio to the SM contribution (upper right) show destructive
interference between the multiplet contribution and the SM well below the
threshold at $m(W^\pm Z)\approx 2M_F=1200$\,GeV. The EFT description follows
this behavior until around $m(W^\pm Z)\approx 1000$\,GeV, after which it clearly
deviates from the full-model prediction. In this region of reliable EFT
description the BSM effects are tiny, however, deviating from the SM by at most
1\%, which renders their observation hopeless. The figures also depict the
contribution from the combined dimension-8 $T_i$ operators only, which well
coincides with the full EFT description. This exacerbates the observation in
\sct{sec:VBS-XS} that the contribution of the dimension 6 operators is
sub-dominant and irrelevant for VBS in practice. 
The lower panels of \fig{fig:VBFNLOWZs} show the transverse momentum
distribution, in which the
destructive interference is not visible anymore, due to the migration of
excess events at higher invariant mass into the low transverse mass region.
The lower right panel reproduces the binning of the events in Fig. 6
of \citere{Sirunyan:2020gyx}. Comparison of the last
two bins with the data (which are compatible with the SM) reveals that the
full fermion model in \fig{fig:VBFNLOWZf} cannot be excluded, whereas
scalar model at $J_S=6$ in \fig{fig:VBFNLOWZs} should have been seen with
considerable significance. Similarly, the sizable signals of the
non-unitarized EFT's in the last (overflow) bin are excluded. Also
shown are unitarized versions of the EFT description,
following the unitarization procedure of \citere{Perez:2018kav}, which stays
closer to the full model description in both the invariant and the transverse
mass distributions.

\begin{figure}[tb!]
\begin{center}
\includegraphics[width=0.5\textwidth]{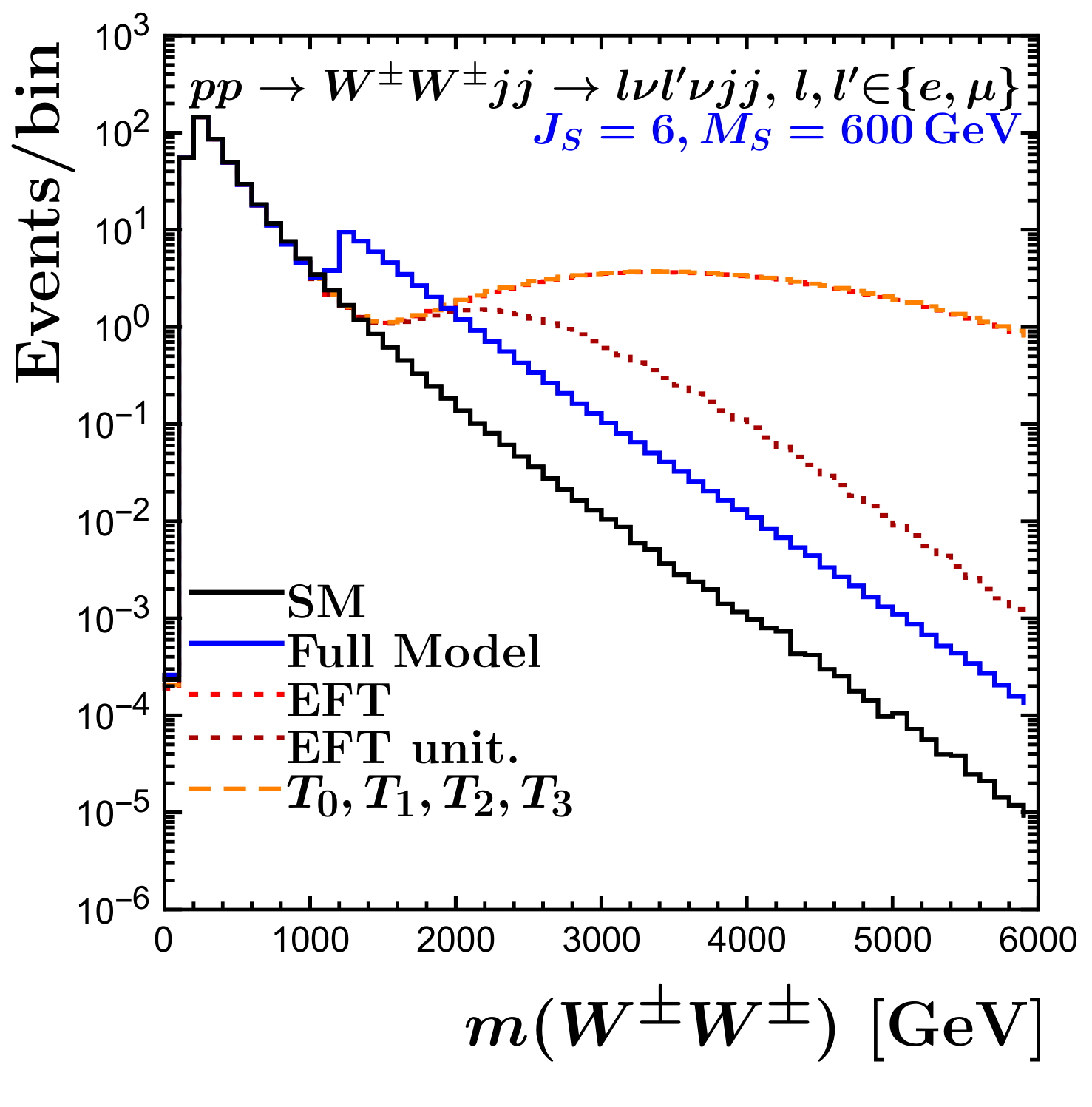}\hfill
\includegraphics[width=0.5\textwidth]{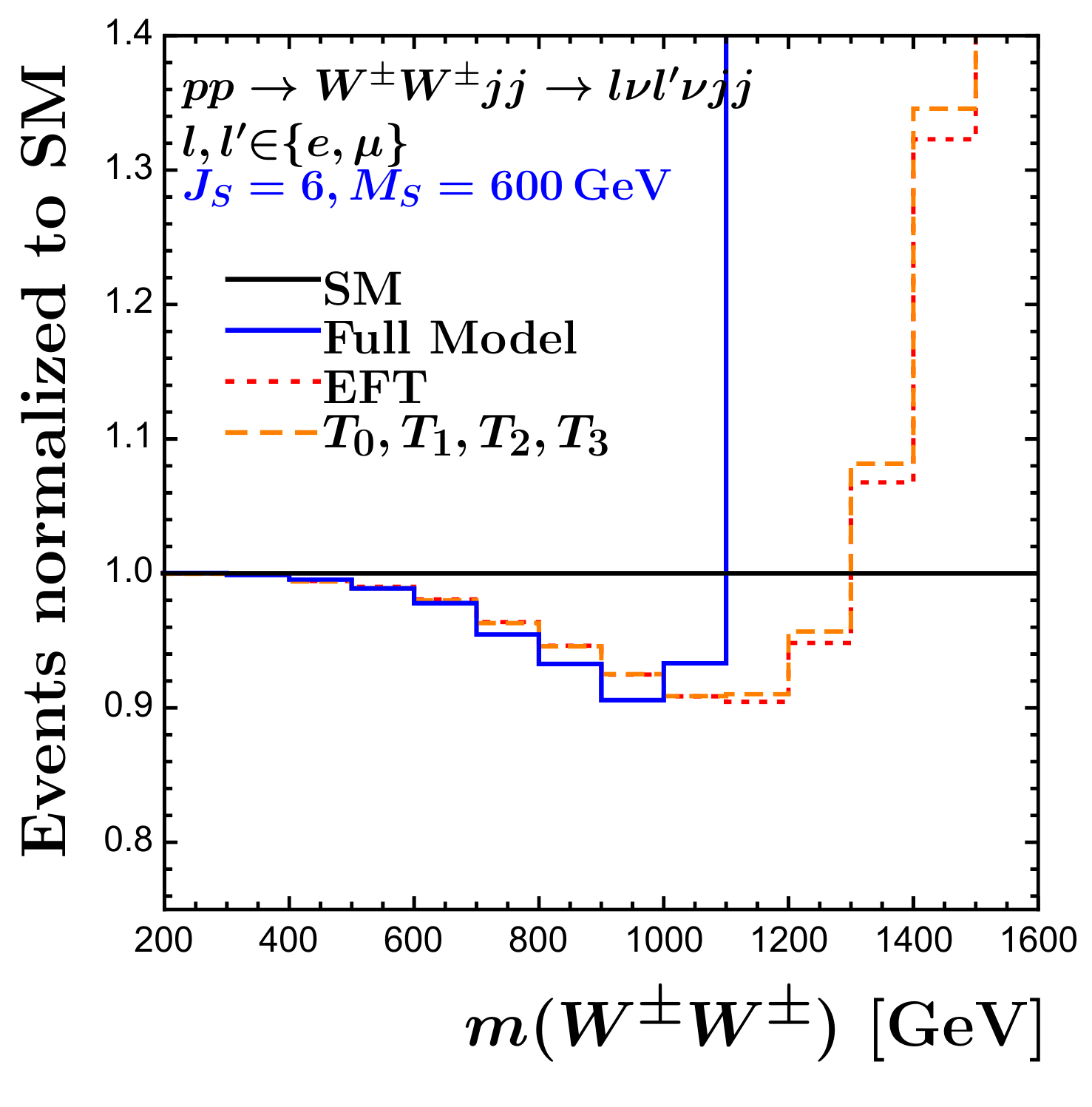}\\
\includegraphics[width=0.5\textwidth]{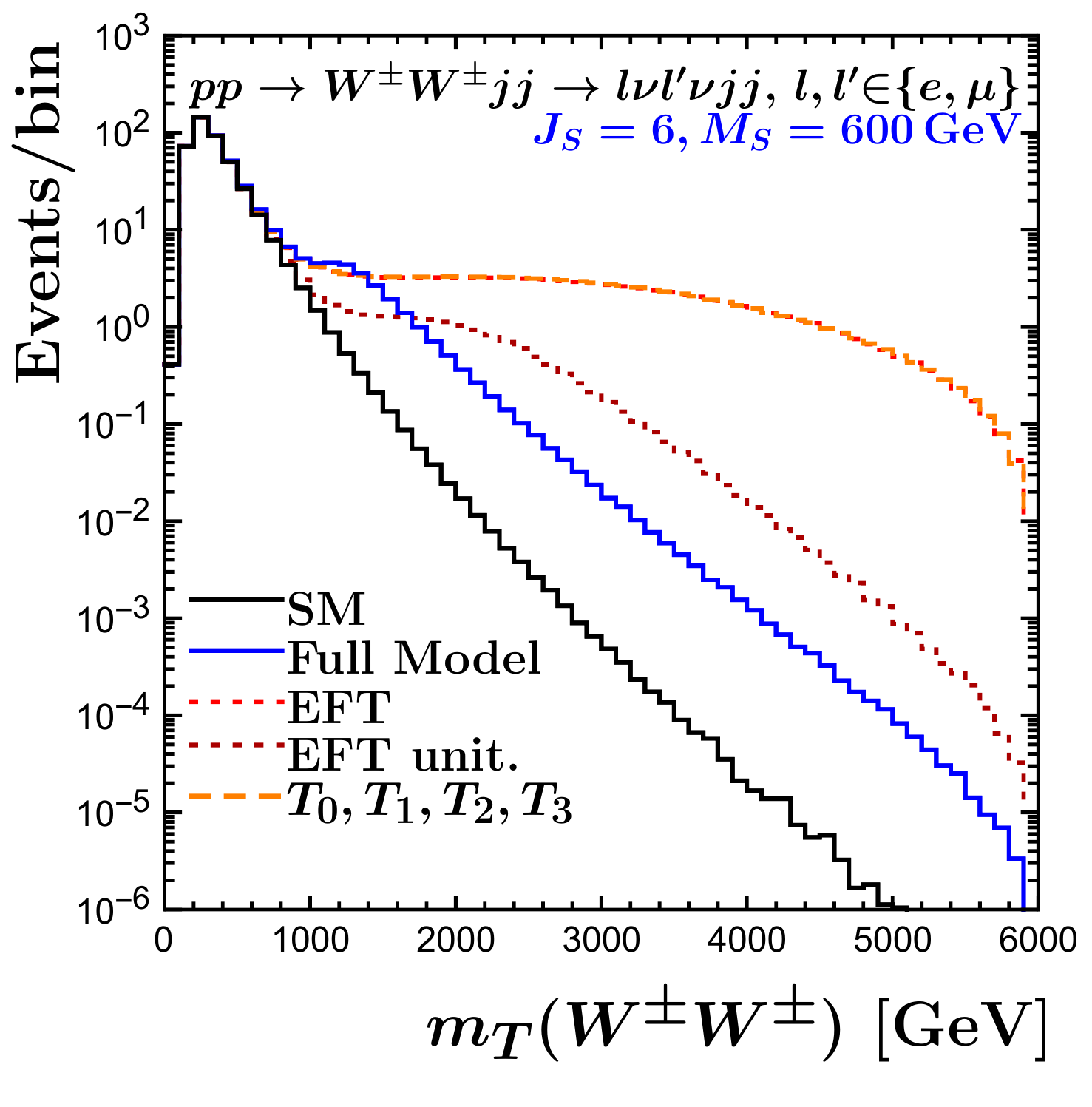}\hfill
\includegraphics[width=0.5\textwidth]{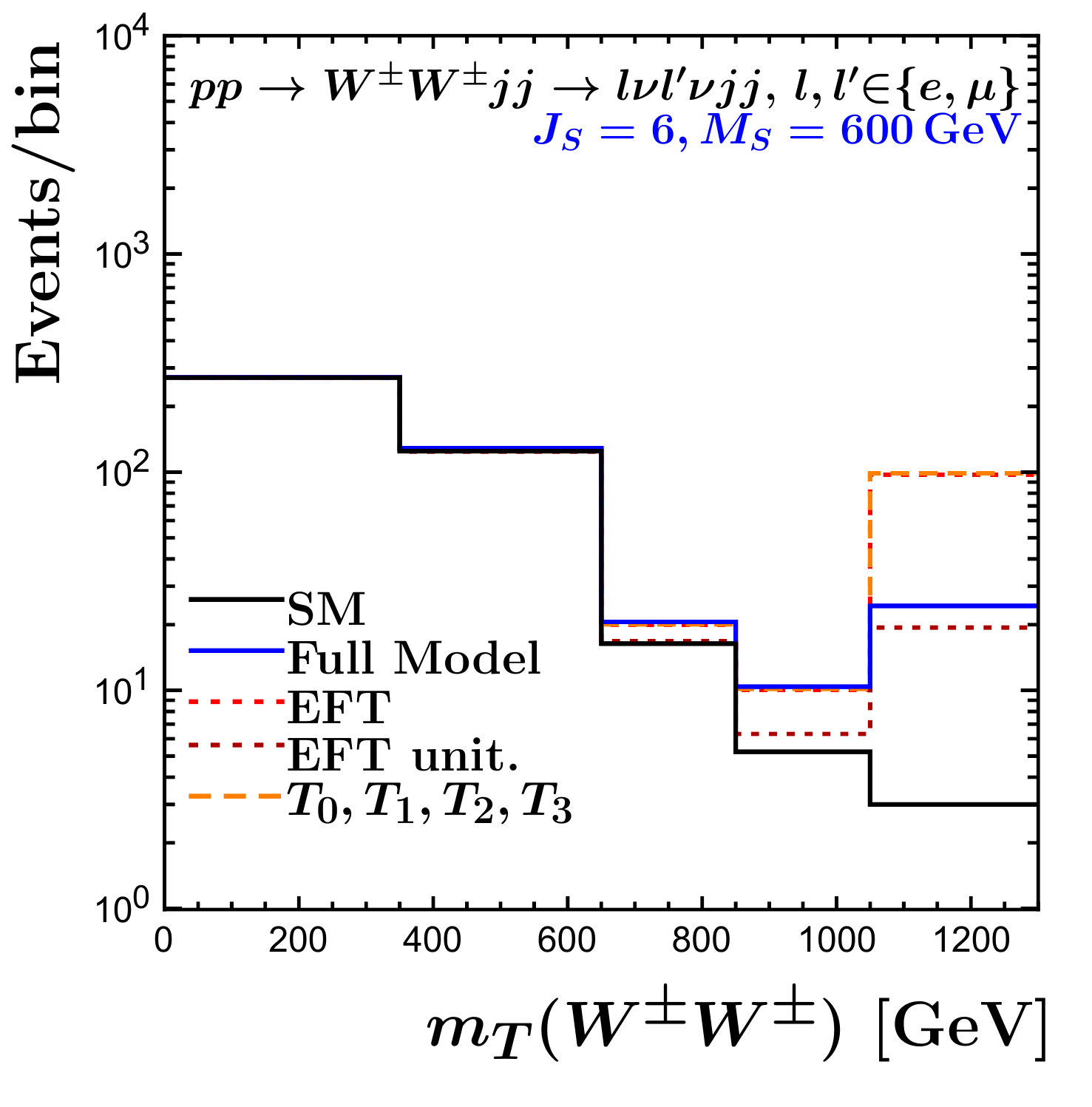}
\end{center}
\caption{$WW$ invariant mass (upper row) and $ll p_T^{\rm miss}$ transverse mass
  (lower row) distributions for VBS production of same-sign $WWjj$ events at
  the LHC. Three panels
  show expected event numbers per bin for a run 2 integrated luminosity of
  $137$\,fb$^{-1}$ while the upper right panel is normalized to the SM. Bin
  width in the two left panels is $100$\,GeV, while on the lower right bin
  size is chosen as in \citere{Sirunyan:2020gyx}. The considered scenario is
  a scalar multiplet with $J_S=6$ and $M_S=600$\,GeV.}
\label{fig:VBFNLOWWs}
\end{figure}

\begin{figure}[tb!]
\begin{center}
\includegraphics[width=0.5\textwidth]{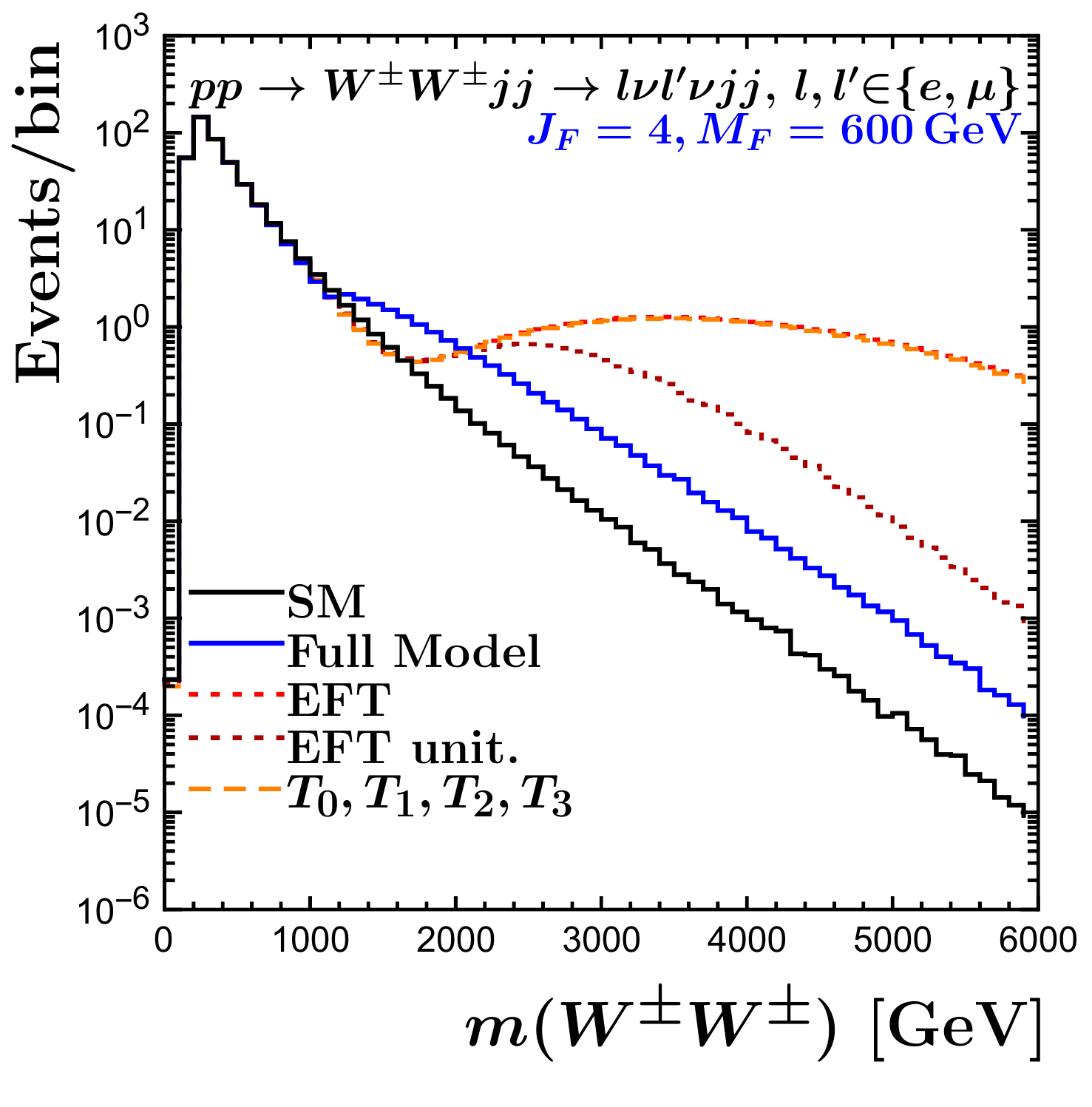}\hfill
\includegraphics[width=0.5\textwidth]{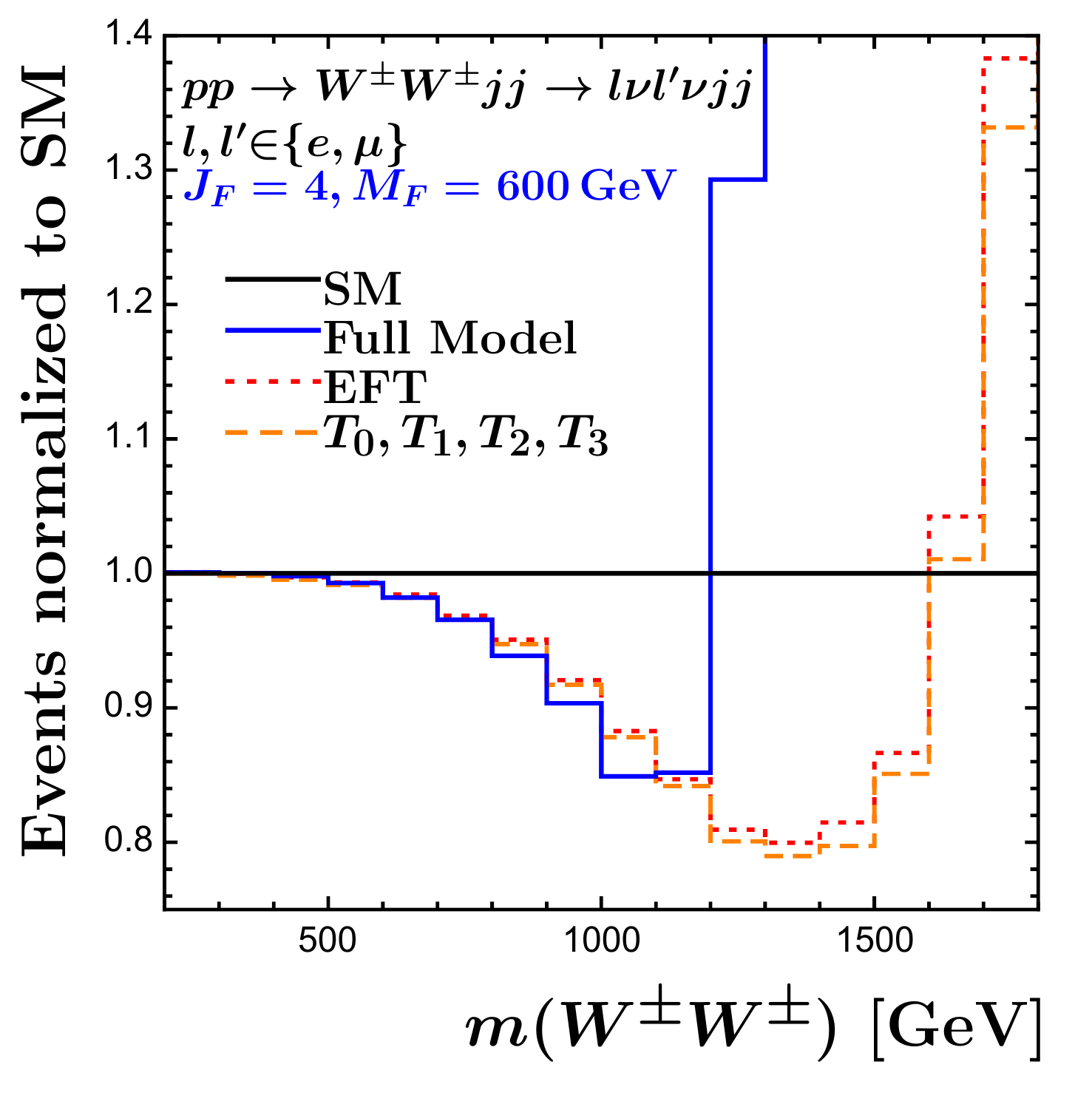}\\
\includegraphics[width=0.5\textwidth]{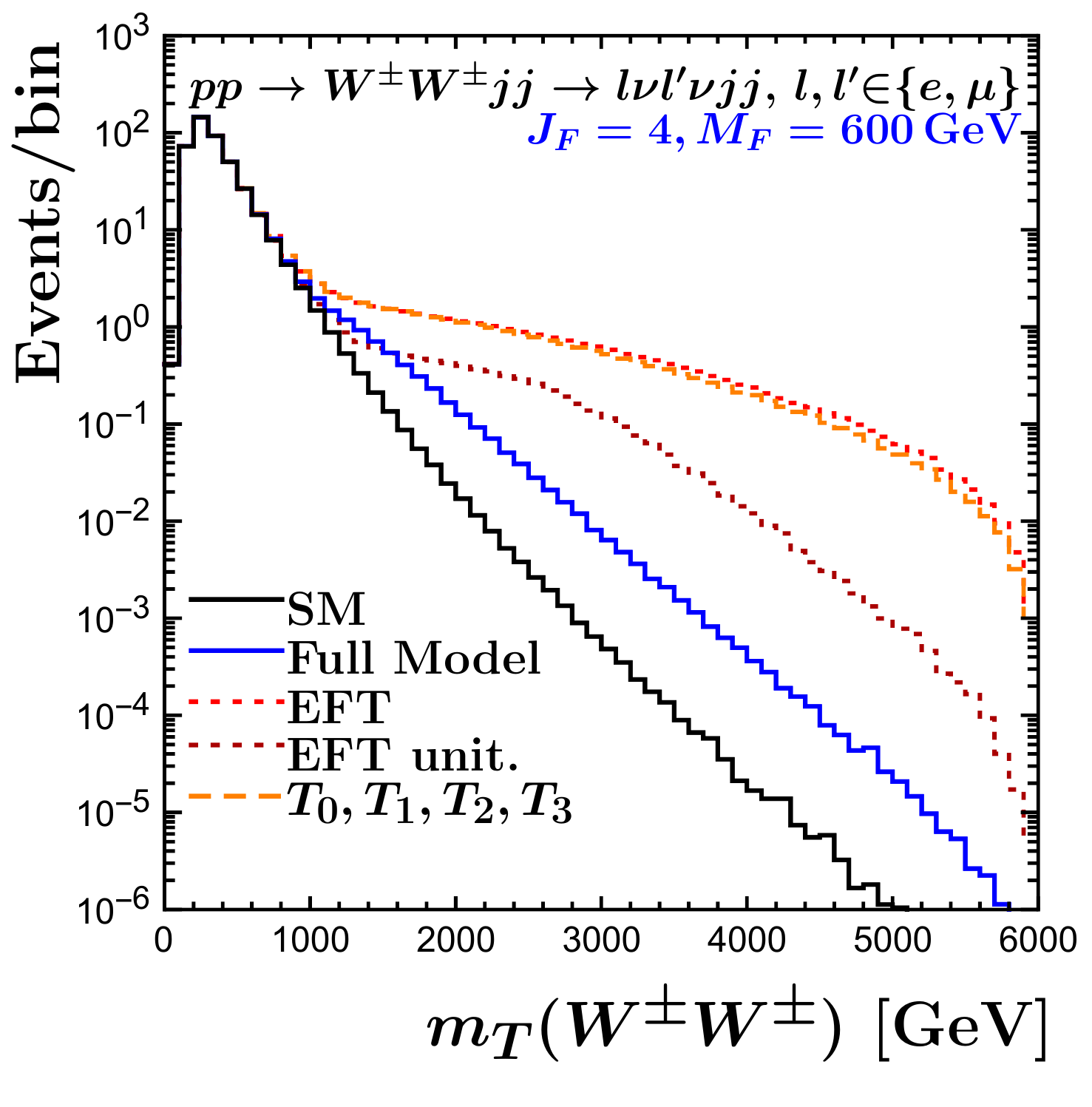}\hfill
\includegraphics[width=0.5\textwidth]{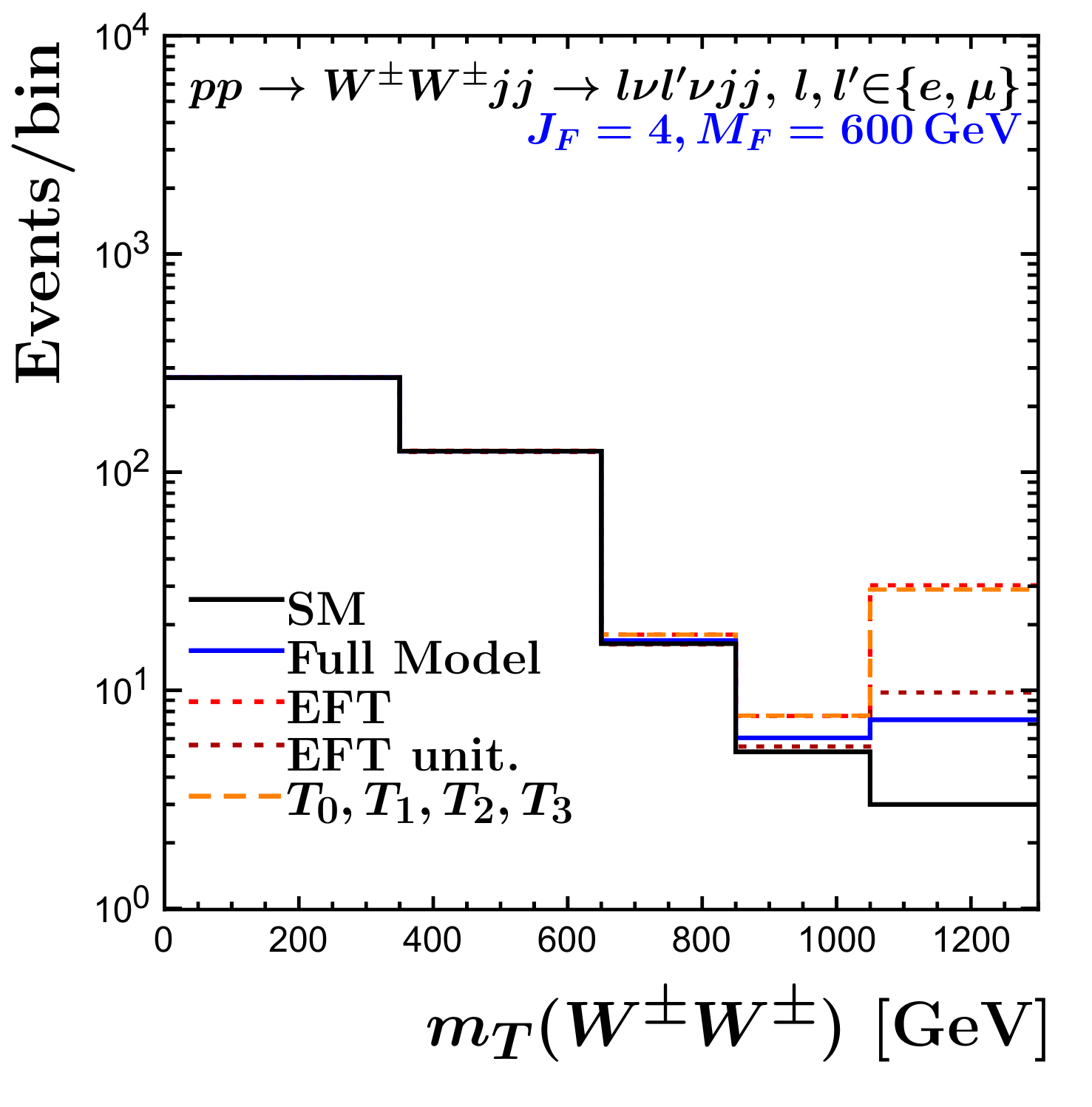}
\end{center}
\caption{Same as \fig{fig:VBFNLOWWs}, but for a fermionic multiplet with
  $J_F=4$ and $M_F=600$\,GeV.}
\label{fig:VBFNLOWWf}
\end{figure}

We continue with a description of the $W^\pm W^\pm$ final state, for which we
depict the scalar case in \fig{fig:VBFNLOWWs} and the fermionic case
in \fig{fig:VBFNLOWWf}.
The destructive interference in the invariant mass distribution is more
pronounced than in the $W^\pm Z$ case. 
Unfortunately, the actually observable transverse mass distribution suffers
from two final-state neutrinos which leave the detector unobserved. This again
leads to the migration of a substantial fraction of high-energy events in
the invariant-mass distribution to lower values of the transverse mass, which
wipes out the destructive interference signal. This migration also leads to
significant deviations between the full-model description and the EFT setup
for all values of the transverse mass, even far below threshold,
$m_T(W^\pm W^\pm)\ll 2M_R$.
Lastly, in the transverse-mass distribution also the unitarized EFT case
is far from both the full-model description and the EFT curve, as the
different behavior at high invariant-mass has a clear impact at lower
values of the transverse mass. The choice of the unitarization
procedure, which is arbitrary to a considerable extent, unfortunately
impacts the transverse-mass distribution substantially.
We note that the  di-lepton invariant mass distribution would show a behavior
very similar to the transverse mass, i.e. throughout the whole range of
masses the full model, the EFT description and its unitarized version differ.
Again the lower right panel shows the binning as performed in the left panel
of Fig. 6 in \citere{Sirunyan:2020gyx}.
In the next-to-last bin, from $850$ to $1050$\,GeV in $m_T(W^\pm W^\pm)$, all
curves for the $J_F=4$ fermionic case are compatible with the data, whereas
the measurement, with a single event in the $m_T(W^\pm W^\pm)>1050$\,GeV bin, 
is visibly under tension with all non-SM curves depicted in the lower right
panels of \fig{fig:VBFNLOWWs} and \fig{fig:VBFNLOWWf}.
It is again apparent that it is much easier to exclude the EFT
description compared to the full model or the unitarized EFT description.

\subsection{Constructive interference in $ZZjj$ production}

Within the {\tt VBFNLO} framework we next consider the electroweak process
\begin{align}\nonumber
 pp\to ZZjj\to l'^+ l'^- l^+ l^- jj
\end{align}
with $l, l'\in \{e,\mu\}$, which was used in
\citeres{Sirunyan:2017fvv,Sirunyan:2020alo} to bound aQGCs.
For a qualitative comparison we again use a cut-flow similar to the
experimental analysis, namely
\begin{align}\nonumber
  p_T^l&>20\,\text{GeV}\,, &\qquad   |\eta^e|&<2.5\,,&
  \qquad |\eta^\mu|&<2.4\\\nonumber
  40\,\text{GeV}<m_{ll}&<120\,\text{GeV}\,,&\qquad |\eta^j|&<4.7\,, &
  \qquad p_T^j&>30\,\text{GeV}\\
  \qquad |\Delta R(j,l)|&>0.3\,, &\qquad
  m_{jj}&>400\,\text{GeV}\,,&\qquad |\Delta\eta_{jj}|&>2.4\,.
 \label{eq:cutflowzz}
\end{align}
We generate the flavor combination $l'\neq l\in \{e,\mu\}$ and multiply
our results with a factor of $2$, thus ignoring the problem of correctly
assigning the leptons to the two $Z$ boson candidates in $4e$ or $4\mu$ events.
Note that the actual experimental analysis in \citere{Sirunyan:2017fvv} is
using a boosted decision tree, which also takes into account the 
$z_l^*$ variable of \eqn{eq:Zeppenfeld} to enhance the fraction of
vector boson scattering events over QCD background events.
The more recent analysis in \citere{Sirunyan:2020alo} is moreover extracting
the bounds not from a VBS-enhanced region, but from a larger phase space
region with only $m_{jj}>100$\,GeV. We stick to the VBS-enhanced region
defined above for our qualitative discussion, because otherwise triple gauge
boson production processes would have to be considered as well.
As a consequence we cannot directly compare against
\citere{Sirunyan:2020alo}, which would anyhow need a detailed
Monte Carlo simulation of all underlying, dominant background processes.

\begin{figure}[tb!]
\begin{center}
\includegraphics[width=0.33\textwidth]{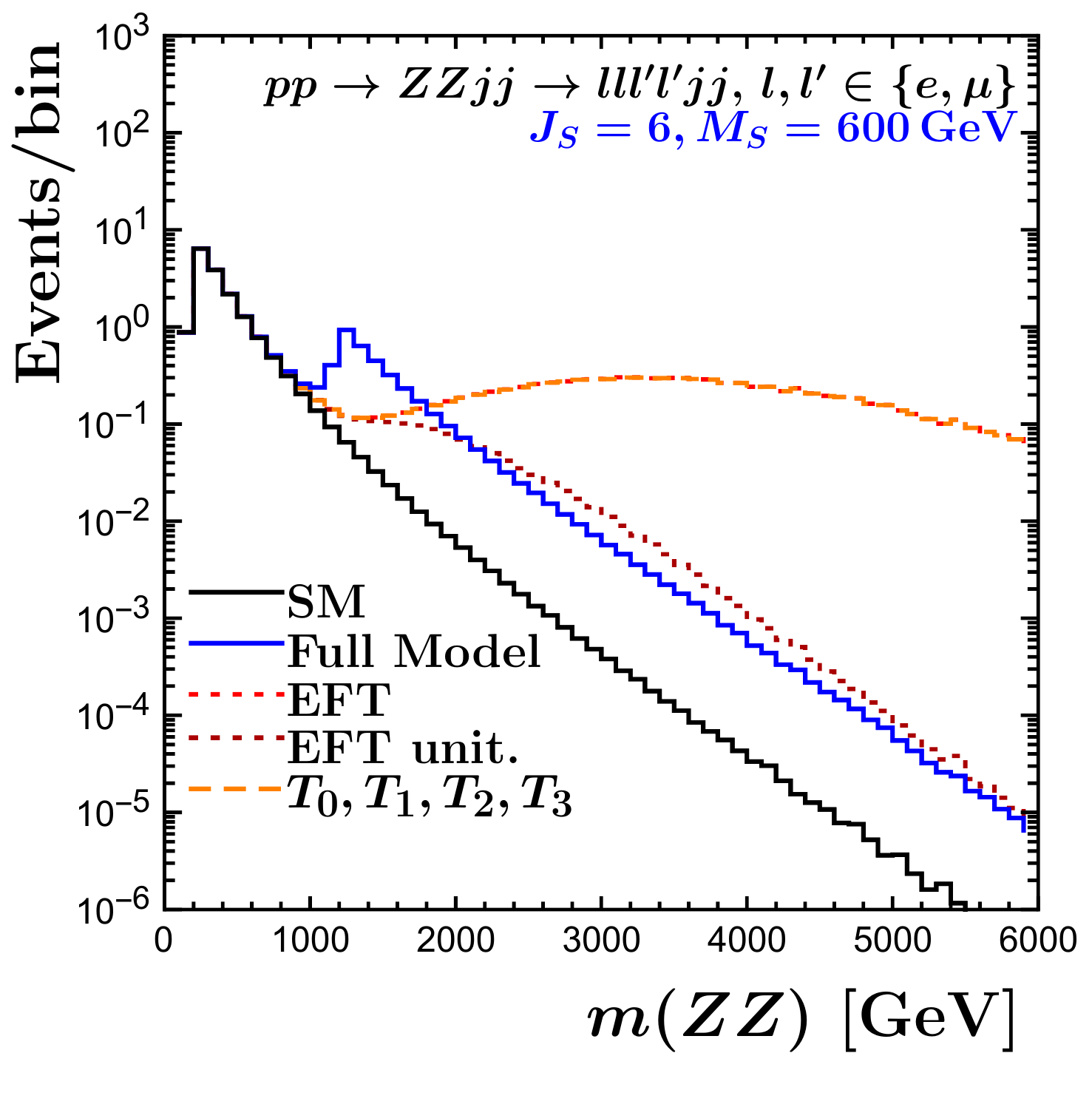}\hfill
\includegraphics[width=0.33\textwidth]{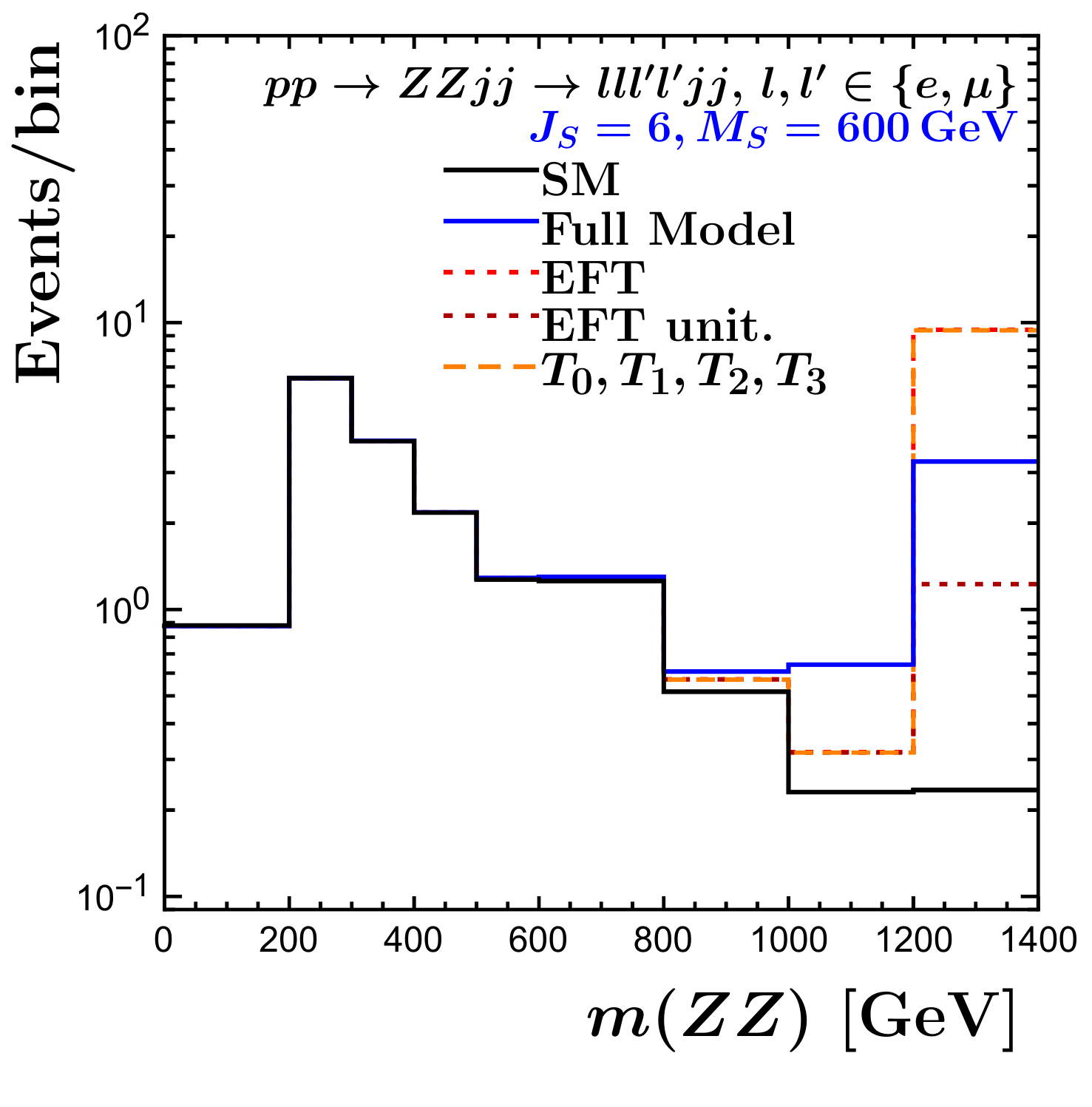}\hfill
\includegraphics[width=0.33\textwidth]{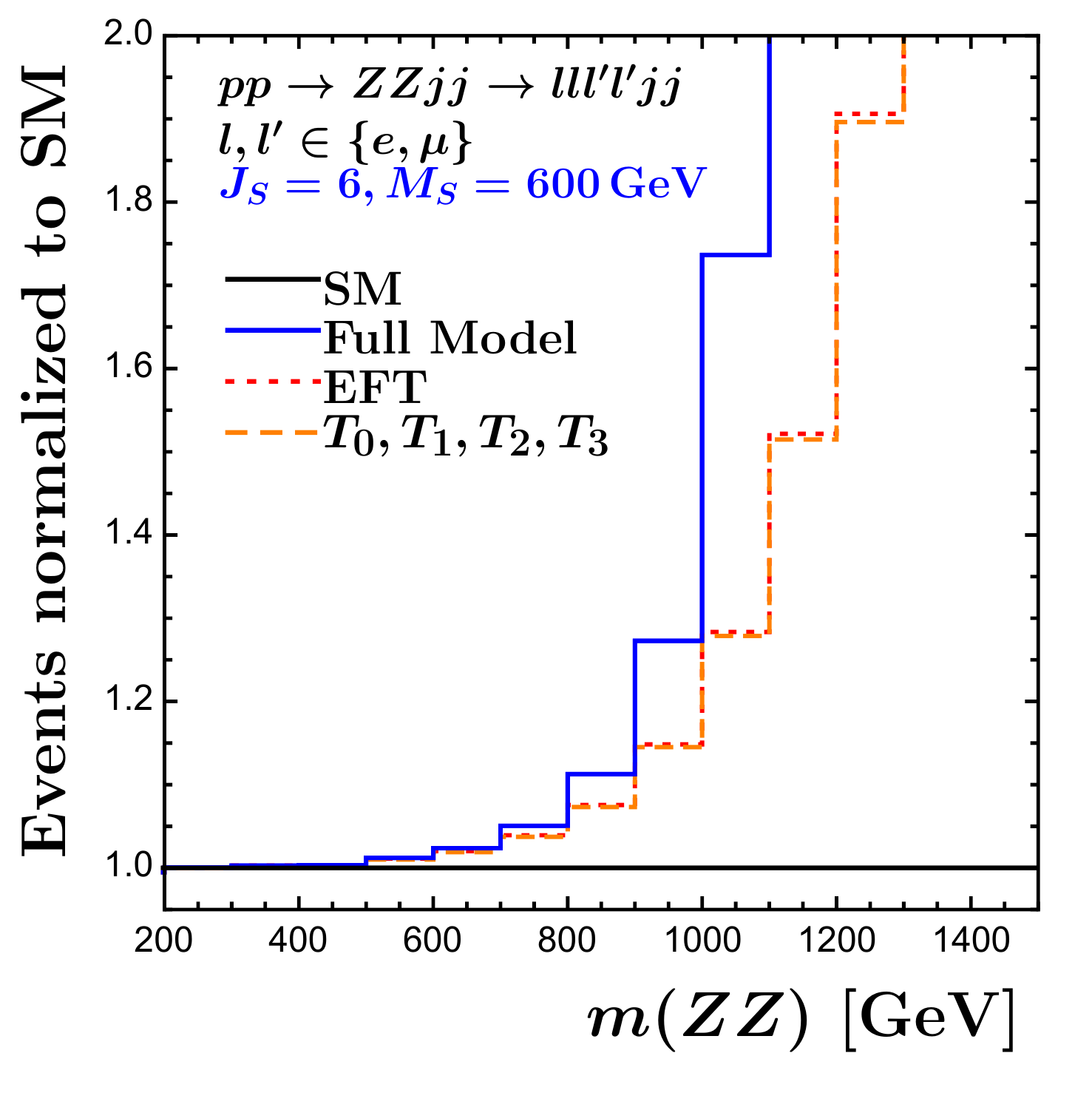}
\end{center}
\caption{
  Diboson invariant mass distributions for $ZZjj$ VBS events at the LHC.
  Shown are events per $100$\,GeV bin (left), events per bin, sized as in
  \citeres{Sirunyan:2017fvv,Sirunyan:2020alo} (middle), each for an integrated
  luminosity of $137$\,fb$^{-1}$, and cross section normalized to the SM
  (right). The considered scenario is a fermionic multiplet with
  $J_S=6$ and $M_S=600$\,GeV. VBS cuts of \eqn{eq:cutflowzz} are applied.}
\label{fig:VBFNLOZZs}
\end{figure}

In \fig{fig:VBFNLOZZs} we show the $ZZ$ invariant-mass distribution
for the scalar case with $J_S=6$ and $M_S=600$\,GeV which, in contrast to
$WW$ or $WZ$ production, is fully accessible experimentally for $ZZ\to 4l$.
While the left panel is
for a constant 100~GeV bin width, the central panel uses the binning
employed in \citeres{Sirunyan:2017fvv,Sirunyan:2020alo}, and the ratio
to the SM is shown in the right panel. In contrast to the same-sign $WW$
and $W^\pm Z$ final states the interference with the SM is constructive
(see right panel) and the peak itself is more pronounced (see left panel).
We emphasize again that a direct comparison to the experiments is non-trivial
due to different regions in the phase space of the two jets. Still both the
full model as well as the unitarized EFT description yield about three extra
events in the last bin ($m(ZZ)>1200$\,GeV), which seems well compatible with
the experimental data. The qualitative behavior of the $J_F=4,\, M_F=600$~GeV
model is similar to the scalar model shown. However, deviations from the SM
are somewhat smaller, with a single excess event expected in the last bin.

%%%%%%%%%%%%%%%%%%%%%%%%%%%%%%%%%%%%%%%%%%%%%%%%%%%%%%%%%%%%%%%%%%%%%%%%%%%%%%%

\subsection{Implications for experimental analysis}
\label{sec:implications}

The various cases discussed in the previous sections differ in details like
constructive vs. destructive interference below threshold, or how large a
threshold peak may be expected. However, there are a number of generic
observations which can be made and which are important for BSM searches in
VBS at the LHC.
\begin{enumerate}
\item One finds a very limited energy range, $m_{VV}\lesssim 1.3 M_R$ for
  the diboson invariant masses, where the \eft{} description (even at
  dimension-8 level) adequately approximates the underlying UV-complete
  model. Within this validity range of the \eft{} description, cross section
  deviations from the SM stay below 10\% even for isospins as large as
  $J_F=4$ or $J_S=6$, which are at the edge of perturbative behavior of the
  electroweak interactions. For less extreme isospin
  choices, deviations are even smaller, scaling like $J_R^5$.
\item In spite of small BSM effects in the \eft{} validity range, the
  overall BSM signal can be sizable. Enhancements by a factor of 10 in
  invariant mass or transverse mass distributions are possible, starting
  in the threshold region of the new physics. However, they require large
  isospin, $J_R$, close to the perturbativity limit. For more ``reasonable''
  multiplet assignments, the cross section increase at high  $m_{VV}$ is
  substantially smaller. For example, three replicas of $J_F=2,\,M_F=600$~GeV
  Dirac fermions lead to an increase by about a factor 1.3 around 
  $m_{W^\pm W^\pm}\approx 3$~TeV in the $W^\pm W^\pm jj$ cross section, as
  compared to the factor 7 visible in \fig{fig:VBFNLOWWf}.
\item The non-unitarized \eft{} description gives a completely wrong account
  of the BSM physics at high $VV$ invariant mass. Due to the migration of the
  (fake) huge excess of high energy events to lower values of e.g.
  measurable transverse mass, this wrong description completely spoils most
  distributions. Non-unitarized \eft{} descriptions clearly should
  not be used.
\item Unitarized \eft{} approximations of the BSM physics fare somewhat
  better and may provide a qualitative description. However, agreement with
  the full model (as e.g. in the $ZZ$ invariant mass distribution of
  \fig{fig:VBFNLOZZs} above 2~TeV) is only accidental, and would not
  occur for smaller isospin representations.
\item The BSM signal in the complete model is most pronounced around pair
  production threshold. Thus searches (in particular for less extreme isospin
  choices) should be optimized for modest increases in rate at
  intermediate diboson invariant masses, around $m_{VV}\approx 2M_R$, instead
  of looking for huge rate increases in the highest energy bins.
\item In the threshold region, $1.5 M_R\lesssim m_{VV} \lesssim 3 M_R$,
  the \eft{}
  description strongly underestimates the BSM signal of our toy model. This
  intermediate energy range thus provides an attractive opportunity to search
  for extra isospin multiplets in VBS. Because of the growth of amplitudes
  with $J_R^5$ in VBS vs. $J_R^3$ in dilepton production (as discussed in
  \sct{sec:WCbounds}), VBS can actually be competitive with and complementary
  to analyses of Drell-Yan dilepton events in finding BSM hints in the
  $(J_R,M_R)$ parameter plane. 
\end{enumerate}  

The above observations were made for a toy model with a large SU$(2)_L$
multiplet of otherwise inert heavy scalars or fermions. Many modifications
of the model can be contemplated. For example, the new heavy matter fields
might come in multiplets of an additional (confining) gauge interaction, which
would add non-perturbative effects and resonances, similar to the formation
of heavy quarkonium states, like the $J/\psi$ or $\Upsilon$, in QCD. Borrowing
from quark-hadron duality in QCD, however, we would still expect the above
features to approximately hold when smearing $VV$ invariant mass distributions
over energy intervals which are larger than the spacing of the resonances
which are induced by the new confining interaction. In contrast, going to
higher SU$(2)_L$ multiplets than contemplated in e.g. \fig{fig:argand-circle},
i.e. entering the non-perturbative realm for electroweak interactions, might
lead to new features, not discussed in this paper. Even then, the bounds
established by the $T_u$-model unitarization of the dimension-8 \eft{} would
still present upper limits for observable cross sections in VBS.

\section{Discussion and conclusions}
\label{sec:conclusions}

In this paper we have investigated possible sources of anomalous quartic
gauge boson couplings and their impact on vector boson scattering processes
at the LHC. The $T$-operators of \eqn{eq:toperators} arise naturally, at the
one-loop level, in any BSM extension with extra fields which carry weak
isospin, and they modify the scattering of transversely polarized weak bosons.
Conversely, the appearance of field strength tensors in the $T$-operators
signifies that they must be loop-induced~\cite{Arzt:1994gp}. Within the setting
of renormalizable field theories, and ignoring the possibility of an extended
electroweak gauge group,\footnote{Embedding the SM SU$(2)_L$ in a larger
  non-abelian gauge group merely leads to additional heavy
  isospin $\onehalf$ vector bosons, which do not share the high
  multiplicity enhancement factors, $\sim J_R^5$, discussed in this paper
  for large $J_R$  matter multiplets.}
the extra scalar and spin $\onehalf$ matter fields, which we have
considered, constitute the most general source of these operators. We have
avoided operators involving the SM Higgs-doublet field, like the $M_0$- or
$S_0$-operators of \eqn{eq:EFTintro}, by only considering isospin and 
hypercharge representations which cannot form gauge invariant Yukawa type
interactions with the Higgs doublet field (and possibly SM fermions) and
by assuming small $H^\dagger H\Phi^\dagger \Phi$ couplings in the UV-complete
model.

Relaxing these
model constraints will not significantly alter our results for the
$T$-operators, and transverse VBS in general, as long as extra
interactions do not induce very large mass splitting within the new matter
multiplets. Our toy model represents a variant of
natural dark matter models in the spirit of \citere{Cirelli:2005uq}. However,
we have only considered relatively modest masses, of order 1~TeV, which
would provide only a fraction of the observed dark matter in the universe.
In a less constrained UV-complete model, in particular when allowing mixing
with SM matter fields, direct collider searches for the extra multiplets as
well as their dark matter impact would be strongly affected by additional
interactions, leading to a vast and rich phenomenology. We were not interested
in such issues here but rather have concentrated on the generic loop-induced
effects of extra matter multiplets for weak boson interactions.

The one-loop effects of generic, degenerate scalar or Dirac multiplets have been
calculated up to weak boson four-point functions, in the zero-hypercharge limit.
In addition, the \eft{} representation of these results was derived,
including all necessary operators up to energy dimension 8. Both the
full calculation as well as the \eft{} low-energy approximation were
implemented for on-shell VBS in the SU$(2)_L$ limit of the \sm{}.
In addition, an approximated version was added into
{\tt VBFNLO}~\cite{Arnold:2008rz,Baglio:2014uba} for the estimation of
its impact in current experiments at the LHC. Assuming a single multiplet
for simplicity, we have tuned the parameters of the multiplet, i.e. its
isospin $J_R$ and its mass $M_R$, to produce sizable, but not yet excluded
deviations from the \sm{} in VBS. Surprisingly, present VBS constraints are
of comparable strength as bounds from dilepton production at the LHC, and
considerably stronger than existing constraints from aTGCs as measured via
vector boson pair production.

This somewhat surprising result (given the much higher event rate for Drell-Yan
or $q\bar q \to VV$ production at the LHC as compared to VBS) is due to the
fact that large isospin multiplets induce one-loop effects in weak boson
two- and three-point functions which rise as $J_R^3$ only, while a $J_R^5$
rise is found for four-point functions and, thus, also aQGCs. In addition,
an accidental cancellation between the two relevant dimension-6 operators
leads to a particularly small value for the $\lambda$-aTGC (see
\eqn{eq:lambda_par}). Degenerate, non-mixing matter multiplets at one-loop
level thus provide another example that dimension-8 operators can be
more important than dimension-6 operators in LHC phenomenology. One should
keep in mind, however, that for small isospin multiplets (like
$J_R=\onehalf$ or $1$) the one-loop BSM effects on weak boson
vertex-functions are tiny, and models with large isospin matter fields
are required to produce noticeable effects in VBS.

An upper bound on reasonable values of $J_R$ is provided by unitarity
considerations, combined with perturbativity of the model. As shown in
\sct{sec:hel_proj}, the $J_R^5$ growth of the one-loop VBS amplitude leaves
the allowed range of the Argand circle for fermion multiplets with
$J_F\gtrsim 5$ and scalars with $J_S> 6$. 
We have therefore considered a fermion or scalar with the largest isospin
$J_R$ which is compatible with the unitarity bounds (i.e. $J_F=4$ and $J_S=6$)
and a mass value such that at least one of the model's Wilson coefficients
has a value in the ball-park of current \eft{} operator bounds obtained from
aQGC measurements at the
LHC~\cite{Sirunyan:2020gyx,Sirunyan:2017fvv,Aaboud:2016ffv,SummaryWebpage}.
Both particle species show comparable impact in VBS processes.

Turning to individual VBS processes, we observe constructive interference
with the \sm{} for the case of $ZZ$ production and destructive interference
for $WZ$ and same-sign $WW$ scattering. However, the characteristic
destructive interference in $WZ$ and same-sign $WW$ scattering is only visible
as a function of the invariant mass of the diboson pair,
which, unfortunately, is not a kinematic variable that is directly accessible
in purely leptonic decays involving neutrinos. These interference effects,
which remain at the 10\% level even for isospins at the perturbativity limit,
are thus hard to detect. Large BSM signals are possible in VBS at and above
pair production threshold of the new particles, however, and such effects
should be searched for at the LHC in the 1 to 2~TeV region.

As discussed in the last section, an \eft{} approximation to the loop effects
is only justified in the interference region, well below threshold. 
Dissecting the effects of the individual dimension-8 operators further,
one finds large destructive interference between the different ${T_i}$
operators in $WZ$ and same-sign $WW$ scattering for the fermionic case.
This implies that the experimental bounds on individual Wilson coefficients
for these operators are only of limited use and tend to overly restrain
the allowed parameter space of our fermionic model. 

Because the validity region of the \eft{} approximation is confined to
$m_{VV}\lesssim 1.3 M_R$, where the BSM effects are still small, the \eft{}
is not the appropriate tool to search for extra large multiplets of Dirac
fermions or scalars. Rather a full one-loop simulation of the extra matter
fields should be used. An approximate implementation in {\tt VBFNLO}, as
discussed in \sct{sec:VBFNLO} is available upon request and a full
implementation is envisaged for the future.

\subsection*{Acknowledgments}
This research is partially supported by the Deutsche Forschungsgemeinschaft
(DFG, German Research Foundation) under grant 396021762 - TRR 257.

\clearpage

\vfill

{\footnotesize
\bibliographystyle{utphys}
\bibliography{VBS_EFTvsModel}
}

\end{document}